\documentclass[12pt]{report}
\usepackage[nonstop]{feynmf}
\usepackage{amssymb,amsmath}

\begin{document}

\def\lf{16\pi^2}
\def\er{eqnarray}

\begin{titlepage}
\begin{center}
{\LARGE\bf Marginal Deformations of ${\cal N}=4$ SYM and of its
Supersymmetric Orbifold Descendants \\ }
\end{center}

\vskip 1.5in

\centerline{\bf Thesis for the M.Sc. degree } %
\centerline{\bf submitted to the} \centerline{\bf Scientific
Council
  of the Weizmann Institute of Science}

\vskip 1.5in

 \centerline{\bf by} \centerline{\bf Shlomo S.Razamat}
 \centerline{\bf Departament of Particle Physics}
 \centerline{\bf Weizmann Institute of Science}

\vskip .5in \centerline{\bf Advisor:} \centerline{\bf Dr. Ofer
Aharony}

\vskip 1.5in

\centerline{\Large\bf  December 2001}

\end{titlepage}

\newpage
\pagenumbering{roman}

\vskip .5in
 {\LARGE\bf Acknowledgments}
\vskip 1in

I would like to thank Dr. Ofer Aharony for his constant support
and guidance throughout the preparation of this work. I would also
like to thank all the students of the Physics Department with whom
I had pleasure of discussing Physics.

\newpage

$ $

\newpage

\tableofcontents

\clearpage

\newpage

$ $\\
$ $\\\addcontentsline{toc}{chapter}{Abstract}
 \vskip .5in
 {\LARGE\bf Abstract}
 \vskip 1in

  There was a great burst of research and interest in ${\cal N}=4$ Super
  Yang-Mills in recent years. Most of the interest is due to the
  relation of this theory to the {\bf AdS/CFT} correspondence (see
  ~\cite{a:OA},~\cite{a:Maldac}). ${\cal N}=4$ SU(N) SYM appears in this context as an effective
  description of {\bf N} coincident D3-branes.
  It was argued (~\cite{a:SilKach},~\cite{a:quiver},~\cite{a:quiver2}) that after orbifolding the
  configuration above one can obtain effective descriptions in
  terms of conformal field theories with less supersymmetries
  (${\cal N}=2,{\cal N}=1,{\cal N}=0$). There is a great interest in looking on marginal
  deformations of such theories. In the {\bf AdS/CFT} ~\cite{a:Maldac}
  correspondence such deformations on the field theory side correspond
  to moduli of the string theory. For instance, as we will see (see
  ~\cite{a:SilKach}), in the ${\cal N}=2$ $Z_k$ orbifold theory there are {\bf k} marginal deformations which don't
   break the ${\cal N}=2$ SUSY. These deformations correspond to the string
   coupling and the ALE blow-up modes on the string side.

    In this work we investigate marginal deformations, which keep at least ${\cal N}=1$ SUSY unbroken, of ${\cal N}=4$ SYM and
    its orbifold ${\cal N}=2,1$ descendants. There are well known
    \textit{exactly} marginal deformations
  of this type for the ${\cal N}=4$ SYM (~\cite{a:Leigh}), in our work
    we argue that these are the only \textit{exactly} marginal
  deformations of this theory. It was also argued
  (~\cite{a:N_plan1},~\cite{a:N_plan2}) that the planar diagram
  contribution in the orbifold theory is the same as in the ${\cal
  N}=4$
  theory, leading to a conclusion that in the large {\bf N} limit many correlation functions in these
  theories coincide up to some gauge coupling rescaling. One could expect that the orbifold theories possess similar
  \textit{exactly} marginal deformations. We will see that this is actually the
  case,
  and we get additional \textit{exactly} marginal operators from the twisted
  sectors. In some cases we find that the dimension of the space
  of \textit{exactly} marginal deformations at low orders in perturbation theory is
  smaller than the general analysis implies.
   Another interesting observation which is made is about
  gauge theories with gauge group  SU(N=3). Here we find
  a very large number of \textit{exactly} marginal deformations. These deformations
  can not be directly related to the string theory because the
  correspondence is well understood only in the large {\bf N} limit.

\newpage
\pagenumbering{arabic}

\chapter{General Introduction}

 \section{AdS/CFT}
   The aim of this work is investigating marginal deformations of
   a specific class of
  supersymmetric field theories. These theories appear in the framework
  of the {\bf AdS/CFT} correspondence. Here we will review the general idea of
  this correspondence and make a link to the specific research
  presented in the thesis.

  It was already since t'Hooft's~\cite{a:thooft} work anticipated
  that string theories are linked to gauge theories. The
  general idea of t'Hooft was based on two simple facts. The first fact is  that pure gauge theories consist of
  fields in the adjoint representation of the gauge group which
  appears in the product of fundamental and anti-fundamental
  representations (for SU(N)), and thus every adjoint index can be described by
  two indices - one fundamental and one anti-fundamental. In the
  Feynman diagrams we can describe each index by a line, so the
  propagators in pure gauge theories can be represented by two
  lines.
   The
  other simple fact is that in the double line notation we can attach for any
   Feynman diagram
  to every index loop a surface. Thus, we can view a diagram
  as a decomposition of some closed surface (for vacuum or gauge
  invariant diagrams). The main result here is that if we take the
  large {\bf N} limit with fixed $g_{YM}^2N$ then the {\bf N}
  dependence of any diagram will now be determined by
  the topology
  of the surface which it decomposes: the power of {\bf N} we get is
   $(2-2{\bf g})$, where {\bf g} is the genus of the surface.
  For a sphere {\bf g}=0, for a torus {\bf g}=1 etc, {\bf g} simply counts the handles of the
  surface (In other words: when we
  say that a diagram has a topology of surface ${\bf{\cal G}}$, we mean
  that it can be drawn on surface ${\bf{\cal G}}$ in double line
  notation without any line crossings).

  When the number of colors, {\bf N}, is taken
  to infinity\footnote{{\bf N} is taken to infinity while keeping ${\bf g_{YM}^2N}$ fixed.},
   one can expand the path integral in a power series in
  ${\bf {1\over N}}$, such that the leading contribution is of the planar diagrams.
   So we get a power series with powers being linear functions of
   the genera of oriented closed surfaces, exactly like in oriented closed string
   theory ( If we add matter fields in (anti)fundamental
   representation we get open surfaces, leading to an open string theory-like expansion, and if we look at gauge group SO(N)
    for example we get an unoriented string theories-like expansion, because
    here the adjoint is a product of two fundamentals). We stress that we
    don't see from here any well defined string theory appearing, but
    only that the perturbation series is very similar to the
    perturbation series of string theory, with the string coupling constant
    being ${\bf {1\over N}}$.

   It was Maldacena's work~\cite{a:Maldac} that for the first time translated
   t'Hooft's idea of similarity between large N gauge theories and
   string theories to a definite, although still conjectured, relation between a subclass of conformal field theories
   and a class of well defined string theories.
    Maldacena's conjecture was based on the
   following observation. In superstring theories appear
   solitonic, non-perturbative, objects called Dq-branes. These
   objects have at least two descriptions:

   \begin{itemize}
   \item In string-perturbative language they are defined as manifolds ( extended in q directions) on which
     an open string can end.

   \item In the supergravity language, which is
   supposed to describe the low energy limit of string theories,
   they are defined as extended (in q directions) black hole solutions.
   \end{itemize}

   We now look at a system of N coincident D3-branes. In the
   string-perturbation theory
   language, in the low energy limit, the physics of the system is
   described by ${\cal N}=4$ SYM with U(N) gauge group\footnote{ The U(1) part is free so we will discuss essentially
   only the SU(N) part.} on the brane and by supergravity
   in the bulk, with these two systems decoupled. It is well known
   that in order for field theory perturbation theory to
   work ${\bf g_{YM}^2N}$ should be much smaller than one. In the
   supergravity language we will have some black hole solution, which in
   the near horizon limit is described by ${\bf AdS_5\times S^5}$
   geometry. Here again we can describe our physics by two
   decoupled systems: supergravity in the bulk and the type IIB string
   theory on ${\bf AdS_5\times S^5}$. The supergravity solution is valid
   only if the radius of curvature is much larger than the string
   scale, which leads us to demand large {\bf N}, since the radius in string units of {\bf AdS} and the radius of the sphere are both
   proportional to ${\bf (g_{string}N)^{1\over 4}}$. Thus, we see that the same object is
   described on one hand by field theory and supergravity and on the other hand by
   string theory and supergravity. This led Maldacena to conjecture
   that :

     ${\cal N}=4$ d=4 SU({\bf N}) SYM is equivalent to type IIB string theory on
   ${\bf AdS_5\times
     S^5}$ in the large {\bf N} limit.

     There is also a stronger conjecture that these theories describe
     the same physics for every value of {\bf N}.

   There are many different indirect checks of this conjecture. One such check is the striking property of S-duality.
   S-duality relates two theories, one with small and the other with large coupling. Both ${\cal N}=4$ SYM and
   type IIB string theory are believed to be self dual under the
   S-duality.

     The ${\bf AdS/CFT}$ correspondence relates expectation values in
     string theory to coupling constants in the field theory. For
     instance we get from the correspondence that
     ${\bf g_{YM}^2\propto g_{string}}$, and from string theory we know that
     ${\bf g_{string}}$ is related to the vacuum expectation value of the
     dilaton field. Thus we conclude that changing the gauge coupling on
     the field theory side, which is done by adding some marginal
     operator, is equivalent to changing the expectation value of
     the dilaton field on the string theory side. The
     marginal operators of the field theory are related to some
     moduli of the string theory\footnote{We need operators to be marginal in
     order not to spoil the conformal properties of the field
     theory.}. In general, scalar supergravity fields $\phi$ which live in {\bf AdS} couple
    to operators ${\cal O}$ which live on the boundary of {\bf AdS} via
    $\int_{\mathbb{R}^{3,1}}\phi_0{\cal O}$, where $\phi_0$ is a
    restriction of $\phi$ to the boundary (up to some power of the radial coordinate). The dimension of ${\cal
    O}$, $\Delta$, is related to the mass $m^2$ of the scalar field by:

    \begin{eqnarray}
      m^2=\Delta(\Delta-d).
    \end{eqnarray}

    Here {\bf d} is the dimensionality of the space-time which the field theory lives
    in.
    We see that the massless, massive and tachyonic fields on the
   supergravity side correspond to marginal, irrelevant and relevant
   operators, respectively, on the field theory side.

    The classification of operators to marginal, relevant and
    irrelevant in this way is meaningful before we deform our theory
    with them. After we deform our theory with these oprators
    the conformal dimensions of operators can receive corrections (via the anomalous
    dimensions). The  marginality of an operator, as defined in the
    previous paragraph, can not
    assure that it will remain marginal after deforming the theory:
    the operators can be exactly marginal, marginally
    relevant or marginally irrelevant.

    On the field theory side adding an irrelevant operator strongly affects
     the UV limit of the theory. Thus, because usually we define
     field theories in the UV and then flow to the IR, it does not
     make sense to discuss theories with irrelevant deformations.
     On the other hand, relevant deformations affect weakly the UV
     limit but break the conformal invariance. Finally, the exactly marginal operators
     keep the conformal properties of the theory.

     On the string theory side giving a VEV to a massive field will change
     significantly the behavior on the boundary of {\bf AdS},
     which is equivalent to demanding a new UV description on the
     field theory side. The tachyonic fields will go to zero
     on the boundary, thus this deformation will affect only the
     interior and asymptotically we will still have an {\bf AdS} background.
     Giving a VEV to a massless field (if it corresponds to
     an \textit{exactly} marginal operator) will always leave us with
     an {\bf AdS} factor.

   Thus we see that by finding \textit{exactly} marginal operators on the field
   theory side we can learn about the moduli of string theory.

   Adding additional operators to the theory will in general change
   the supergravity background. The ${\bf AdS_5}$ space has as
   its symmetry group SO(2,4), which is exactly the conformal group
   in four dimensions (the boundary of  ${\bf AdS_5}$ is four dimensional). Thus, if we demand conformality, this factor
   will remain even after deforming the original theory. The
   second factor (the five-sphere) is related to the SU(4) global
   symmetry of the SYM in some sense, and thus can be and will be
   deformed after deforming the original theory, if we break some
   supersymmetry\footnote{The breaking of supersymmetry here is
   inevitable, because essentially there is only one renormalizable,
   consistent ${\cal N}=4$ theory in d=4 which is the SYM theory. Thus, by adding
   additional operators, other than the change of the gauge coupling, we always break the
   ${\cal N}=4$ SUSY.}. Thus, we can say
   that if we deform the original ${\cal N}=4$, d=4 SYM by some marginal
   operators, then on the string theory side we have to deform the
   supergravity solution: ${\bf AdS_5\times S^5\to AdS_5\times M}$, where
   {\bf M} is some five dimensional compact manifold (For marginal
   deformations of ${\cal N}=4$ SYM from supergravity side see for
   example~\cite{a:petrini}), and sometimes we will also have to turn on some fields.

    It is really striking that two so different mathematical tools
    like string theory and field theory may describe the same
    object.

    To summarize, learning the marginal deformations of the field
   theory side can contribute to a better understanding of string
   theories in general and of the {\bf AdS/CFT} correspondence in
   particular.

   Another motivation comes from pure field theoretic
   considerations. New {\bf CFT}s can be used as UV fixed points,
   leading in the IR to various field theories, including perhaps phenomenologically interesting theories.

   \newpage

   \section{Marginal operators}

    In this work we discuss the marginal deformations of field
    theories coming from a system of N coincident D3-branes.  The whole
    discussion is done from the field theory perspective. The
    theories we discuss have ${\cal N}=4,2,1$ supersymmetry.

   An operator is \textit{exactly} marginal if upon adding it to the original conformal theory all the $\beta$-functions
    still vanish. Generally if we have {\bf p} couplings in the theory
    we also have {\bf p} $\beta$-functions. The conditions for the theory
    to be conformal are:

\begin{eqnarray}\label{set}
 0&=&\beta_{g_1}(g_i,h_j)\nonumber\\
 &\vdots&\nonumber\\
 0&=&\beta_{g_n}(g_i,h_j)\nonumber\\
 0&=&\beta_{h_1}(g_i,h_j)\nonumber\\
 &\vdots&\nonumber\\
0&=&\beta_{h_k}(g_i,h_j)
\end{eqnarray}

(Here $h_i$s are the couplings and $g_i$s are the gauge couplings
of the system.) We have {\bf n+k} equations in {\bf n+k}
variables.
    Thus, in general, we expect
    to have isolated, if at all we will have any, solutions of
    this system of equations.

    However, in
    supersymmetric field theories we have several simplifications.
    The first one is that from nonrenormalization of the
    superpotential in supersymmetric theories we get a relation between the anomalous dimensions
    of the fields and the coupling
    associated with the superpotential term(see~\cite{a:west}).
    For a superpotential $W={1\over6}Y^{ijk}\Phi_i\Phi_j\Phi_k$ we get:

\begin{equation} 
\beta_Y^{ijk}= Y^{p(ij}\gamma^{k)}_p =
Y^{ijp}\gamma^k_p+(k\leftrightarrow i)+(k\leftrightarrow j).
\end{equation}

Here $\gamma^k_p$ is the anomalous dimension related to the
$\langle\Phi^{\dag k}\Phi_p\rangle$ Z factor.
 The second simplification is the relation between the gauge
 coupling and the anomalous dimensions - the NSVZ
 $\beta$-function
 (~\cite{a:NSVZ1},~\cite{a:NSVZ2},~\cite{a:NSVZ3}),

\begin{equation}
\beta_g = {{g^3}\over{\lf}}\left[ {{Q- 2r^{-1}Tr\left[\gamma
C(R)\right]}
    \over{1- 2C(G)g^2{(\lf)}^{-1}}}\right].
\end{equation}

 The symbols appearing here will be defined later.
 We conclude that in a supersymmetric field theory, in order to find \textit{exactly} marginal
deformations we have to solve a set of linear equations in the
anomalous dimensions. These equations can be linearly dependent,
giving a manifold of solutions ~\cite{a:Leigh}. The equations
(~\ref{set}) become:

\begin{eqnarray}\label{set1}
 0&=&\beta_{g_1}(\gamma_l,g_i,h_j)\nonumber\\
 &\vdots&\nonumber\\
 0&=&\beta_{g_n}(\gamma_l,g_i,h_j)\nonumber\\
 0&=&\beta_{h_1}(\gamma_l,g_i,h_j)\nonumber\\
 &\vdots&\nonumber\\
0&=&\beta_{h_k}(\gamma_l,g_i,h_j)
\end{eqnarray}

 Here usually we will get that the righthand sides of these equations
 depend only on $\gamma$s which will greatly simplify our job.

 In order to find \textit{exactly} marginal
directions we have to solve a set of linear equations, to find the
possible values for the anomalous dimensions ($\gamma$s) such that
all $\beta$s vanish. Then, by loop calculations we calculate the
dependence of the $\gamma$s on the couplings and other parameters
of the theory, and finally we impose the conditions from the first
step on the $\gamma$s and see if they can be satisfied. This will
be the strategy in our search for \textit{exactly} marginal
deformations throughout this work.

  When solving the set of linear equations (~\ref{set1}) we can get
  possible solutions which will be ruled out from the loop
  calculations\footnote{ We can count on the loop calculations
  only in the weak coupling regime. Thus, we can not rule out these
  solutions from appearing in the strong coupling regime.}. We will see
  examples of this below.

\newpage

\chapter{${\cal N}=4$ theory}

First we review some basic properties of ${\cal N}=4$ SYM with
gauge group SU(N).
 In ${\cal N}=0$ language the theory contains six
scalar fields, four Weyl fermions and a real vector field. All
fields are in the adjoint representation of $SU(N)$. In ${\cal
N}=1$ language the six scalars can be coupled to form three
complex scalars which together with three Weyl fermions form three
chiral superfields $\Phi_i$, while the vector and the remaining
Weyl spinor can be joined to form a vector superfield $V$. The
Lagrangian in ${\cal N}=1$ language is then:

\begin{eqnarray}\label{CDFval}
  L& =& \int{d^4\theta
  {\sum}_{i=1}^{3}{Tr(e^{-gV}\bar{\Phi}_{i}e^{gV}{\Phi}^i})}+\frac{1}{64g^2}
\int{d^2{\theta}Tr(W^{\alpha}W_{\alpha})}\nonumber\\
& &+\left(\frac{ig\sqrt{2}}{3!}\int{d^2{\theta}{\epsilon_{ijk}Tr({\Phi}^i[{\Phi}^j,{\Phi}^k])}
    +h.c.} \right).
\end{eqnarray}

Here traces are taken in the fundamental representation of SU(N).

This is a pure Yang Mills theory with sixteen supercharges, in
particular for $U(1)$ gauge group this theory becomes free. The
$\beta$-function of the gauge coupling vanishes identically (at
one loop it's a trivial consequence of having three chiral
superfields in the adjoint representation), thus it is a conformal
theory. It is believed to be exactly self S-dual. This symmetry of
the theory exchanges the strong coupling regime with a weak
coupling regime, and the perturbative, electric, degrees of
freedom with non-perturbative, magnetic degrees of freedom.

 In string theory we get ${\cal N}=4$ SYM with SU(N) gauge group by
putting N D3-branes in type IIB string theory together. In this
picture we have six "vibrational" modes of the branes ( which are
related to the six transverse directions to the brane) which
become six scalars, which in turn when joined in pairs comprise
the ${\cal N}=1$ scalar part of three complex chiral
supermultiplets. The possibility of the fundamental string to end
on one of the N branes gives an SU(N) gauge group and puts the
scalars (as well as the other fields) in the adjoint
representation. To all these integer spin fields we have fermionic
counterparts, and all in all we get an SU(N) gauge group in 4d
with three chiral and one vector multiplets. In type IIB
superstrings we have 32 supercharges, D-branes are BPS states and
thus they break half of the supersymmetry. Finally we have 16
supercharges in d=4 which give us ${\cal N}=4$ supersymmetry.
Three ${\cal N}=1$ chiral multiplets and the vector multiplet in
${\cal N}=4$ language give an ${\cal N}=4$ vector multiplet. Thus
to summarize we get on the D-branes ${\cal N}=4$ pure Yang Mills
with SU(N) gauge group.

 Another, related, way to obtain ${\cal N}=4$ SYM in d=4 is~\cite{a:red} to look at pure
 ${\cal N}=1$ SYM in d=10 and then do the dimensional reduction procedure
 to d=4. In d=10 we had only the vector multiplet, six scalar components
 of which lose their vector nature after the reduction. They can
 be coupled in pairs to form complex scalars which will be the
 scalars of three chiral ${\cal N}=1$, d=4 multiplets. Of course in this
 procedure we have a global $SO(6)\sim SU(4)$ symmetry, which becomes the ${\cal R}$ symmetry of
 ${\cal N}=4$.

 There is extensive literature on this field theory, in particular
 regarding its finiteness and the exact S-duality of this theory.
 There is also research concerning the relevant deformations of
 ${\cal N}=4$ (~\cite{a:mass1},~\cite{a:mass2},~\cite{a:mass3} for example). Relevant deformations break
 the conformal invariance by introducing a scale to the theory. We will
 be interested only in marginal deformations throughout this work.

\newpage
\section{Marginal deformations}

In this section we investigate some \textit{exactly} marginal
deformations of ${\cal N}=4$ SYM. There are essentially only three
types of marginal deformations which one can add to the lagrangian
above\footnote{There are also relevant deformations, inserting
mass terms for the fields, and they were discussed
in~\cite{a:mass1},~\cite{a:mass2}.}. The obvious deformation is
just changing the gauge coupling constant, the two other types are
superpotentials of the form \footnote{We will assume everywhere,
unless stated otherwise, that the couplings are real. The
extension to complex couplings is trivial $\to$ the actual manifold of
fixed points is a complex manifold of same dimensions.}:

\begin{eqnarray}
  \frac{i\delta\lambda\sqrt{2}}{3!}\epsilon_{ijk}Tr({\Phi}^i[{\Phi}^j,{\Phi}^k])\nonumber\\
  \frac{h_{ijk}}{3!}Tr({\Phi}^i\left\{{\Phi}^j,{\Phi}^k\right\}),
\end{eqnarray}

where $h_{ijk}$ is totally symmetric. These operators are marginal
(by power counting), obey gauge invariance and preserve ${\cal
N}=1$ supersymmetry. So by adding these operators we get ${\cal
N}=1$ SQCD. What has to be determined is under what conditions
these marginal deformations are \textit{exactly} marginal, i.e.
the $\beta$-functions vanish to all orders in perturbation theory.

 ${\cal N}=1$ SQCD was analyzed for
general superpotentials and general simple gauge group
G(~\cite{a:beta} and references therein). We will briefly
summarize the general results:

We write the superpotential as:

\begin{equation}
  W={1\over6}Y^{ijk}\Phi_i\Phi_j\Phi_k.
\end{equation}

We assume that the gauge group is simple and that there are no
gauge singlets. The $\beta-function$ of Y can be written in terms
of the anomalous dimensions:

\begin{equation} \label{betaY}
\beta_Y^{ijk}= Y^{p(ij}\gamma^{k)}_p =
Y^{ijp}\gamma^k_p+(k\leftrightarrow i)+(k\leftrightarrow j).
\end{equation}

The one loop gauge $\beta$-function and the anomalous dimensions
are given by:

\begin{equation}\label{1loop}
\lf\beta_g^{(1)}=g^3Q,\quad\hbox{and}\quad
\lf\gamma^{(1)i}_j=P^i_j,
\end{equation}

where we have defined:

\begin{equation}
  Q=T(R)-3C(G),\quad\hbox{and}\quad
  P^i_j={1\over2}Y^{ikl}Y_{jkl}-2g^2C(R)^i{}_j,
\end{equation}

and:

\begin{equation}
T(R)\delta _{AB} = Tr(R_A R_B),\quad C(G)\delta _{AB} = f_{ACD}f_{BCD}
\quad\hbox{and}\quad C(R)^i_j = (R_A R_A)^i_j.
\end{equation}

$R_A$ is the representation of the chiral superfields. A,B,C,D are
indices in the adjoint representation. For the gauge coupling we
use the NSVZ
$\beta-function$~\cite{a:NSVZ1},~\cite{a:NSVZ2},~\cite{a:NSVZ3}:

\begin{equation}\label{NSVZ_1}
\beta_g =
{{g^3}\over{\lf}}\left[ {{Q- 2r^{-1}Tr\left[\gamma C(R)\right]}
    \over{1- 2C(G)g^2{(\lf)}^{-1}}}\right],
\end{equation}

(here $r=\delta_{AA}$) which at one loop gives (as
in(~\ref{1loop})):

\begin{equation}
\lf\beta _g^{(1)}=g^3Q.
\end{equation}

Now we have set the general stage and return to the specific
marginally deformed ${\cal N}=4$ theory. The superpotential can be
rewritten in the form:

\begin{eqnarray}
  W&=&(-\lambda\sqrt {2}f_{abc}\epsilon_{ijk}+d_{abc}h_{ijk}){1\over6}
  \Phi _i^a\Phi _j^b\Phi _k^c, \\
  Y^{ijk}_{abc}&=&(-\lambda \sqrt{2}f_{abc}\epsilon
  _{ijk}+d_{abc}h_{ijk}),
\end{eqnarray}

where:

\begin{eqnarray}
  d_{abc}\equiv Tr\left[T_a\left\{T_b,T_c\right\}\right].
\end{eqnarray}

Here $T_a$ are the generators of the Lie algebra of G in the
fundamental representation of the group and
$\lambda=g+\delta\lambda$. The groups for which $d_{abc}$ is not
vanishing are only $SU(N\geq 3)$ or $E_8$. Here we discuss only
the SU(N) case.

$\Phi_i$ are in the adjoint of SU(N):
\begin{equation}
  R_A=
\left(\begin{array}{ccc}
T^{adj}_A & 0 &0  \\
0 &T^{adj}_A  &0  \\
0 & 0 &T^{adj}_A
\end{array}\right)
\end{equation}

Here $T^{adj}_A$ are the adjoint representation matrices. From
here we can calculate the parameters appearing in the general
setup above:

\begin{eqnarray}
  C(G)\delta_{AB}&=&f_{ACD}f_{BCD}\equiv C_1\delta_{AB}\\
  T(R)\delta^A_B&=&Tr(R_AR_B)=3Tr(T^a_AT^a_B)=3C_1\delta^A_B(=-3f_{ACD}f_{BDC})\nonumber\\
  C(R)^{Ai}_{Bj}&=&(R_DR_D)^{Ai}_{Bj}=C_1\delta^A_B\delta^i_j\nonumber\\
  r&=&N^2-1.\nonumber
\end{eqnarray}

$C_1$ depends on the normalization of the Lie algebra generators,
for the moment we will keep it arbitrary which will not affect our
results. The one loop gauge $\beta-function$ is proportional to Q:

\begin{equation}
Q=T(R)-3C(G)=3C_1-3C_1=0.
\end{equation}

So the gauge $\beta-function$ vanishes at one-loop. The gauge
$\beta-function$ vanishes at one loop in general gauge theories
with three chiral superfields in the adjoint representation.

 First we do the general Leigh-Strassler analysis~\cite{a:Leigh}.
 We have here:

 \begin{eqnarray}
   \beta_g&\propto&Tr\gamma\nonumber\\
   \beta_{\delta\lambda}&\propto&Tr\gamma
 \end{eqnarray}

So in general we have 10 $h_{ijk}$s, $\delta\lambda$ and the gauge
coupling, total of 12 couplings, we have to demand that
$Tr\gamma=0$ and $\beta_{h_{ijk}}=0$ giving a total of 11
conditions. So we expect a one dimensional manifold of the fixed
points which we have already in ${\cal N}=4$ and it is
parameterized by the gauge coupling, with
$\delta\lambda=h_{ijk}=0$. But we can do a more complicated thing.
 If we assume also that $\gamma$ is proportional to identity
matrix\footnote{ There are also other restrictions we can make on
the $\gamma$s and get the same dimensionality of the manifold of
fixed points, but they all are related by the global SU(3)
symmetry we have here.} we get $\beta_{h_{ijk}}\propto Tr\gamma$.
So we will have 12 couplings, one condition $Tr\gamma=0$ and 8
conditions for $\gamma_i^j\propto\delta^j_i$, giving a total of
12-8-1=3 free parameters. So we expect to have a three dimensional
manifold of exactly marginal deformations. We will see below that
we essentially get only these three marginal directions.

Now we continue with the perturbation theory analysis. For SU(N) :
$d_{acd}d_{bcd}=2\frac{N^2-4}{N}C_2^3\delta_{ab}$, where
$C_2\delta_{ab}\equiv Tr(T_aT_b)$. So we can write:

\begin{equation}
  P^{ai}_{bj}=(2C_1(\lambda^2-g^2)\delta_{ij}+\frac{N^2-4}{N}C_2^3h^{(2)}_{ij})\delta_{ab}.
\end{equation}

Here $h^{(2)}_{ij}\equiv h_{ilm}h^*_{jlm}$. And finally we get the
one-loop anomalous dimensions and $\beta-functions$:

\begin{eqnarray}
  \gamma^{(1)ai}_{bj}&=&\frac{1}{\lf}(2C_1(\lambda^2-g^2)\delta_{ij}+\frac{N^2-4}{N}C_2^3h^{(2)}_{ij})\delta_{ab}\\
  \beta^{(1)ijk}_{abc}&=&\frac{1}{\lf}\left\{6C_1(\lambda^2-g^2)Y^{ijk}_{abc}+\frac{N^2-4}{N}C_2^3\left(-\sqrt{2}\lambda
  f_{abc}\tilde{h}_{ijk}+d_{abc}h^{(3)}_{ijk}\right)\right\}\\
h^{(3)}_{ijk}&\equiv&h^*_{plm}(h_{ijp}h_{klm}+h_{kjp}h_{ilm}+h_{ikp}h_{jlm})\\
\tilde{h}_{ijk}&\equiv&\epsilon_ {ijp}h^{(2)}_{kp}+\epsilon
_{pjk}h^{(2)}_{ip}+\epsilon _{ipk}h^{(2)}_{jp}
\end{eqnarray}

$\tilde h_{ijk}$ is totally antisymmetric, thus because (i j k)
run over (1 2 3), $\tilde h_{ijk}$ has only one independent
component:

\begin{eqnarray}
\tilde{h}_{123}&=&\epsilon_{123}h^{(2)}_{33}+\epsilon_{123}h^{(2)}_{22}+\epsilon_{123}h^{(2)}_{11}=Tr(h^{(2)})\\
\tilde{h}_{ijk}&=&Tr(h^{(2)})\epsilon_{ijk}.
\end{eqnarray}

Now we can look separately on the part of the $\beta-function$
proportional to $f_{abc}$ and on the part proportional to
$d_{abc}$:

\begin{eqnarray}
\beta^{(1)}_{\lambda}&=&\frac{\lambda}{\lf}\left\{6C_1(\lambda^2-g^2)
    +\frac{N^2-4}{N}C_2^3Tr(h^{(2)})\right\}\\
\beta^{(1)}_{ijk}&=&\frac{1}{\lf}\left\{6C_1(\lambda^2-g^2)h_{ijk}
  +\frac{N^2-4}{N}C_2^3h^{(3)}_{ijk}\right\}.
\end{eqnarray}

When we constrain ourselves only to the case of
$h_{123},h_{111}=h_{222}=h_{333},\lambda$ non zero (which is the
only case where we will get \textit{exactly} marginal deformations
as we will see later), we get:

\begin{equation}
h^{(3)}_{ijk}=h_{ijk}Tr(h^{(2)}).
\end{equation}

And in this case:

\begin{equation}
  \frac{\beta_{\lambda}}{\lambda}=\frac{\beta_{ijk}}{h_{ijk}}.
\end{equation}

So if we are looking for fixed points we have only one condition
on four couplings, and thus we have a three dimensional manifold
of fixed points in the coupling constants space~\cite{a:Leigh}.

\section{RG flow analysis}

Here we will analyze the $\beta-functions$ obtained in the
previous section. The equations will simplify if we rescale the
coupling constants:

\begin{equation}
g\to\frac{\sqrt{C_1}}{4\pi}g \quad
\lambda\to\frac{\sqrt{C_1}}{4\pi}\lambda  \quad\hbox{and}\quad
h_{ijk}\to\frac{\sqrt{C_2^3}}{4\pi}\sqrt{\frac{N^2-4}{N}}h_{ijk}.
\end{equation}

The $\beta-functions$ become:

\begin{equation}
\beta_g=-\frac{2g^3}{1-2g^2}Tr\gamma, \quad
\beta_{\lambda}=\lambda Tr\gamma.
\end{equation}

Here the trace is taken only over the SU(3) indices and not over
gauge indices. From these $\beta-functions$ we can obtain a
differential equation:

\begin{equation}
  -\frac{1}{2g^3}dg+\frac{1}{g}dg=\frac{d\lambda}{\lambda}.
\end{equation}

This can be easily solved to give:

\begin{equation}
  \frac{\lambda}{\lambda_0}=\frac{ge^{\frac{1}{4g^2}}}{g_0e^{\frac{1}{4g_0^2}}}.
\end{equation}

This result means that the RG flow lines in the $\lambda-g$ plane
are exactly known (to the extent that we can count on the NSVZ
$\beta-function$). It is easy to convince oneself that there is no
line with the couplings going to zero in the UV, except the
trivial case when one of the couplings is constantly zero. This
implies that there is no choice of coupling constants for which
this theory is asymptotically free.

Another interesting question is the existence of fixed points. In
order to have a fixed point we have to satisfy $Tr\gamma=0$ which
implies at one loop that:

\begin{equation} \label{condFix}
  Tr(h^{(2)})=-6(\lambda^2-g^2)
\end{equation}

And we can substitute this into $\beta_{ijk}$ to get another
condition:

\begin{equation}
Tr(h^{(2)})h_{ijk}=h^{(3)}_{ijk}. \label{condFix_h}
\end{equation}

 We will argue that these conditions can be satisfied (in the
limit $g\to 0$ ) only if the anomalous dimensions matrix is
proportional to identity matrix.

First we don't assume any special property of $\gamma$. By
multiplying ~(\ref{condFix_h}) on both sides by $h^*_{ijk}$ we
get:

\begin{\er}
  3Tr((h^{(2)})^2)&=&(Tr(h^{(2)}))^2.\label{cond1}
\end{\er}

We denote:

\begin{\er}
  h^{(2)}&\equiv&
  \left(\begin{array}{ccc}
a &b  &c  \\
d &e  &f  \\
g &h  &k
\end{array}\right)\\
\end{\er}

And then:

\begin{\er}
(Tr(h^{(2)}))^2&=&(a+e+k)^2\\
Tr((h^{(2)})^2)&=&(a^2+e^2+k^2)+2(bd+cg+hf).
\end{\er}

So ~(\ref{cond1}) implies:

\begin{equation}\label{cond2}
  (a-e)^2+(a-k)^2+(k-e)^2+6(cg+bd+fh)=0
\end{equation}

But remembering that $h^{(2)}$ is hermitian:

\begin{equation}
c=g^* \quad f=h^* \quad b=d^*,
\end{equation}

we get that the only possibility for ~(\ref{cond2}) to hold is if:

\begin{equation}
  a=e=k, \quad h=f=b=g=c=d=0,
\end{equation}

which implies: $ h^{(2)}_{ij}=\alpha^2\delta_{ij}$ and $\gamma$
which is proportional to identity matrix. So the theory at weak
coupling only has fixed points when the anomalous dimensions
matrix is proportional to identity matrix.

If the anomalous dimensions matrix is proportional to the identity
then:

\begin{equation}
  \gamma^i_j\equiv \rho \delta^i_j.
\end{equation}

From here and from ~(\ref{betaY}) we obtain:

\begin{equation}
\beta_{ijk}=Tr\gamma\cdot h_{ijk}.
\end{equation}

The one loop $\gamma$ implies further that:

\begin{equation}
h^{(2)}_{ij}=\alpha^2\delta_{ij}, \quad
\alpha^2\equiv\frac{1}{3}\sum_{i,j,k}{|h_{ijk}|^2}.
\end{equation}

So the condition ~(\ref{condFix_h}) is automatically satisfied and from ~(\ref{condFix}) we get:

\begin{equation}\label{fixed}
  \alpha^2=-2(\lambda^2-g^2).
\end{equation}

The fixed points we found are essentially IR stable fixed points,
we have:

\begin{equation}
  Tr\gamma=3(2(\lambda^2-g^2)+\alpha^2),
\end{equation}

and the condition for a fixed point is $Tr\gamma=0$. From the
$\beta-functions$ we calculated we see that if we increase one of
the couplings $\lambda,h_{ijk}$, $Tr\gamma$ becomes positive thus
decreasing these couplings and increasing the gauge coupling in
IR, till we get again zero. And the same if we decrease the
couplings. Thus we can conclude that in the weak coupling limit
all fixed points that exist imply diagonal $\gamma$ and are IR
stable. Consequently nothing is known of the UV behavior of the
theory, and we can unambiguously define it only at the conformal
fixed points. (A simple calculation shows also that the fixed
points we find are IR stable even if we go out of the special
$\gamma$ regime. )

All the calculations in this section were done based on the one
loop anomalous dimensions. An interesting question is how the
results are altered by higher loop calculations. Again
following~\cite{a:beta} general formulae, we get for couplings
such that $\gamma^{(1)}=0$ $ $\footnote{At three loop level the
calculation is renormalization scheme dependent ( the above result
is in $\overline{MS}$).  As described in~\cite{a:beta} one can
redefine the coupling constant or equivalently change the
renormalization scheme and get that $\gamma$ vanishes also at
three loop order exactly, including the non-planar graphs. }  :

\begin{\er}
  \gamma^{(2)i}_j&=&-2(2(\lambda^2-g^2)+\alpha^2)(2\lambda^2+g^2+\alpha^2)\delta^i_j\nonumber\\
  \gamma^{(3)i}_j&=&2\kappa
  g^2(2(\lambda^2-g^2)+\alpha^2)(2(2\lambda^2+\alpha^2)+3g^2)\delta^i_j.
\end{\er}

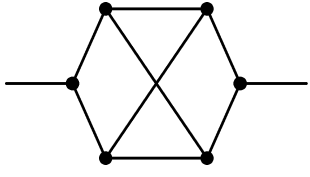
\begin{figure}[htbp]
  \begin{center}
   \begin{fmfgraph}(40,40)
     \fmfpen{thin}
     \fmfleft{f1,v1,f2} \fmfright{f3,v2,f4}
     \fmf{phantom,tension=1}{f1,i1} \fmf{phantom,tension=1}{f3,i2}
     \fmf{phantom,tension=1}{f2,i3} \fmf{phantom,tension=1}{f4,i4}
     \fmf{vanilla,tension=1}{v1,o1} \fmf{vanilla}{o1,i1} \fmf{vanilla}{i1,i2}
     \fmf{vanilla}{i2,o2} \fmf{vanilla,tension=1}{o2,v2}
     \fmf{vanilla}{o1,i3} \fmf{vanilla}{i3,i4} \fmf{vanilla}{i4,o2}
     \fmf{vanilla,tension=0.01}{i1,i4} \fmf{vanilla,tension=0.01}{i2,i3}
     \fmfdot{i1} \fmfdot{i2} \fmfdot{i3} \fmfdot{i4}
     \fmfdot{o1} \fmfdot{o2}
   \end{fmfgraph}
    \caption{Typical non-planar graph appearing in three loop}
    \label{fig:diagram}
  \end{center}
\end{figure}

Here $\kappa=6\zeta(3)$ and in $\gamma^{(3)}$ we have omitted
terms proportional to $\gamma^{(1)}$ in the general formulae. We
have omitted contributions from the non-planar diagrams
(figure~(\ref{fig:diagram})) coming only from the superpotential
(We do it under the assumption that  $g,\lambda \gg \alpha$).
These extra contributions are proportional to:

\begin{equation}
  \Delta^{ij}_{ax}\equiv {\kappa\over 4}
  Y^{*ikl}_{abc}Y^{kmn}_{bed}Y^{lrs}_{cfg}Y^{*pmr}_{hef}Y^{*qns}_{odg}Y^{jpq}_{xho}.
\end{equation}

\begin{fmffile}{mfmaq}
  \unitlength=1mm
\end{fmffile}

$\Delta$ is proportional to at least one power of $\alpha^2$
(because in the ${\cal N}=4$ theory $\Delta$ vanishes), thus the
condition above is necessary for neglecting it. So we see that up
to third order $\gamma$ is proportional to
$2(\lambda^2-g^2)+\alpha^2$ which is the one loop
$\gamma-function$, and the only deviation from it comes from the
non-planar graph (which first appears at three loops). So the
fixed points we find at one loop are essentially correct fixed
points even up to three loops, assuming $\lambda ,g\gg \alpha$. We
will encounter the same assumption also in the S-duality context
and it simply means that we are close to the ${\cal N}=4$ theory.

 The one-loop solution can be easily argued to extend to
 all-loops. Our condition for zero $\beta$s is $\gamma^i_j=0$. Now
 if we restrict ourselves only to the case of
 $h_{111}=h_{222}=h_{333}\equiv h\neq 0$ and $h_{123}\equiv h'\neq 0$ (which is
 consistent with the one loop solution), with all other
 h's vanishing, then from the symmetry of the interactions we see that
 $\gamma^i_j$ is proportional to the identity($\equiv\gamma\delta^i_j$). Thus let's parameterize
 our solutions by $\lambda$, h and h'. The vanishing of $\gamma$
 will give an equation of the form
 $g=g(\lambda,h,h')$. We can expand this relation as
 a power series in its arguments. Assume we have determined
 $g(\lambda,h,h')$ up to (n-1)th order\footnote{(n-1)th order in $\lambda^2,h^2,h^{'2}$.}. Now we determine it
 to n'th order. We define
 $\tilde\gamma^{(n)}\equiv\sum_{i=2}^{n}\gamma^{(i)}$ where we
 insert $g(\lambda,h,h')$ such that this expression will be of n'th
 order. This expression consists only of already determined
 quantities, since the undetermined n'th order of $g(\lambda,h,h')$
 appears only at one loop. Thus we get:

 \begin{eqnarray}
    \tilde\gamma^{(n)}-{1\over16\pi^2}2C_1(g^2)^{(n)}=\gamma^{(n)}.
 \end{eqnarray}

Thus from the demand of vanishing $\gamma$ we get
$(g^2)^{(n)}=({1\over16\pi^2}2C_1)^{-1}\tilde\gamma^{(n)}$, and we
can extend our
    solution to any order of perturbation theory (see~\cite{a:Reduct} , and a slightly different approach ~\cite{a:Luc}).

\newpage

\section{S-duality}

 S-duality was first conjectured by Montonen and Olive~\cite{a:MO} (see ~\cite{a:intro_S} for a review with references) as a symmetry
 of Yang-Mills theory interchanging magnetic and electric
 degrees of freedom. S-duality involves a conjecture that a theory
 with weak coupling constant is equivalent to another theory with a
 strong one. This conjecture can be meaningful only if the notion
 of a coupling constant for a theory is well defined. In general field
 theory coupling constants depend on the energy scale at which
 experiments are done, namely the couplings flow. The cases when the
 couplings are independent of a scale happen when the theory possesses
 exact conformal invariance, even at a quantum level. So in order to
 have a Montonen-Olive duality a theory is expected to be
 conformal. The only finite quantum theories known include some amount
 of supersymmetry. The ${\cal N}=4$ SYM under discussion is believed to possess
 the Montonen-Olive duality, and moreover to be selfdual in this sense.

 The MO conjecture can be schematically written as:

\begin{equation}
\left\langle
...\right\rangle=\int{D[\phi](...)e^{-\frac{1}{g^2}S[\phi]}}=\int{D[\tilde\phi](...)e^{-g^2S[\tilde\phi]}}.
\end{equation}

This duality is hard to check because it involves strong coupling.
There are several checks of this duality, mainly through some BPS
arguments - counting of degrees of freedom of some stable
configurations in both cases. As we mentioned above the ${\cal
N}=4$ SYM theory is believed to be selfdual. We can write the
lagrangian as:

\begin{equation}\label{lag}
  L={1\over g^2}Tr(-{1\over 4}F_{\mu\nu}^2+{\cal D_\mu}\phi^I{\cal D^\mu}\phi^I+[\phi^I,\phi^J]^2+fermions)
\end{equation}

Under the SL(2,$\mathbb{Z}$)\footnote{ In its complete form, the
S-duality transformations involve also the $\theta$ term of the
lagrangian (by taking it to $\theta+2\pi$). If we define
$\tau\equiv {4\pi i\over g_{YM}^2}+{\theta\over 2\pi}$, the
S-duality acts as $\tau\to{a\tau+b\over c\tau+d}$ (ad-bc=1 and
$a,b,c,d\in\mathbb{Z}$). This is the SL(2,$\mathbb{Z}$)
transformation.} transformation $g^2\to {1\over g^2}$ and for the
theory to remain invariant the operators appearing in the
lagrangian have also to undergo some kind of transformation. It is
convenient to define the lowest components of chiral primary
operators of the ${\cal N}=4$ superconformal algebra to be:

\begin{equation}\label{modO}
{\cal O}^{(p)}_{I_1I_2\ldots I_p}\equiv N (g_{YM}^2 N)^{-p/2}
Tr(\phi^{\{I_1}\phi^{I_2}\ldots \phi^{I_p\}}).
\end{equation}

The other single trace (gauge invariant) chiral operators of the
theory can be obtained from these by the supersymmetry generators.
These operators are SL(2,$\mathbb{Z}$) invariant~\cite{a:Intri}.
This is consistent with the U(N=1) case where the theory is free
and thus nothing depends on the coupling, and the normalization we
chose is just scaling the fields such that the coupling doesn't
appear in the lagrangian. The action of the ${\cal N}=4$
supersymmetry algebra may be schematically written as:

\begin{\er}
&[Q_\alpha^A, \phi^I]& \sim \lambda_{\alpha B}\nonumber\\
&\{Q_\alpha^A, \lambda_{\beta B}\}& \sim (\sigma^{\mu
\nu})_{\alpha \beta} F_{\mu \nu} + \epsilon_{\alpha
\beta} [\phi^I, \phi^J]\\
&\{Q_\alpha^A, {\bar \lambda}_{\dot{\beta}}^B\}& \sim
({\sigma}^\mu)_{\alpha \dot{\beta}} {\cal D}_\mu \phi^I\nonumber\\
&[Q_{\alpha}^A, A_{\mu}]& \sim
({\sigma}_\mu)_{\alpha\dot{\alpha}}{\bar
  \lambda}_{\dot{\beta}}^A \epsilon^{\dot{\alpha}\dot{\beta}}.\nonumber
\end{\er}

We wish to extend our discussion in the previous sections to the
strong coupling limit using the S-duality of ${\cal N}=4$. We wish
to learn how the marginal deformations we found look like at
strong coupling. As we mentioned we believe in S-duality in ${\cal
N}=4$ SYM, so this transition has to assume that we are somehow
very close to the ${\cal N}=4$ theory. We will assume that
${\delta\lambda\over g}\ll 1$ and ${h\over g}\ll 1$. Now we have
to obtain the transformation of the operators we add to the
lagrangian under the SL(2,$\mathbb{Z}$). Namely we want to find
$\bar\lambda$ and $\bar h$ such that if we add operators
$\lambda{\cal O},h{\cal O}'$ in the weak coupling limit, we have
to add $\bar\lambda{\cal O},\bar h\bar{\cal O}'$ in the strong
coupling limit. So we first calculate what scalar operators are we
actually adding to the ${\cal N}=4$ action.

By adding the superpotential:

\begin{equation}
  Y^{ijk}_{abc}=(-\delta\lambda\sqrt{2}f_{abc}\epsilon_{ijk}+d_{abc}h_{ijk})
\end{equation}

we actually change the F term and thus the scalar potential. The F
term now is equal to:

\begin{equation}
F^i_a=-{1\over 2}(-\lambda\sqrt{2}f_{abc}\epsilon_{ijk}+d_{abc}h_{ijk})\Phi_j^{*b}\Phi_k^{*c}
\end{equation}

So after integrating out F we get the F-term contribution to the
scalar potential:

\begin{equation}
 -{1\over 4} Y^{ijk}_{abc}Y^{ilm}_{aed}\phi_j^{*b}\phi_k^{*c}\phi_l^{e}\phi_m^{d}
\end{equation}

The D term contribution is like in ${\cal N}=4$:

\begin{equation}
  -{1\over 2}(f_{abc}\phi^{*i}_a\phi^i_b)^2.
\end{equation}

Now if we input the exact expression for $Y^{ijk}_{abc}$ in the F term
contribution we get:

\begin{\er}
  & &({1\over 2}g^2f_{abc}f_{aed}\epsilon_{ijk}\epsilon_{ilm}+{1\over
  2}\delta\lambda^2f_{abc}f_{aed}\epsilon_{ijk}\epsilon_{ilm},\nonumber\\
  &+&{1\over 2}(\delta\lambda+\delta\lambda^*) gf_{abc}f_{aed}\epsilon_{ijk}\epsilon_{ilm}+{1\over
  4}d_{abc}d_{aed}h^*_{ijk}h_{ilm},\\
  &-&\left\{{1\over\sqrt{2}\cdot 2}(g+\delta\lambda)f_{aed}d_{abc}
  \epsilon_{ilm}h^*_{ijk}+c.c\right\})\phi_j^{*b}\phi_k^{*c}\phi_l^{e}\phi_m^{d}.\nonumber
\end{\er}

The fermionic couplings can be obtained from here by
supersymmetry.

 The term proportional to $g^2$ with the D term gives the
usual ${\cal N}=4$ scalar potential $Tr[\phi^I,\phi^J]^2$. The
remaining terms are the perturbation scalar potential. We would
like to bring them to the form ~(\ref{modO}) or its supersymmetric
descendants. By using the following equations:

\begin{\er}
  {1\over C_2}Tr[T_a,T_b][T_e,T_d]&=&-f_{abc}f_{edc}\nonumber\\
  Tr\left\{T_a,T_b\right\}[T_e,T_d]&=&{1\over C_2}id_{abc}f_{edc}\\
  Tr\left\{T_a,T_b\right\}\left\{T_e,T_d\right\}&=&{4\over N^2}\delta_{ab}\delta_{ed}+{1\over C_2^3}d_{abc}d_{edc}\nonumber
\end{\er}

we get that: the $\delta\lambda g$ term is proportional to an
operator of the form $Tr[\phi,\phi]^2$ which is part of an
operator of the form $Q^4{\cal O}^{(2)}$. The $g\cdot h$ term is
proportional to an operator of the form
$Tr[\phi,\phi]\left\{\phi,\phi\right\}$ which is of the form
$Q^2{\cal O}^{(3)}$. All the other terms, which are not linear in
the deformations, can be obtained by the supersymmetry
considerations. From here we can get the transformations of our
couplings. First however we have to get to the form ~(\ref{lag})
and this is done by scaling all the fields by factor of $g$.

\begin{eqnarray}
 \delta\lambda\cdot g\cdot Tr[\phi,\phi]^2&\to&{\delta\lambda\over
 g^3}\cdot
 Tr[\tilde\phi,\tilde\phi]^2\to{\delta\lambda\over g^3}\cdot Q^4\tilde \phi^2\to{\delta\lambda\over
 g}\cdot Q^4{\cal O}^{(2)}\nonumber\\
h\cdot g\cdot Tr[\phi,\phi]\left\{\phi,\phi\right\}&\to&{h\over
 g^3}\cdot
 Tr[\tilde\phi,\tilde\phi]\left\{ \tilde\phi,\tilde\phi \right\}\to{h\over g^3}\cdot Q^2\tilde \phi^3\to h\cdot Q^2{\cal O}^{(3)}
\end{eqnarray}

 Now we finally can write down the transformations of the couplings
$g^2\to{1\over g^2}$:

\begin{\er}\label{trans}
{\delta\lambda\over g}Q^4{\cal
 O}^{(2)}\to{\delta\tilde\lambda g}Q^4{\cal O}^{(2)}\Rightarrow &\hspace{1cm}&
 \delta\lambda\to{\delta\lambda\over g^2}=\delta\tilde\lambda\\
 {h}Q^2{\cal
 O}^{(3)}\to{\tilde h}Q^2{\cal O}^{(3)}\Rightarrow &\hspace{1cm}&
 h\to h=\tilde h\nonumber
\end{\er}

Here we got the coupling constant transformations only from the
terms linear in the deformations $\delta\lambda$ and $h$, the
other, quadratic terms, are automatically transformed in the right
way by supersymmetry transformations.

In ~(\ref{fixed}) we got the condition for a fixed point, so now
we can apply this equation to the strong coupling limit using the
S-duality:

\begin{\er}
  \alpha^2&=&-2(\lambda^2-g^2)\to\alpha^2\cong 4\delta\lambda  g\hspace{1cm}g\to 0\nonumber\\
  \alpha^2&\cong& {4\delta\lambda\over g^3}\hspace{5cm}g\to \infty
\end{\er}

The second line is the condition for having a conformal theory at
large coupling. In order to use S-duality we have to assume that
the strong coupling quantities satisfy ${1\over
g}\gg\delta\lambda,\alpha$ and of course to use the perturbation
theory ${1\over g}\ll 1$. From the conditions above we see that
the result is consistent with these demands.

 The only knowledge we have about the fixed points, as
discussed above, is in the small coupling region and in the limit
where $g\to \infty$ and $\delta\lambda g\ll 1$. We established
that in the regions of our knowledge there are no UV fixed points,
namely there is no line going in the UV to zero couplings, or no
asymptotically free regime.

\newpage

 \chapter{${\cal N}=2$ theory}

 It is possible to reduce the number of supersymmeties of the d=4
 ${\cal N}=4$ SYM, which we dealt with in the previous chapter, via the
 orbifolding
 procedure(~\cite{a:quiver},~\cite{a:quiver2},~\cite{a:SilKach}, see also discussion in
 ~\cite{a:quiver3} and for supergravity
 part~\cite{a:Gukov},~\cite{a:Oz}). In this chapter we will
 concentrate on the case with ${\cal N}=2$ and in the next one we will
 deal with ${\cal N}=1$ theories.

 We will look on N coincident D3-branes at the $\mathbb{Z}_k$ orbifold
 singularity of an $A_{k-1}$ ALE space. If we consider the low energy theory on D-branes in
 the bulk and not on the singularity then essentially we are back to
 the non-orbifold case.
  The orbifold group acts on
 the $\mathbb{C}^3$ (we put the $\mathbb{R}^6$ coordinates in pairs, for instance
 $(Z_1,Z_2,Z_3)=(X_4+iX_5,X_6+iX_7,X_8+iX_9)$ assuming that the branes lie in
 0123 directions)
  perpendicular space of the D3-branes as:

\begin{eqnarray}
   Z^i\to \omega^{a_i}Z^i,
\end{eqnarray}

where $\omega\equiv e^{2\pi i\over k}$ and
$(a_1,a_2,a_3)\equiv(1,0,-1)$\footnote{The vector
$\overrightarrow{a}$ has to satisfy $\sum_i a_i=0$(mod k) in order
that the orbifold action will be part of SU(3) and not U(3), in
other words to preserve supersymmetry. So if we choose all the
components to be non zero we have $\mathbb{Z}_k\subset SU(3)$ thus
leaving us with one supersymmetry. If we choose one of the
components zero then we can have (n,0,-n)(mod k) case. This case
is equivalent to (1,0,-1) case, and the important fact is that in
this case $\mathbb{Z}_k\subset SU(2)$ and we have ${\cal N}=2$
supersymmetry. So there is only one choice giving an ${\cal N}=2$
theory here. If we put two components of $\overrightarrow{a}$ to
zero we will have to take the remaining one to zero too, thus
remaining in the ${\cal N}=4$ case. If we consider orbifolds
$\mathbb{R}^6/\mathbb{Z}_k$ rather than
$\mathbb{C}^3/\mathbb{Z}_k$ we can have non-supersymmetric
theories~\cite{a:N_sus_orb1}~\cite{a:N_sus_orb2}.}.

 To see\footnote{There are other possible choices for the orbifold
 to act on the brane indices ~\cite{a:quiver3}.}
  how the orbifold can act on the D-branes we put kN
 D3-branes on the covering ALE space and group them in N sets of
 k branes. We put each set of the D-branes in the regular
 representation of the $\mathbb{Z}_k$ $\to$
 $\gamma^i_j(g)\equiv\delta^i_j\omega^i$ where {\bf $g$} is the generator
 of $\mathbb{Z}_k$. As we see the regular representation is
 reducible and is the direct sum of k one dimensional irreducible
 representations (${\cal R}_r=\bigoplus_{n=0}^{k-1}{\cal R}_n$), parameterized by the integer n, given by $\omega^n$ ($\omega$
 defined above). We define the N set indices by I and the
 brane indices within each set are $i\in (0,1,..,k-1)$.

 From here we conclude that the projection on
 the gauge vector fields is (We write the gauge fields in double
 index notation - upper fundamental,lower anti-fundamental):

 \begin{eqnarray}
   A^{I,i}_{J,j}=\omega^{i-j}A^{I,i}_{J,j}
 \end{eqnarray}

 And the projection on the bosonic components of the chiral multiplets
 ( which are related to the transverse 6 directions ) is:

 \begin{eqnarray}
   Z^{l I,i}_{J,j}=\omega^{i-j+a_{l}}Z^{l I,i}_{J,j}
 \end{eqnarray}

This projection comes from acting on the Chan-Paton indices as
well as on the space-time index.

Thus we see that in our case the fields which survive the
 projection are: gauge fields $A^{I,i}_{J,j}$ where both indices lie in
 the same irreducible representation of $\mathbb{Z}_k$
 (as defined above), giving a total of k copies of U(N).
 The bosonic part of the matter fields which survive are:
 $Z^{2I,i}_{J,j}$ survives if $i-j=0$, exactly as for the gauge
 field $\to$ we get fields in the adjoint. $Z^{1I,i}_{J,j}$
 survives when we have $i-j+1=0$, thus these fields are in the
 fundamental of the {\bf i}'th gauge group and the antifundamental of
 the {\bf (i+1)}'th gauge group. $Z^{3I,i}_{J,j}$
 survives when we have $i-j-1=0$, thus these fields are in the
 fundamental of the {\bf (i+1)}'th gauge group and the antifundamental of
 the {\bf i}'th gauge group.

  The same projection can be made for the
 fermions. Giving a total of three types of chiral super fields:

 \begin{itemize}
  \item For each U(N) group an adjoint field which is a singlet of the other groups
  $\to$ denote it by {\bf $\Phi_i$}.
  \item Fields which are in the fundamental of the i'th U(N) and in the
  anti-fundamental of the (i+1)'th U(N) $\to$ denote them by {\bf
  $Q_i$}.
  \item Fields which are in the fundamental of the (i+1)'th U(N) and in the
  anti-fundamental of the i'th U(N) $\to$ denote them by {\bf $\tilde
  Q_i$}.
 \end{itemize}

  So to summarize, we have an ${\cal N}=2$
theory with gauge group\footnote{The U(1) factors are expected to
decouple in the IR so we will deal with an $SU(N)^k$ group.}
$U(N)^k$ and hypermultiplets in the representations:

\begin{equation}
  (N,\bar N,1,..,1)\oplus(1,N,\bar N,1,..,1)\oplus\cdots \oplus(\bar
  N,1,..,1,N).
\end{equation}

\begin{figure}[htbp]
\begin{fmffile}{quiqy}
\unitlength=1mm
\begin{center}
\begin{fmfchar}(50,50)
    \fmfpen{thin}
    \fmfleft{v1,u1} \fmfright{v2,u2}
    \fmf{phantom}{v1,i1} \fmf{phantom}{v2,i1}
    \fmf{phantom,tension=20}{u1,i2} \fmf{vanilla}{i1,i1} \fmf{phantom,tension=20}{u2,i3}
    \fmf{plain_arrow,left=.1}{i1,i2} \fmf{plain_arrow,left=.1}{i2,i3}
    \fmf{plain_arrow,left=.1}{i3,i1}
    \fmf{plain_arrow,left=.1}{i2,i1} \fmf{plain_arrow,left=.1}{i3,i2}
    \fmf{plain_arrow,left=.1}{i1,i3}
    \fmfblob{100}{i1} \fmfblob{100}{i2} \fmfblob{100}{i3}
    \fmf{vanilla}{i2,i2} \fmf{vanilla}{i3,i3}
\end{fmfchar}
\caption{$\mathbb{C}^3/\mathbb{Z}_3$ (1,-1,0) quiver diagram}
\label{quiv}
\end{center}
\end{fmffile}
\end{figure}
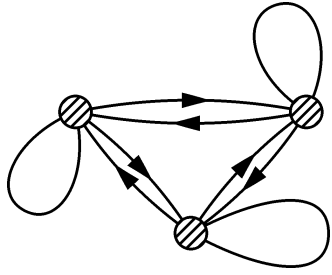

The matter content of this theory can be summarized in a so called
"quiver" diagram, where vertices represent the gauge groups, and
oriented lines represent chiral multiplets in the fundamental of
the group to which they point and the antifundamental of the
second group. Unoriented lines are lines pointing in both
directions $\to$ lines in fundamental and anti-fundamental of the
 groups which they connect, i.e. they represent fields in the adjoint if they end on same group vertex . In
figure~\ref{quiv} we have an example for the $\mathbb{Z}_3$
orbifold theory.
\\

 The superpotential of this theory can be easily read off from the
 ${\cal N}=4$ superpotential by restricting it only to $\mathbb{Z}_k$
 invariant fields. The easier way, however, is: we know that the
 theory has ${\cal N}=2$ supersymmetry, and we know the gauge group and
 matter content $\to$ from here the theory is uniquely defined by supersymmetry considerations
 \footnote{Of course we have here no mass terms.}.

The ${\cal N}=2$ theory vector multiplet consists of an ${\cal
N}=1$ vector multiplet V and an ${\cal N}=1$ chiral multiplet
$\Phi$, both in the adjoint representation of the group. The
hypermultiplet consists of chiral and antichiral fields $(\tilde
Q^{\dag},Q)$, both transforming under the same representation. The
general lagrangian is:

\begin{equation}
  \sum_i\int{d^4\theta(Q_i^{\dag}e^{-2gV}Q_i+\tilde Q_i^{\dag}e^{2gV}\tilde
  Q_i)}+\int{d^2\theta\sqrt{2}g\tilde Q_i\Phi Q_i}+\cdots
\end{equation}

Where dots mean the ${\cal N}=2$ pure SYM action. The ${\cal N}=2$
theories have only one coupling\footnote{For each simple factor of
the gauge group.} , the gauge coupling. In our case we denote for
any one of the k SU(N)'s the chiral field of the vector multiplet
by $\Phi^i$ and the vector field $V^i$. The hypermultiplets will
be denoted by $Q^i$ and $\tilde Q^i$ where each such field has two
indices one in N and one in $\bar N$ of i'th and (i+1)'th gauge
group. ${\cal N}=2$ theories are renormalized only at one-loop.
The general one-loop gauge $\beta-function$ is:

\begin{equation}\label{1loopB}
  \beta(g) \propto -{1 \over {16\pi^2}} \left[ 3C_2(G)-\sum_A
  T_A(R)\right].
\end{equation}

In ${\cal N}=2$ SQCD we have two chiral multiplets from each
hypermultiplet ( the antichiral one can be treated as the complex
conjugate of a chiral) and one chiral multiplet coming from the
vector multiplet, so if the chirals are in the fundamental of
SU(N) (like in our case) $T(R)={1\over 2}$ and $C_2=N$. For the
adjoint representation $T(R)=N$, so from (~\ref{1loopB}) we get
for m hypermultiplets:

\begin{equation}\label{1loopB2}
  \beta(g) \propto -{1 \over {16\pi^2}} \left[ 3N-N-m\right]=-{1 \over {16\pi^2}} \left[
  2N-m\right].
\end{equation}

In our case we effectively have 2N hypermultiplets for every gauge
group: $Q^i_{\alpha\tilde\beta}$ is N chirals of the i'th SU(N)
(and also N antichirals of the (i+1)'th SU(N)), and
$Q^{(i-1)}_{\alpha\tilde\beta}$ is N antichirals of the i'th
SU(N). So m=2N and the $\beta$-functions vanish $\to$ the theory
is finite for any value of the couplings.

\newpage

\section{Marginal deformations}

Now we consider the marginal deformations of the theory which
preserve some ${\cal N}=1$ supersymmetry. First, obviously we can
change the couplings of the ${\cal N}=2$ superpotentials to be
different from the gauge couplings and change the gauge couplings
themselves. Another option is to add a superpotential for the
$\Phi$'s: $h^id_{abc}\Phi_i^a\Phi_i^b\Phi_i^c$ (where
$d_{abc}\equiv Tr\left[T_a\left\{T_b,T_c\right\}\right]$, $T_a$ in
the fundamental of SU(N)). We can add also a superpotential for
Q's: $s^{(i)abc}_{lmn}Q^{(i)l}_aQ^{(i)m}_bQ^{(i)n}_c$ (and
similarly for $\tilde Q$). Here the {\bf s} couplings are
constrained from gauge invariance to satisfy:

\begin{equation}\label{condGaugeInv}
  (T^\alpha)^a_es^{(i)ebc}_{lmn}+(T^\alpha)^b_es^{(i)aec}_{lmn}+(T^\alpha)^c_es^{(i)abe}_{lmn}=0.
\end{equation}

The same condition we get also for the lower indices.
 These constrains
we can satisfy only for $SU(N=3)$, because otherwise $N\otimes
N\otimes N$ doesn't include gauge singlets. Obviously the {\bf s}
couplings have to be symmetric. The only possible choice for the
{\bf s} couplings is thus:
$s^{ijk}_{abc}\propto\epsilon^{ijk}\epsilon_{abc}$.

The terms of the form $p^{(i)lbc}_{amn}\tilde
Q^{(i)a}_lQ^{(i)m}_bQ^{(i)n}_c$ and $q^{(i)lmc}_{abn}\tilde
Q^{(i)a}_l\tilde Q^{(i)b}_mQ^{(i)n}_c$ are ruled out by the
constraints (~\ref{condGaugeInv}), since $N\otimes\bar N\otimes N$
doesn't include gauge singlets.

 In k=3 case we also can have operators of the form ${\kappa\over 3!}
 Q_1Q_2Q_3$ (and similarly for $\tilde Q$).

So to summarize, we have the following marginal deformations of
our finite theory:

\begin{eqnarray}
  W_1&=&{1 \over 6}(\alpha^i\tilde Q_i\Phi^i Q_i+\delta^i Q_i\Phi^{i+1}\tilde Q_i)\nonumber\\
  W_2&=&{1 \over 6}h^id_{abc}\Phi_i^a\Phi_i^b\Phi_i^c\nonumber\\
  W_3&=&{\rho_i\over
 3!}\epsilon_{lmn}\epsilon^{abc}Q^{(i)l}_aQ^{(i)m}_bQ^{(i)n}_c\nonumber\\
  W_4&=&{\tilde\rho_i\over
 3!}\epsilon_{lmn}\epsilon^{abc}\tilde Q^{(i)l}_a\tilde
  Q^{(i)m}_b\tilde Q^{(i)n}_c\\
  W_5&=&{\kappa\over 3!}Q_1Q_2Q_3\nonumber\\
  W_6&=&{\tilde\kappa\over 3!} \tilde Q_1\tilde Q_2\tilde Q_3\nonumber
\end{eqnarray}

Where the $W_3,W_4$ can be added only for $SU(N=3)$ and $W_5$,
$W_6$ only for k=3. So for general N and k the only marginal
deformations are changing the couplings of the ${\cal N}=2$
superpotential and adding a superpotential for the $\Phi_i$. As we
will see later, there are also additional interactions in the k=2
case.

Now we can engage in the search for fixed points. The one-loop
gauge $\beta$ function will still vanish, because it is not
affected by the superpotential.

 The $Q$'s and $\tilde Q$'s appear symmetrically in the case of turning on only
 $W_1$ and $W_2$, thus we expect
their $\gamma$-functions to be the same. Moreover obviously the
fields appearing here don't mix in renormalization (for $k>2$),
thus all the anomalous dimensions are diagonal.  Now we wish to
obtain the $\beta$-functions via the anomalous dimensions.

For the gauge $\beta$-functions we have the NSVZ
formula(~\ref{NSVZ_1}), in which for a gauge group (i) we have to
take only the $\Phi_i$ and $Q_i,Q_{i-1},\tilde Q_i,\tilde Q_{i-1}$
anomalous dimensions into account. For the superpotential
couplings we have from the ${\cal N}=1$ non-renormalization
theorem:

\begin{equation}
\beta_Y^{ijk}= Y^{p(ij}\gamma^{k)}_p =
Y^{ijp}\gamma^k_p+(k\leftrightarrow i)+(k\leftrightarrow j)
\end{equation}

Where we arrange the superpotential to the form:
$W={1\over6}Y^{ijk}S_iS_jS_k$. Here the indices run over the gauge
indices as well as over the field indices. From this we get:

\begin{eqnarray}\label{betas}
  \beta_{g_i}&=&-{2g^3\over 16\pi^2}{N\over
  1-{2Ng^2\over16\pi^2}}({1\over2}(\gamma_{Q_i}+\gamma_{\tilde Q_i}+\gamma_{Q_{i-1}}+\gamma_{\tilde Q_{i-1}})+\gamma_{\Phi_i})\nonumber\\
  \beta_{h_i}&=&3h_i\cdot \gamma _{\Phi_i}\nonumber\\
  \beta_{\alpha_i}&=&\alpha_i(\gamma_{Q_i}+\gamma_{\tilde Q_i}+\gamma_{\Phi_i})\\
  \beta_{\delta_i}&=&\delta_i(\gamma_{Q_i}+\gamma_{\tilde
  Q_i}+\gamma_{\Phi_{i+1}})\nonumber
\end{eqnarray}

From here we see that if we want the $\beta$-functions to vanish a
sufficient condition is the vanishing of the $\gamma$-functions.
We also notice that
$\beta_{g_i}\propto{\beta_{\alpha_i}\over\alpha_i}+{\beta_{\delta_{i-1}}\over\delta_{i-1}}$.
Thus essentially we only have to worry about the vanishing
superpotential coupling $\beta$-functions, and the gauge
$\beta$-function will vanish automatically. The gauge couplings
contribute with negative sign to the anomalous dimensions and the
other couplings with positive sign (at one-loop), so because of
the signs of the $\beta$-functions the fixed points we get in this
way are IR-stable fixed points.

\newpage
\subsection{General k case}\label{gen_kk}

Let's first analyze the general expressions(~\ref{betas})
following ~\cite{a:Leigh}. The  interactions we have here, as
shown above, are symmetric in respect to $Q_i$ and $\tilde Q_i$,
thus $\gamma_{Q_i}=\gamma_{\tilde Q_i}$. First we analyze the case
without $h_i$'s. We see that in order that the $\beta$'s will
vanish all $\gamma_{\Phi_i}$'s have to be equal $\to$ $\forall
i:\gamma_{\Phi_i}\equiv \gamma$. And also all the $\gamma_{Q_i}$'s
have to be equal and equal to $-{1\over2}\gamma$. So we have 2k
conditions on the parameters, 3k couplings and one additional
parameter {\bf $\gamma$}. So totally we expect for a $3k+1-2k=k+1$
dimensional manifold of fixed points. This is one dimension more
than the space of the ${\cal N}=2$ theory.

If the $h_i$'s don't vanish then we have to have
$\gamma_{\Phi_i}=0$ and so don't have the additional parameter. In
this case we have 4k couplings and 2k conditions and so we expect
a 2k dimensional manifold of fixed points.

Now we will see if we get the expected results from loop
calculations. The one-loop anomalous dimensions are\footnote{For
one loop calculation we can use the results of~\cite{a:beta},
although here we have a non-simple gauge group. This is because at
 one loop every diagram contains at most one gauge group (out of
k), thus we can simply sum the contributions from every gauge
group.}:

\begin{eqnarray}
  \gamma_{\Phi_i}&=&{1 \over 16\pi^2}\left\{{1\over 8}{N^2-4\over N}
    h_i^2+{N\over 4} ((\delta_{i-1}^2+\alpha_i^2)-8g_i^2)\right\}.\\
  \gamma_{Q_i}&=&{1 \over 16\pi^2}{N^2-1\over4N}\left\{(\delta_i^2+\alpha_i^2)-4(g_i^2+g_{i+1}^2)\right\}.\nonumber
\end{eqnarray}

In particular we see that in the case that all $h_i$ vanish and
$\alpha_i=\delta_{i-1}=2g_i$ the anomalous dimensions vanish, and
this is exactly the ${\cal N}=2$ case.

\addcontentsline{toc}{subsection}{ \quad \quad \quad
 $\bullet$ Vanishing $h_i$s}
\begin{itemize}
\item {\bf Vanishing $h_i$s}
\end{itemize}

 First we put all $h_i$'s to zero.
 Let's define: $B_i\equiv
\delta_{i-1}^2-4g_i^2$, $A_i\equiv\alpha_i^2-4g_i^2$ and $16\pi^2{4N\over
N^2-1}\gamma\to\gamma$. Then, the requirement of vanishing
 $\beta$-functions becomes :
\begin{eqnarray}\label{one_loop_sol}
  B_I+A_I&=&{N^2-1\over N^2}\gamma,\\
  B_{I+1}+A_I&=&-{1\over2}\gamma.\nonumber
\end{eqnarray}

From here, by subtracting the first line from the second and
summing over i, we get that $\gamma=0$. Thus in the one loop
precision the ${\bf \gamma}$ parameter has to vanish. As we will
see later this is not necessarily true for higher loop
calculations. The case of vanishing {\bf $\gamma$} is the case of
vanishing of all anomalous dimensions.
 We see that in this case the condition for having zero $\beta$-functions is:
  $\forall i \left\{B_i\equiv X=-A_i\right\}$ where {\bf X} is a
parameter. From here we find a family of solutions parameterized
by {\bf X} and the gauge couplings:

\begin{eqnarray}\label{sol}
  \delta_{i-1}^2&=&X+4g_i^2\\
  \alpha_i^2&=&4g_i^2-X.\nonumber
\end{eqnarray}

We see that the parameter {\bf X} is constrained to be:
$-\min_{i}\left\{4g_i^2\right\}\leq X\leq
\min_{i}\left\{4g_i^2\right\}$. The case X=0 is the case of ${\cal
N}=2$ SUSY.

Thus to summarize, we get a {\bf k+1} dimensional solution. We
expected a {\bf k+1} dimensional manifold from the Leigh-Strassler
analysis, but the {\bf +1} was due to the $\gamma$ parameter, here
at one loop we find that $\gamma=0$ but nevertheless we get a {\bf
k+1} dimensional space of solutions. The question is whether the
vanishing of $\gamma$ extends to higher loops and whether we can
find the parameter {\bf X} at higher loops.

 We will prove now that the non-vanishing {\bf X} solution doesn't
 disappear at higher loops. First we will represent a general solution as a
 function of the gauge couplings and the {\bf X} parameter. The
 procedure we use here is similar to the coupling constant
 reduction procedure described in ~\cite{a:Reduct}.

 We define the most general solution for $\alpha_i$ and
  $\delta_i$ depending on {\bf X}, {\bf $g_i$}, and consistent with the one
  loop analysis
  and with the ${\cal N}=2$ case (which is known to be exactly
  conformal):

\begin{eqnarray}\label{gen_sols}
  \delta_{i-1}^2&=&4g_i^2+X(1+\sum_{m,j,l_s}a^{(i)m}_{l_1...l_j}X^mg_{l_1}^2...g_{l_j}^2)\\
  \alpha_i^2&=&4g_i^2-X(1+\sum_{m,j,l_s}b^{(i)m}_{l_1...l_j}X^mg_{l_1}^2...g_{l_j}^2)\nonumber
\end{eqnarray}

Here {\bf a}, {\bf b} are some constants and $m+j>0$.
  Now assume we have computed the {\bf a} and {\bf b} parameters in these solutions
  up to (n-1)'th order in $g^2$, {\bf X}. We look at the n'th order.
  First we calculate the $\gamma_{Q_i}$ and $\gamma_{\Phi_i}$. We
  write them as:

  \begin{eqnarray}\label{demand}
\gamma_{Q_i}^{(n)}=\gamma_{Q_i}^{(n)(1-loop)}+\gamma_{Q_i}^{(n)(2..n-loops)}\\
\gamma_{\Phi_i}^{(n)}=\gamma_{\Phi_i}^{(n)(1-loop)}+\gamma_{\Phi_i}^{(n)(2..n-loops)}.\nonumber
  \end{eqnarray}

 We define:

  \begin{eqnarray}
  \tilde B_i^{(n)}&\equiv&\Delta\delta_{i-1}^2=X\cdot(\sum_{m,j,l_s,m+j=(n-1)}a^{(i)m}_{l_1...l_j}X^mg_{l_1}^2...g_{l_j}^2)\\
  \tilde A_i^{(n)}&\equiv&\Delta\alpha_i^2=-X\cdot(\sum_{m,j,l_s,m+j=(n-1)}b^{(i)m}_{l_1...l_j}X^mg_{l_1}^2...g_{l_j}^2)\nonumber
\end{eqnarray}

 The $\gamma_{Q_i}^{(1-loop)}$ and $\gamma_{\Phi_i}^{(1-loop)}$
  have a special structure, giving:

\begin{eqnarray}
  \tilde B_{i+1}^{(n)}+\tilde A_i^{(n)}&=&\gamma_{Q_i}^{(n)(1-loop)}\\
  \tilde B_i^{(n)}+\tilde A_i^{(n)}&=&{N^2-1\over N^2}\gamma_{\Phi_i}^{(n)(1-loop)}\nonumber
\end{eqnarray}

Now we parameterize the remaining contributions to $\gamma$s as:

\begin{eqnarray}
    \gamma_{Q_i}^{(n)(2..n-loops)}&\equiv&T_i^{(n)}+S_{i+1}^{(n)}-{1\over2}\tilde\gamma^{(n)}\\
    {N^2-1\over N^2}\gamma_{\Phi_i}^{(n)(2..n-loops)}&\equiv&T_i^{(n)}+S_i^{(n)}+{N^2-1\over
    N^2}\tilde\gamma^{(n)}\nonumber
\end{eqnarray}

Where the different quantities are defined as :

\begin{eqnarray}
 -k({1\over2}+{N^2-1\over N^2})\tilde\gamma^{(n)}&\equiv&\sum_i{\gamma_{Q_i}^{(n)(2..n-loops)}-{N^2-1\over
N^2}\gamma_{\Phi_i}^{(n)(2..n-loops)}}\nonumber\\
 \Delta X^{(n)}&=& S_1^{(n)}(\equiv 0)\\
  \gamma_{Q_i}^{(n)(2..n-loops)}-{N^2-1\over
N^2}\gamma_{\Phi_i}^{(n)(2..n-loops)}&=&S^{(n)}_{i+1}-S^{(n)}_i-({1\over2}+{N^2-1\over
N^2})\tilde\gamma^{(n)}\nonumber
\end{eqnarray}

$\Delta X^{(n)}$ is just a redefinition of X, so we can set it to
zero without any loss of generality, and the $T_i^{(n)}$s are
automatically determined from above. We see that the definitions
above are well and uniquely defined.

 The crucial point is that in order to calculate the one loop
 contribution to the n'th order we use the n'th order components of
 (~\ref{gen_sols}), and for two loops we use the (n-1)'th order of
 (~\ref{gen_sols}),..., for n'th order we use the first order of
 (~\ref{gen_sols}). Thus because we have already determined
 (~\ref{gen_sols})up to n-1'th order,
 $\gamma_{\Phi_i}^{(n)(2..n-loops)}$,
 $\gamma_{Q_i}^{(n)(2..n-loops)}$
 depend only on already determined quantities. The yet
 undetermined quantities appear only in one loop.

From the $\beta$-function analysis we know that
 $\gamma_{Q_i}=-{1\over2}\gamma$ and $\gamma_{\Phi_i}=\gamma$.
 Thus the demand (~\ref{demand}) is translated to:

 \begin{eqnarray}
\tilde A_i^{(n)} &=& -T_i^{(n)}\\
\tilde B_i^{(n)} &=& -S_i^{(n)}\nonumber\\
 \tilde\gamma^{(n)} &=& \gamma^{(n)}\nonumber
\end{eqnarray}

Here, in the first two lines, we are defining the yet undetermined
{ \bf a}'s and {\bf b}'s. And in the third line we compute the
$\gamma$ parameter. We see that it does not have to be zero in
higher loops.

 Again this procedure is well defined and unique, and can be
 extended to any order in perturbation theory.

   We see that we found a solution to our problem which obviously
exists in any order of perturbation theory. So we have proven that
there exists a {\bf k+1} dimensional manifold of fixed points
parameterized by the gauge couplings and the {\bf X} parameter (or
equivalently the $\gamma$ parameter if it is non zero starting
from some order in perturbation series), in all orders of
perturbation theory.

\addcontentsline{toc}{subsection}{ \quad \quad \quad $\bullet$
Non-vanishing $h_i$s}\label{CCC}
\begin{itemize}
\item {\bf Non-vanishing $h_i$s}
\end{itemize}

 Now we turn our attention to the non zero $h_i$ case. In this
case we are constrained to have $\gamma_{\Phi_i}=0$ or
equivalently $\gamma=0$. At one-loop we get: (define
$C_i\equiv{1\over  8}{N^2-4\over N}h_i^2$)

\begin{eqnarray}\label{1_loop_C}
  B_i+A_i&=&-C_i\\
  B_{i+1}+A_i&=&0\nonumber
\end{eqnarray}

From here:

\begin{eqnarray}
  B_{i+1}-B_i=C_i.
\end{eqnarray}

But from the cyclic nature of our couplings
($\alpha_{k+1}=\alpha_1$ etc.) we get
$\sum_i\left\{B_{i+1}-B_i\right\}=0$ $\to$ $\sum_i C_i=0$, but
this is impossible unless all $h_i$ vanish because $C_i$ is
positive definite.  So we conclude that there are no fixed points
with non vanishing $h_i$. So at one loop level we get no new
marginal directions in this case.

 We now proceed in search of all loop solutions like we did above.
The general expressions
     for $\alpha_i$ and $\delta_i$ now depend also on $h_i$s, and
     we proceed as before:

  \begin{eqnarray}
\gamma_{Q_i}^{(n)}=\gamma_{Q_i}^{(n)(1-loop)}+\gamma_{Q_i}^{(n)(2..n-loops)}\\
\gamma_{\Phi_i}^{(n)}=\gamma_{\Phi_i}^{(n)(1-loop)}+\gamma_{\Phi_i}^{(n)(2..n-loops)}\nonumber
  \end{eqnarray}

  The $\gamma_{Q_i}^{(1-loop)}$ and $\gamma_{\Phi_i}^{(1-loop)}$
  have a special structure, giving:

\begin{eqnarray}
  \tilde B_{i+1}^{(n)}+\tilde A_i^{(n)}&=&\gamma_{Q_i}^{(n)(1-loop)}\\
  \tilde B_i^{(n)}+\tilde A_i^{(n)}+C_i^{(n)}&=&{N^2-1\over N^2}\gamma_{\Phi_i}^{(n)(1-loop)}\nonumber
\end{eqnarray}

Now we parameterize the remaining contributions to $\gamma$s as:

\begin{eqnarray}
    \gamma_{Q_i}^{(n)(2..n-loops)}&\equiv&T_i^{(n)}+S_{i+1}^{(n)}\\
    {N^2-1\over N^2}\gamma_{\Phi_i}^{(n)(2..n-loops)}&\equiv&T_i^{(n)}+S_i^{(n)}-C_i^{(n)}\nonumber
\end{eqnarray}

Where the different quantities are obtained from here as :

\begin{eqnarray}
 \sum_iC_i^{(n)}&=&\sum_i{\gamma_{Q_i}^{(n)(2..n-loops)}-{N^2-1\over
N^2}\gamma_{\Phi_i}^{(n)(2..n-loops)}}\nonumber\\
 \Delta X^{(n)}&=& S_1^{(n)}(\equiv 0)\\
  \gamma_{Q_i}^{(n)(2..n-loops)}-{N^2-1\over
N^2}\gamma_{\Phi_i}^{(n)(2..n-loops)}&=&S^{(n)}_{i+1}-S^{(n)}_i+C_i^{(n)}\nonumber
\end{eqnarray}

$\Delta X^{(n)}$ is just a redefinition of {\bf X}, so we can set
it to zero without any loss of generality, and the $T_i^{(n)}$s
are automatically determined from above. We see that the
definitions above are well and uniquely defined, except for one
caveat. The first equation above cannot always be satisfied:
$C_i\equiv\sum_nC_i^{(n)}\geq 0$, implying that in the lowest
order where $C_i^{(n)}$ is not zero it has to be positive, thus in
that order $\sum_i{\gamma_{Q_i}^{(n)(2..n-loops)}-{N^2-1\over
N^2}\gamma_{\Phi_i}^{(n)(2..n-loops)}}$ has to be positive. At one
loop and two loops this quantity is zero, and in Appendix A we
make a calculation and find that at three loops this is also zero,
thus implying that no $h_i$s can be turned on at this order.

 If this quantity has the right sign we proceed like in the previous case to obtain a solution, by
 demanding that $\gamma_{Q_i}=\gamma_{\Phi_i}=0$:

 \begin{eqnarray}
\tilde A_i^{(n)} &=& -T_i^{(n)}\\
\tilde B_i^{(n)} &=& -S_i^{(n)}\nonumber
\end{eqnarray}

Here  we are defining the yet undetermined {\bf a}'s and {\bf
b}'s.

Again this procedure is well defined and unique, and can be
 extended to any order in perturbation theory, considering the caveat above is
 satisfied.

 In the calculation of the first non-vanishing contribution to the
 $\gamma$ parameter we will take advantage of the fact that if the {\bf
 X}
 parameter is zero then we are in the ${\cal N}=2$ case and we know that
 $\gamma=0$, and that following the analysis of (~\cite{a:N_plan1},~\cite{a:N_plan2}) we know that
 the theory we are concerned with (if all the gauge couplings are equal) is the same as the ${\cal N}=4$ theory up to the
 non-planar diagrams. Thus, we expect contributions to $\gamma$ only from these diagrams.\\
  The two-loop $\gamma$-function vanishes here if the one-loop $\gamma$-function vanishes so it doesn't
  bring any new features. At three loops however we get several
  non-planar diagrams . As we see in Appendix~\ref{Ap} the contribution
  proportional to  {\bf X} will vanish because it always comes in a $X(g_i^4-g_{i+1}^4)$
  combination, and the graphs with three gauge interactions don't include {\bf
  X}
  parameter, and thus are not interesting due to the ${\cal N}=2$ case. The
  explicit calculations ( see Appendix A ) show that the
  contribution of the rest of the diagrams is zero at three loops, implying that the $\gamma$ parameter is
  zero and thus there is no possibility of satisfying (in three loop precision):

\begin{eqnarray}
\sum_iC_i&=&\sum_i{\gamma_{Q_i}-{N^2-1\over N^2}\gamma_{\Phi_i}}.
\end{eqnarray}

   Thus there are two possibilities: either we have
non zero $\gamma$
  at higher loops, or the  $\gamma$ parameter is strictly zero.

 So to summarize, in the general {\bf k} case there are always {\bf k+1} \textit{exactly} marginal directions which can
 be parameterized by the {\bf k} gauge couplings and a parameter {\bf X}.
  If at higher loops $\sum_i{\gamma_{Q_i}-{N^2-1\over
  N^2}\gamma_{\Phi_i}}$ is turned on with the correct sign, then
  we obtain another {\bf k-1} exactly marginal deformations as
  anticipated from the Leigh-Strassler analysis.

\newpage
\subsection{k=3 case}\label{s:k_3}

If we consider the special case of k=3 we can add additional
marginal operators: ${\kappa\over 3!} Q_1Q_2Q_3$ and
${\tilde\kappa\over 3!} \tilde Q_1\tilde Q_2\tilde Q_3$. Again
first we deal with vanishing $h_i$'s. We get two additional
$\beta$-functions:

\begin{eqnarray}
  \beta_{\kappa}=\kappa(\gamma_{Q_1}+\gamma_{Q_2}+\gamma_{Q_3})\nonumber\\
  \beta_{\tilde\kappa}=\tilde\kappa(\gamma_{\tilde Q_1}+\gamma_{\tilde
  Q_2}+\gamma_{\tilde Q_3}).
\end{eqnarray}

From here and (~\ref{betas}) we get that in order that the
$\beta$-functions will vanish we have to demand
$\gamma_{\Phi_i}=0$ for all i, and $\gamma_{Q_i}=-\gamma_{\tilde
Q_i}\equiv\gamma_i$, $\sum_i\gamma_i=0$. Thus we have 3k+2
couplings, k parameters $\gamma_i$ and 3k+1 conditions, so the
expected dimension of the manifold of fixed points is
3k+2+k-(3k+1)=k+1=4. We get the same prediction as for general k
and thus we expect the solution to be exactly as there.

The extra interactions of the k=3 case will add a term of the form
${\kappa^2N\over 2}(\equiv D)$ to the one-loop $\gamma_{Q_i}$ and
${\tilde\kappa^2N\over 2}(\equiv\tilde D)$ to the one-loop
$\gamma_{\tilde Q_i}$, and won't effect the one-loop
$\gamma_{\Phi_i}$. Thus now the general condition for vanishing of
the $\beta$-functions becomes:(defining:
$\gamma_i\to16\pi^2{4\over N}\gamma_i$)

\begin{eqnarray}\label{k3}
  B_i+A_i&=&0\nonumber\\
  B_{i+1}+A_i+D&=&\gamma_i\\
  B_{i+1}+A_i+\tilde D&=&-\gamma_i.\nonumber
\end{eqnarray}

 By subtracting the
first equation from the second and the third ones and summing over
i we get two conditions: $k\tilde D=-\sum_i\gamma_i=0$ and
$kD=\sum_i\gamma_i=0$. So we see that $\kappa$ and $\tilde\kappa$
have to be zero at one loop precision because D and $\tilde D$ are
positive definite. Subtracting the second equation from the third
we get: $2\gamma_i=D-\tilde D=0$. We see that all $\gamma_i$'s
have to vanish $\to$ for all i $\gamma_i=0$. Again, as in the
previous subsection, we can write a general solution:

\begin{equation}
  B_{i+1}=B_i.
\end{equation}

From here we see that also all $A_i$ have to be equal. So we get
that a solution is parameterized by the gauge couplings and by
$B_1$. Thus the dimension of the manifold of fixed points is
k+1=4. We get the expected result. Which is exactly the same as
the general {\bf k} solution.

 Now we turn on non vanishing $h_i$'s. Again we will have to demand
$\gamma_{\Phi_i}=0$, as was the case also in the vanishing $h_i$'s
case, which implies also $\gamma_{Q_i}+\gamma_{\tilde
  Q_i}=0$. From here we get k parameters corresponding to the value of each
$\gamma_{Q_i}\equiv\gamma_i$, with $\sum_i\gamma_i=0$. Thus we
have 4k+2 couplings, k parameters and 3k+1 conditions giving a
naive expectation for the dimension of the manifold of fixed
points 4k+2+k-(3k+1)=2k+1=7. The one loop analysis gives: ($C_i$
defined in previous subsection)

\begin{eqnarray}
  B_i+A_i&=&-C_i\nonumber\\
  B_{i+1}+A_i+D&=&\gamma_i\\
  B_{i+1}+A_i+\tilde D&=&-\gamma_i\nonumber
\end{eqnarray}

By subtracting the second equation from the third we get: $\tilde
D-D=-2\gamma_i$, thus as in the previous case all $\gamma_i$ have
to be equal (to some $\gamma$). But $\sum_i\gamma_i=0$, so
$\gamma=0$. By subtracting the first equation from the second and
summing over i we get: $kD=\sum_iC_i$. And the general solution
is:

\begin{equation}
  B_{i+1}=B_i+C_i-D
\end{equation}

So, if we choose $B_1\equiv X$ the solution is:

\begin{eqnarray}
 B_i&=&X+\sum_{n=1}^{i-1}C_n-(i-1)D\nonumber\\
 A_i&=&-B_i-C_i=-(X+\sum_{n=1}^iC_n-(i-1)D).
\end{eqnarray}

Obviously we have a constraint on D: $\sum_iC_i=kD$. For the
$\alpha$s and $\delta$s we get:

\begin{eqnarray}
  \delta_{i-1}^2&=&X+4g_i^2+\sum_{n=1}^{i-1}C_n-(i-1)D\\
  \alpha_i^2&=&4g_i^2-X-(\sum_{n=1}^iC_n-(i-1)D).\nonumber
\end{eqnarray}

 It is parameterized by {\bf X}, gauge couplings, $h_i$s and D subject to
 a
 constraint  $\to$ giving a total of (1+3+3+1)-1=7 dimensional manifold of
 marginal deformations.

 Here again we can extend the solution to higher loops using the
 procedure we described in general k case, again here we will have:

\begin{eqnarray}
  \tilde B_{i+1}^{(n)}+\tilde A_i^{(n)}+D^{(n)}&=&\gamma_{Q_i}^{(n)(1-loop)}\\
  \tilde B_{i+1}^{(n)}+\tilde A_i^{(n)}+\tilde D^{(n)}&=&\gamma_{\tilde Q_i}^{(n)(1-loop)}\nonumber\\
  \tilde B_i^{(n)}+\tilde A_i^{(n)}+C_i^{(n)}&=&{N^2-1\over N^2}\gamma_{\Phi_i}^{(n)(1-loop)}\nonumber
\end{eqnarray}

Now we parameterize the remaining contributions to $\gamma$s as:

\begin{eqnarray}
    \gamma_{Q_i}^{(n)(2..n-loops)}&\equiv&T_i^{(n)}+S_{i+1}^{(n)}-D^{(n)}+\tilde\gamma_i^{(n)}\\
    \gamma_{\tilde Q_i}^{(n)(2..n-loops)}&\equiv&T_i^{(n)}+S_{i+1}^{(n)}-\tilde D^{(n)}-\tilde\gamma_i^{(n)}\nonumber\\
    {N^2-1\over N^2}\gamma_{\Phi_i}^{(n)(2..n-loops)}&\equiv&T_i^{(n)}+S_i^{(n)}-C_i^{(n)}\nonumber
\end{eqnarray}

Where the different quantities are defined as :

\begin{eqnarray}
 \sum_iC_i^{(n)}-kD^{(n)}&\equiv&\sum_i{\gamma_{Q_i}^{(n)(2..n-loops)}-{N^2-1\over
N^2}\gamma_{\Phi_i}^{(n)(2..n-loops)}}\nonumber\\
\sum_iC_i^{(n)}-k\tilde D^{(n)}&\equiv&\sum_i{\gamma_{\tilde
Q_i}^{(n)(2..n-loops)}-{N^2-1\over
N^2}\gamma_{\Phi_i}^{(n)(2..n-loops)}}\nonumber\\
 \Delta X^{(n)}&=& S_1^{(n)}(\equiv 0)\\
S^{(n)}_{i+1}-S^{(n)}_i+C_i^{(n)}-D^{(n)}+\tilde\gamma_i^{(n)}&=&
\gamma_{Q_i}^{(n)(2..n-loops)}-{N^2-1\over
N^2}\gamma_{\Phi_i}^{(n)(2..n-loops)}\nonumber\\
2\gamma_i^{(n)}&=&\gamma_{Q_i}^{(n)(2..n-loops)}-\gamma_{\tilde
Q_i}^{(n)(2..n-loops)}-(D^{(n)}-\tilde D^{(n)}).\nonumber
\end{eqnarray}

We can consider solution to be defined by $h_i$s, {\bf X} and the
gauge couplings, and then all other quantities can be expressed in
terms of these 7 ones as we see above. The first two expressions
can be seen as defining $D$ and $\tilde D$ ( we can set
$C_i^{(n)}=0$ for $n>1$, setting nonzero value is just
redefinition of $h_i$s), the fifth equation is definition of
$\tilde\gamma_i$, and the fourth is definition of $S_i$, again
$T_i$ can be defined from here.

 The solution as before is (demanding $\gamma_{Q_i}=-\gamma_{\tilde Q_i}=\gamma_i,\gamma_{\Phi_i}=0$):

 \begin{eqnarray}
\tilde A_i^{(n)} &=& -T_i^{(n)}\\
\tilde B_i^{(n)} &=& -S_i^{(n)}\nonumber\\
\tilde\gamma_i^{(n)}&=&\gamma_i^{(n)}\nonumber
\end{eqnarray}

(We can see that $\sum_i\tilde\gamma_i^{(n)}=0$ as required for
$\gamma_i$).

 Again, we have proven existence of an all orders solution.
In the k=3 case the switching on of the $h_i$'s doesn't produce
any new conditions, so unlike the general k case here the case of
$h_i=0$ and the case $h_i\neq 0$ can be treated together, thus
giving that in this case the space of fixed points is {\bf 7}
dimensional.

Here the {\bf k+1=4} directions are as in the general {\bf k}
case, and {\bf 1} deformation coming from $\kappa$ and
$\tilde\kappa$ is due to the $h_{iii}$ interactions of the ${\cal
N}=4$ theory which survive orbifolding\footnote{So we know that it
is \textit{exactly} marginal at least in the large {\bf N}
limit.}. We get {\bf 2} additional \textit{exactly} marginal
deformations which we don't see in the general case and do not
come from the ${\cal N}=4$ theory $\to$ this is a prediction of
our analysis.

\newpage

\subsection{k=2 case}\label{sec_k_2}

In this case the fields $Q_1$ and $\tilde Q_2$ are in the same
gauge group representations, and also the fields $Q_2$ and $\tilde
Q_1$ are. What this implies is that we have here a global
$SU(2)^2$ symmetry rotating these fields and we can write the most
general marginal deformations of this theory as:

\begin{eqnarray}
  W=\Phi_1
\begin{pmatrix}
  Q_1 &
  \tilde Q_2
\end{pmatrix}
\begin{pmatrix}
  \alpha & s \\
  p & \delta
\end{pmatrix}
\begin{pmatrix}
  \tilde Q_1 \\ Q_2
\end{pmatrix}+
  \Phi_2
\begin{pmatrix}
  Q_2 &
  \tilde Q_1
\end{pmatrix}
\begin{pmatrix}
  \alpha' & s' \\
  p' & \delta'
\end{pmatrix}
\begin{pmatrix}
  \tilde Q_2 \\ Q_1
\end{pmatrix}
  \end{eqnarray}

Lets make the Leigh-Strassler analysis. Using the global $SU(2)^2$
symmetry we can diagonalize the anomalous dimensions (apriori we
can have mixing terms here). Thus we will have here additional
$\beta$-functions:

\begin{eqnarray}
  \beta_{s,s'}&\propto&\gamma_{Q_{1,2}}+\gamma_{ Q_{2,1}}+\gamma_{\Phi_{1,2}}=0\nonumber\\
  \beta_{p',p}&\propto&\gamma_{\tilde Q_{1,2}}+\gamma_{\tilde Q_{2,1}}+\gamma_{\Phi_{2,1}}=0
  \end{eqnarray}

  From here and as before we conclude that
  $\gamma_{\Phi_1}=\gamma_{\Phi_2}\equiv \gamma$,
  $\gamma_{Q_1}=\gamma_{\tilde Q_2}\equiv \tilde \gamma$ and
  $\gamma_{Q_2}=\gamma_{\tilde Q_1}\equiv -\gamma-\tilde \gamma$.

So we have here {\bf 10} couplings, {\bf 2} $\gamma$s $\to$ {\bf
12} parameters. We have {\bf 8} anomalous dimensions  $\to$ we
have {\bf 8} constraints. Thus, we expect a {\bf 4} dimensional
manifold of fixed points with non zero new interactions.

This case is a special one of the $\mathbb{Z}_k$ {\bf (a,a,-2a)}
orbifold theory ( with $\mathbb{Z}_{k=2}$ and {\bf a=1}). In the
${\cal N}=1$ chapter we will deal with this case extensively and
so we will postpone our discussion until then. In particular we
will see that actually we get only a {\bf 3} dimensional manifold
of fixed points because of the one-loop structure.

\newpage
\subsection{SU(N=3) case}\label{s:N_3}

 In this case one can add an operator ${\rho_i\over
 3!}\epsilon_{lmn}\epsilon^{abc}Q^{(i)l}_aQ^{(i)m}_bQ^{(i)n}_c$ and ${\tilde\rho_i\over
 3!}\epsilon_{lmn}\epsilon^{abc}\tilde Q^{(i)l}_a\tilde
Q^{(i)m}_b\tilde Q^{(i)n}_c$. Again we begin by analyzing the
 vanishing $h_i$'s case.

\begin{eqnarray}
  \beta_{\rho_i}=3\rho_i\cdot\gamma_{Q_i}\nonumber\\
  \beta_{\tilde\rho_i}=3\tilde\rho_i\cdot\gamma_{\tilde Q_i}
\end{eqnarray}

  So from here and (~\ref{betas}) we
  have for all i $\gamma_{Q_i}=0$, $\gamma_{\tilde Q_i}=0$, $\gamma_{\Phi_i}=\gamma$ and
 $\gamma_{Q_i}+\gamma_{\tilde Q_i}=-\gamma\to\gamma_{\Phi_i}=0$.
 Thus we see that as in the k=3 case adding non zero $h_i$s doesn't
 change the conditions for the $\gamma$s ($\gamma_{\Phi_i}$ has to be zero anyway), so we can consider them
 together.
 We have here 6k couplings and 3k
 conditions, leading naively to a 3k dimensional manifold
 of fixed points.
 The interactions we add affect the one loop $\gamma_{\Phi_i}$ as in
 (~\ref{1_loop_C},) and we
 add a term $2\rho_i^2(\equiv K_i)$ for Q and
 $2\rho_i^2(\equiv\tilde K_i)$ for $\tilde Q$:

\begin{eqnarray}
  B_i+A_i&=&-C_i\nonumber\\
  B_{i+1}+A_i&=&-K_i\\
  B_{i+1}+A_i&=&-\tilde K_i\nonumber
\end{eqnarray}

Again we get here that at one loop there are some conditions:

\begin{eqnarray}\label{ccon}
   \sum_i C_i&=&\sum_i K_i\\
   K_i&=&\tilde K_i\nonumber
\end{eqnarray}

And the one loop solution is:

\begin{eqnarray}
 B_i&=&X+\sum_{n=1}^{i-1}(C_n-K_n)\nonumber\\
 A_i&=&-C_i-B_i=-X+\sum_{n=1}^{i-1}K_n-\sum_{n=1}^{i}C_n
\end{eqnarray}

Implying for $\alpha$s and $\delta$s:

\begin{eqnarray}
 \delta_{i-1}^2&=&X+4g_i^2+\sum_{n=1}^{i-1}(C_n-K_n)\nonumber\\
 \alpha_i^2&=&4g_i^2-X+\sum_{n=1}^{i-1}K_n-\sum_{n=1}^{i}C_n.
\end{eqnarray}

 So the general solution is parameterized by k gauge couplings, the {\bf
 X}
 parameter, k $\rho_i$s and k $h_i$s subject to the condition (~\ref{ccon}) above,
  giving a total of 3k parameters $\to$ we get 3k dimensional
  manifold of fixed points as expected.

In the k=3 case we can have two additional marginal operators (see
section ~\ref{s:k_3}), they will add 2 parameters to our analysis
and no extra conditions. So we will expect to find a 3k+2=11
dimensional manifold of fixed points.

 We can here also extend\footnote{Here essentially because the solution we get at one loop
  is exactly as predicted from the Leigh-Strassler analysis the existence of the solution to all orders
   is guaranteed. Nevertheless we write the all loop extension for completeness.}
  the solution to all orders in
 perturbation theory. We choose to parameterize our solution by k
 gauge couplings, the {\bf X} parameter, k $K_i$s and (k-1) $C_i$s
 ($i\in(1,..,(k-1))$). Again we can write:

\begin{eqnarray}
  \tilde B_{i+1}^{(n)}+\tilde A_i^{(n)}+K_i^{(n)}&=&\gamma_{Q_i}^{(n)(1-loop)}\\
  \tilde B_{i+1}^{(n)}+\tilde A_i^{(n)}+\tilde K_i^{(n)}&=&\gamma_{\tilde Q_i}^{(n)(1-loop)}\nonumber\\
  \tilde B_i^{(n)}+\tilde A_i^{(n)}+C_i^{(n)}&=&{N^2-1\over N^2}\gamma_{\Phi_i}^{(n)(1-loop)}.\nonumber
\end{eqnarray}

Now we parameterize the remaining contributions to $\gamma$s as:

\begin{eqnarray}
    \gamma_{Q_i}^{(n)(2..n-loops)}&\equiv&T_i^{(n)}+S_{i+1}^{(n)}-K_i^{(n)}\\
    \gamma_{\tilde Q_i}^{(n)(2..n-loops)}&\equiv&T_i^{(n)}+S_{i+1}^{(n)}-\tilde K_i^{(n)}\nonumber\\
    {N^2-1\over N^2}\gamma_{\Phi_i}^{(n)(2..n-loops)}&\equiv&T_i^{(n)}+S_i^{(n)}-C_i^{(n)},\nonumber
\end{eqnarray}

where the different quantities are defined as:

\begin{eqnarray}
 \sum_i(C_i^{(n)}-K_i^{(n)})&\equiv&\sum_i{\gamma_{Q_i}^{(n)(2..n-loops)}-{N^2-1\over
N^2}\gamma_{\Phi_i}^{(n)(2..n-loops)}}\nonumber\\
\sum_i(C_i^{(n)}-\tilde K_i^{(n)})&\equiv&\sum_i{\gamma_{\tilde
Q_i}^{(n)(2..n-loops)}-{N^2-1\over
N^2}\gamma_{\Phi_i}^{(n)(2..n-loops)}}\nonumber\\
 \Delta X^{(n)}&=& S_1^{(n)}(\equiv 0)\\
S^{(n)}_{i+1}-S^{(n)}_i+C_i^{(n)}-K_i^{(n)}&=&
\gamma_{Q_i}^{(n)(2..n-loops)}-{N^2-1\over
N^2}\gamma_{\Phi_i}^{(n)(2..n-loops)}\nonumber\\
\gamma_{Q_i}^{(n)(2..n-loops)}-\gamma_{\tilde
Q_i}^{(n)(2..n-loops)}&=&(K_i^{(n)}-\tilde K_i^{(n)}).\nonumber
\end{eqnarray}

The last line can be seen as the definition of $\tilde K_i$ and
the first equation can be seen as the definition of $C_k$. And
finally again we demand:

\begin{eqnarray}
\tilde A_i^{(n)} &=& -T_i^{(n)}\\
\tilde B_i^{(n)} &=& -S_i^{(n)}.\nonumber\\
\end{eqnarray}

In the SU(3) k=3 case the calculation is very similar.

  So to conclude: in general k with SU(3) group we get {\bf 3k}
  \textit{exactly}
 marginal directions, and in k=3, SU(3) we get {\bf 11} \textit{exactly} marginal directions.

\newpage

\section{RG flow analysis}

   As we saw in the ${\cal N}=4$ case the dependence of the $\beta$-
   functions on the anomalous dimensions can teach us about the
   flow lines in the space of the coupling constants.
   The main point is that from (~\ref{betas}) we get
  relations between the couplings which are obeyed during the RG
  flow. We rewrite (~\ref{betas}) here:

\begin{eqnarray}\label{betas_RG}
  \beta_{g_i}&=&-{2g^3\over 16\pi^2}{N\over
  1-{2Ng^2\over16\pi^2}}({1\over2}(\gamma_{Q_i}+\gamma_{\tilde Q_i}+\gamma_{Q_{i-1}}+\gamma_{\tilde Q_{i-1}})+\gamma_{\Phi_i})\nonumber\\
  \beta_{h_i}&=&3h_i\cdot \gamma _{\Phi_i}\nonumber\\
  \beta_{\alpha_i}&=&\alpha_i(\gamma_{Q_i}+\gamma_{\tilde Q_i}+\gamma_{\Phi_i})\\
  \beta_{\delta_i}&=&\delta_i(\gamma_{Q_i}+\gamma_{\tilde Q_i}+\gamma_{\Phi_{i+1}}).\nonumber
\end{eqnarray}

From here we conclude that (We do same the rescaling of couplings
as in the ${\cal N}=4$ section):

\begin{eqnarray}
-{1-2g_i^2\over g_i^3}{\partial g_i\over\partial \ln\mu}& =
&{\partial \ln \delta_{i-1} \over\partial \ln\mu}+{\partial
\ln\alpha_i\over\partial
\ln\mu}\nonumber\\
{\partial \ln \delta_i \over\partial \ln\mu}-{\partial
\ln\alpha_i\over\partial \ln\mu}&=&{1\over3}({\partial \ln h_{i+1}
\over\partial \ln\mu}-{\partial \ln h_i\over\partial \ln\mu})
\end{eqnarray}

So we get :

\begin{eqnarray}\label{rell}
    \alpha_i\delta_{i-1}&\propto& g_i^2exp( {1\over2g_i^2}),\nonumber\\
    ({\delta_i\over \alpha_i})^3&\propto& {h_{i+1}\over h_i}.
\end{eqnarray}

Here the proportionality factors are determined by the initial
conditions and are not changed by the RG flow. In the ${\cal N}=4$
case these constraints on the flow were sufficient to determine
the flow lines. Here however we see that we only have some
relations which have to be obeyed during the flow.

 The expressions above are of explicit non perturbative nature.
 As an example of the use of these relations (~\ref{rell})
 consider this:
 we can use the solutions above to find to what value of {\bf X} the
 couplings will flow from given initial conditions, in terms of the final gauge couplings.
  We have:$\alpha_i\delta_{i-1}={\cal A} g_i^2exp(
 {1\over2g_i^2})$ (Where the initial conditions are encoded in
 ${\cal A}$). So using the one loop solution from general k
 section we obtain: $X^2=16g_i^4-{\cal A}^2g_i^4exp({1\over g_i^2})$.
 Thus we see that if we know the final gauge couplings and the
 initial parameters we know which value of {\bf
 X} we will flow to\footnote{Because the {\bf X} parameter is obtained from perturbation theory
 the argument is only valid if we end up in the small coupling regime.}.

\newpage

\chapter{${\cal N}=1$ theory}

From the discussion in the previous chapter we see that to get
${\cal N}=1$ we need to have an $\overrightarrow{a}$ vector
satisfying: no one of the $a_i$s is zero and
$a_1+a_2+a_3=0(mod{\bf k})$. We get gauge group $SU(N)^k$, and we
have matter in the representations (chiral fields):

\begin{eqnarray}
  \oplus_{i=0}^{k-1}(1,1...1,N_{(i)},1...1,\bar N_{(i+a_1)},1....1,1)\nonumber\\
  \oplus_{i=0}^{k-1}(1,1...1,N_{(i)},1...1,\bar N_{(i+a_2)},1....1,1)\nonumber\\
  \oplus_{i=0}^{k-1}(1,1...1,N_{(i)},1...1,\bar
  N_{(i+a_3)},1....1,1).
\end{eqnarray}

We will denote the fields in our theory by $Q_l^I$ where $I\in
(0,...,k-1)$, $l\in(1,2,3)$. The index {\bf I} denotes the group
SU(N) of which the field is in the fundamental representation, and
{\bf l} denotes the $\overrightarrow{a}$ component.

In this chapter we won't be interested in the potential coming
from the orbifold theory but rather in the most general
superpotential with this matter content. The only superpotential
we can get here is of the type we got in the k=3 case in the
previous chapter, because we don't have here fields in the adjoint
representation. This implies that the only
superpotential\footnote{It is very easy to read out all the
possible interactions (for $SU(N\neq3)$) from the quiver diagrams:
all oriented triangles in the diagram correspond to an interaction
term (see the examples in the next sections).}
 possible here is of the form:

\begin{eqnarray}
  W=h^I_{lmn}Q^I_lQ^{I+a_l}_mQ^{I+a_l+a_m}_n,
\end{eqnarray}

where we have to assume that $a_l+a_m+a_n=0(mod{\bf k})$. The
definition of the couplings $ h^I_{lmn}$ in this way is redundant,
we see that:

\begin{eqnarray}\label{redu}
   h^I_{lmn}= h^{I+a_l}_{mnl}= h^{I+a_l+a_m}_{nlm}.
\end{eqnarray}

Obviously for a general choice of k and the $\overrightarrow{a}$
vector, the only possibility for this to be true is by taking
(l,m,n) to be some permutation of (1,2,3)\footnote{We will see
examples with larger possibilities, in particular an example of
$\mathbb{C}^3/\mathbb{Z}_3$ where the possibilities are much
larger.}, so we will start by constraining our research to this
case.

 From the string theory arguments we have to have here at least one
 exactly marginal direction at large {\bf N}
 parameterized by the string coupling ($g_{string}$) of the dual string theory. We will see now that we have for any k
 another marginal deformation.

 First look at the $\beta$-functions. The one loop gauge
 $\beta$-function is proportional to: $3C_2(G)-\sum_AT_A$. In our
 case $C_2(G)=N$ for SU(N) and $T_A$ is ${1\over 2}$ for
 (anti)fundamental representations of SU(N). For each gauge group
 the matter content above implies that we effectively have 6N
 chiral multiplets\footnote{Work this out for the i'th gauge group: we have
 $Q^i_l,Q^{i-a_1}_1,Q^{i-a_2}_2,Q^{i-a_3}_3$ matter fields
 transforming non trivially under this group. The first one is
 $3\times N$ fields, and the three others give N fields each
 (N being the index of the other gauge group in which they
 transform), giving a total of 6N.}, so the one loop result is
 proportional to
 $3N-{1\over2}6N=0$.

 Further we see that $\langle Q^{\dag I}_lQ^J_l\rangle\propto\delta_{IJ}$
 from gauge symmetry considerations, so the only mixing allowed is
 between the lower indices $\to$ the $\gamma$s can be written as
 $\gamma^I_{lm}$, and obviously $\gamma^I_{lm}$ can be non
 vanishing only if $a_l=a_m$. We conclude from NSVZ that:

 \begin{eqnarray}\label{betta1}
   \beta_{g_I}\propto
   Tr\gamma^I+\gamma^{I-a_1}_{11}+\gamma^{I-a_2}_{22}+\gamma^{I-a_3}_{33}.
 \end{eqnarray}

 For superpotential couplings we have the usual expression coming from the
 general formula (~\ref{betaY}):

 \begin{eqnarray}\label{betta2}
 \beta_{h^I_{lmn}}\propto
 h^I_{pmn}\gamma^I_{pl}+h^I_{lpn}\gamma^{I+a_l}_{pm}+h^I_{lmp}\gamma^{I+a_l+a_m}_{pn}.
 \end{eqnarray}

Obviously if we demand $\gamma=0$ for all $\gamma$s the
$\beta$-functions will vanish (we will see that actually we will
have solutions also for non vanishing $\gamma$s in some cases).

 Let us constrain
  our theory further to get some general result: we will assume that all k gauge couplings are equal and
 that $h^I_{lmn}\equiv h$ for (lmn) an even permutation of (123), and
 $h^I_{lmn}\equiv h'$ for (lmn) an odd permutation of
 (123)\footnote{The superpotential coming from the orbifold is of
 this
 form~\cite{a:quiver3}:
 $\sum_IQ^I_1Q^{I+a_1}_2Q^{I+a_1+a_2}_3-Q^I_2Q^{I+a_2}_1Q^{I+a_1+a_2}_3$.
}. It is easy to see that in this case we have
 two discrete symmetries: by taking $Q^I_l\to Q^I_{l+1}$ we remain with the same theory, and also
 by taking $Q_l^{I}\to Q_l^{I+1}$ we also remain with same theory.
 These symmetries obviously imply that all the $\gamma$ function of the matter fields here are the same.
  And from the demand for vanishing $\beta$-functions we will have to demand
  $\gamma=0$. $\gamma$ is a function of three parameters and thus we
   expect a two dimensional manifold of fixed points.

 Let us look at the one loop contributions to the $\gamma$-functions. We can schematically
 write them as\footnote{We define
${\cal B}={1\over16\pi^2}{N^2-1\over N}$, ${\cal
A}={1\over16\pi^2}{N\over 2}$.}:

 \begin{eqnarray}
   \gamma_{Q_l^I}={\cal A}(h^2+h^{'2})-2{\cal B}g^2=0.
 \end{eqnarray}

 Thus we see that we have the same demand for all the $\gamma$-functions, so we will have a solution:

 \begin{eqnarray}
   h^{'2}(g,h)=2{{\cal B}\over{\cal A}}g^2-h^2.
 \end{eqnarray}

This can easily be seen to generalize to all orders of
perturbation theory (see the next section for the details). Thus
we see that in any ${\cal N}=1$ orbifold theory we have at least
two \textit{exactly} marginal directions at weak coupling,
parameterized by the gauge coupling {\bf g} and the parameter {\bf
h}.

 When the gauge group is SU(N=3) the manifold of fixed points is
 larger. In this case we can add a superpotential of the form:
 $(Q_l^I)^3$. Again from considerations of vanishing of the
 $\beta$-functions we can constrain ourselves to the case where all
 the couplings of the new superpotential are equal (equal to $\rho$).

 Thus again if we constrain ourselves as above (all gauge couplings
 equal and the potential including only {\bf h} and {\bf
 h'} couplings) we have the discrete symmetry and we can schematically
write the anomalous dimensions at one loop as:

\begin{eqnarray}
   \gamma_{Q_l^I}={\cal A}(h^2+h^{'2})+{\cal C}\rho^2-2{\cal
   B}g^2=0.
 \end{eqnarray}

 This leads to a solution:

 \begin{eqnarray}
   h^{'2}(g,h,\rho)=2{{\cal B}\over{\cal A}}g^2-h^2-{{\cal C}\over{\cal
   A}}\rho^2.
 \end{eqnarray}

Thus we see that for $SU(N=3)$ and general k we have at least a
{\bf 3} dimensional manifold of fixed points. Again we have to
stress that the actual manifold for every case can be much larger,
depending on the specific choice of the  $\overrightarrow{a}$
vector.

Now we wish to classify all the possibilities with a larger number
of possible superpotentials. First we denote:
$\overrightarrow{a}\equiv(a,b,-a-b)$, and without any loss of
generality we assume that ${k\over2}\geq a,b>0$.

The possibilities to have other superpotentials than the one
described above are:

\begin{itemize}
 \item {\bf (I)} b+2a=k\\From here we conclude that
 $\overrightarrow{a}\equiv(a,k-2a,a-k)=(a,-2a,a)$.
 \item {\bf (I')} 2a-(a+b)=0\\From here we conclude that
       $\overrightarrow{a}\equiv(a,a,-2a)$, so we see it's
       essentially the same case as above.
 \item {\bf (II)} 3b=k\\From here we conclude: $\overrightarrow{a}\equiv(a,{k\over
3},-a-{k\over3})$.
 \item {\bf (III)} 3(a+b)=k\\From here we conclude: $\overrightarrow{a}\equiv(a,{k\over
3}-a,-{k\over3})$, we see that this case is essentially the same
as the previous one (by taking
$\overrightarrow{a}\to-\overrightarrow{a}$ and $a\to-a$).
\end{itemize}

We see that the three distinct cases above have intersections:
$\overrightarrow{a}\equiv({k\over 3},{k\over 3},{k\over3})$ is in
all of the cases above and $\overrightarrow{a}\equiv({k\over
6},{k\over 6},{2k\over3})$ is in the intersection of {\bf (I)} and
{\bf (III)}. Thus to summarize we have these different cases to
deal with:

\begin{itemize}
 \item general $k$ $\to$ $\overrightarrow{a}\equiv(a,a,-2a)$,
  \item $k=3k'$    $\overrightarrow{a}\equiv(a,{k\over
3}-a,-{k\over3})$,
\end{itemize}

 And two special cases: $\overrightarrow{a}\equiv({k\over 3},{k\over
 3},{k\over3})$, $\overrightarrow{a}\equiv({k\over
6},{k\over 6},{2k\over3})$.

When {\bf a,b} and {\bf k} have a common divisor larger than one (
which we denote by {\bf J}), then essentially we are in the
$\mathbb{Z}_{k\over J}$, ${1\over J}\overrightarrow{a}$ theory. In
particular
 there is no meaning to discussing the $\overrightarrow{a}\equiv({k\over 3},{k\over
 3},{k\over3})$, $\overrightarrow{a}\equiv({k\over
6},{k\over 6},{2k\over3})$ theories for general {\bf k}, they all
are equivalent to the ones with k=3,6 respectively. So we will
assume that {\bf a,b} and {\bf k} have no non trivial common
divisor.

Now we will deal with the specific cases. First with the most
general one, then with the case where two of the
$\overrightarrow{a}$ components equal, then with the k=3k' case
and finally with the two special cases.

\newpage

\section{(a,b,-a-b)}

In this section we consider a theory which does not have any
additional interactions (except the ones which exist in any
orbifold theory). As an example of such a theory we will keep in
mind the $\mathbb{C}^3/\mathbb{Z}_7$ orbifold theory with
$\overrightarrow{a}=(1,2,4)$. We see that the only way here that
$a_l+a_m+a_n=0(mod{\bf k})$ is if $(a_l,a_m,a_n)=(1,2,4)$ (as
sets). This theory can be represented by the following quiver
diagram:
 $ $\\
 $ $\\
\begin{figure}[htbp]
\begin{fmffile}{quiqyI}
\unitlength=1mm
\begin{center}
\begin{fmfchar*}(50,50)
 \fmfpen{thin}
 \fmfsurround{v1,v2,v3,v4,v5,v6,v7}
 \fmf{dots_arrow}{v1,v2,v3,v4,v5,v6,v7,v1}
 \fmf{plain_arrow,left=.1}{v1,v3,v5,v7,v2,v4,v6,v1}
 \fmf{dashes_arrow,left=.1}{v1,v5,v2,v6,v3,v7,v4,v1}
 \fmfblob{150}{v1} \fmfblob{150}{v2} \fmfblob{150}{v3} \fmfblob{150}{v4}
 \fmfblob{150}{v5} \fmfblob{150}{v6} \fmfblob{150}{v7}
\end{fmfchar*}
\caption{$\mathbb{C}^3/\mathbb{Z}_7$ (1,2,4) quiver diagram}
\label{quiv3}
\end{center}
\end{fmffile}
\end{figure}
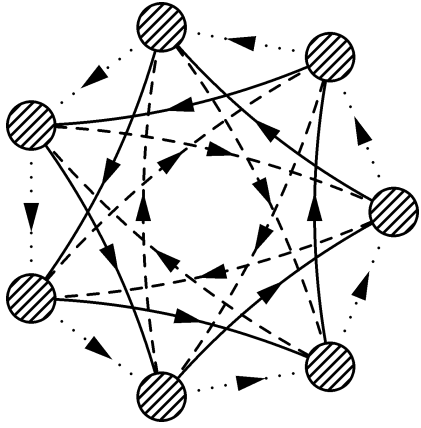

Here the dotted lines represent the $a_l=1$ sector, the plain
lines represent $a_l=2$ sector and the dashed lines represent the
$a_l=4$ sector.

 We use here the notations from the first section of this chapter,
 denoting the fields
  $Q^I_l$. From the redundancy condition (~\ref{redu})
 we can write:

 \begin{eqnarray}
   h^I_{123}&=&h^{I+a}_{231}=h^{I+a+b}_{312}\nonumber\\
   h^I_{132}&=&h^{I+a}_{321}=h^{I-b}_{213}.
 \end{eqnarray}

 Thus we see that we have actually only $2\times k$ independent
 couplings here, $h^I_{123}\equiv h_I$, $h^I_{132}\equiv h'_I$.

 First we do the Leigh-Strassler analysis. From the general
 $\beta$-functions (~\ref{betta1},~\ref{betta2}) we obtain, using the fact that in our
 case there is no mixing between the fields:

 \begin{eqnarray}\label{an_dim}
   \beta_{h_I}&\propto& \gamma_1^I+\gamma_2^{I+a}+\gamma_3^{I+a+b}\nonumber\\
   \beta_{h'_I}&\propto&
   \gamma_1^I+\gamma_3^{I+a}+\gamma_2^{I-b}\nonumber\\
   \beta_{g_I}&\propto&
   \gamma_1^I+\gamma_2^I+\gamma_3^I+\gamma_1^{I-a}+\gamma_2^{I-b}+\gamma_3^{I+a+b}.
 \end{eqnarray}

Now let's define the largest common divisor of k and {\bf a} by
$\alpha$, of k and {\bf b} by $\beta$ and of k with {\bf (a+b)} by
$\gamma$. Let's define ${\cal S}_{a_i}^I$ to be the set of indices
$(I,I+a_i,I+2a_i,...)$.

From these definitions we calculate\footnote{Here $f(g)={1\over
16\pi^2}{2g^3C_1\over 1-{2C_1g^2\over 16\pi^2}}$.}:

\begin{eqnarray}
  \sum_{{\cal S}_{a}^I}{\beta_{h_I}\over h_I}&\propto& \sum_{{\cal
      S}_{a}^I}(\gamma_1^I+\gamma_2^{I})+\sum_{{\cal S}_{a}^{I+b}}\gamma_3^{I+b}\nonumber\\
   \sum_{{\cal S}_{a}^I}{\beta_{h'_I}\over h'_I}&\propto&
   \sum_{{\cal S}_{a}^I}(\gamma_1^I+\gamma_3^{I})+\sum_{{\cal S}_{a}^{I-b}}\gamma_2^{I-b}\nonumber\\
   \sum_{{\cal S}_{a}^I}{\beta_{g_I}\over f(g_I)}&\propto& \sum_{{\cal
      S}_{a}^I}(2\gamma_1^I+\gamma_2^I+\gamma_3^I)+\sum_{{\cal S}_{a}^{I-b}}\gamma_2^{I-b}+\sum_{{\cal
      S}_{a}^{I+b}}\gamma_3^{I+b}.
 \end{eqnarray}

 From here we see that  $\sum_{{\cal S}_{a}^I}{\beta_{g_I}\over f(g_I)}\propto
 \sum_{{\cal S}_{a}^I}{\beta_{h_I}\over h_I}+ \sum_{{\cal
 S}_{a}^I}{\beta_{h'_I}\over h'_I}$, so our system of linear equations
 is dependent. The number of such dependencies is obviously $\alpha$
 (because there are ${k\over\alpha}$ elements in ${\cal S}_{a}^I$).
We can do the same procedure for {\bf b} and {\bf (a+b)}. We will
 get $\sum_{{\cal S}_{b}^I}{\beta_{g_I}\over f(g_I)}\propto
 \sum_{{\cal S}_{b}^I}{\beta_{h_I}\over h_I}+ \sum_{{\cal
 S}_{b}^I}{\beta_{h'_I}\over h'_I}$, $\sum_{{\cal S}_{a+b}^I}{\beta_{g_I}\over f(g_I)}\propto
 \sum_{{\cal S}_{a+b}^I}{\beta_{h_I}\over h_I}+ \sum_{{\cal
 S}_{a+b}^I}{\beta_{h'_I}\over h'_I}$. These three relations are
 not completely independent. Obviously by summing over {\bf I} the
 three relations we get same constraint $\to$ we get from here
 $(\alpha+\beta+\gamma-2)$ relations.

The $\beta_{h_I}$ and $\beta_{h'_I}$ are also not
 completely independent: $\sum_I{\beta_{h_I}\over h_I}=\sum_I{\beta_{h'_I}\over
 h'_I}$, which gives another relation.

  Thus, we
  have $(\alpha+\beta+\gamma-1)$ linear relations
 between the $\beta$-functions.

  Finally, we count our degrees of freedom: we have {\bf k} gauge
  couplings,  {\bf k} $h_I$s and $h'_I$s $\to$ a total of 3k
  parameters. We have {\bf
  $3k-(\alpha+\beta+\gamma-1)$} independent
  equations $\to$ we expect an {\bf
  $(\alpha+\beta+\gamma-1)$} dimensional
  manifold of fixed points. As a byproduct of this the anomalous
  dimensions don't have to vanish.

  Let's now calculate the possible values for the anomalous
  dimensions (without any loss of generality we assume $\gamma\leq\alpha,\beta$) . From (~\ref{an_dim}) we get:

  \begin{eqnarray}
  -\gamma^I_3&=&\gamma_2^{I-b}+\gamma_1^{I-a-b}\nonumber\\
   -\gamma^I_3&=&\gamma_1^{I-a}+\gamma_2^{I-a-b}\nonumber\\
    -\gamma^I_3-\gamma^{I+a+b}_3&=&\gamma_2^{I-b}+\gamma_1^{I-a}+\gamma^I_1+\gamma^I_2.
  \end{eqnarray}

  From here:

  \begin{eqnarray}
\gamma_2^{I-b}+\gamma_1^{I-a-b}&=&\gamma_2^{I-b-a}+\gamma_1^{I-a}\nonumber\\
\gamma_2^{I-b}+\gamma_1^{I}&=&\gamma_2^{I-a-b}+\gamma_1^{I+b}.
  \end{eqnarray}

  And finally we obtain:

  \begin{eqnarray}
     \gamma_1^{I-a-b}-\gamma_1^{I-a}=\gamma_1^{I}-\gamma_1^{I+b}.
  \end{eqnarray}

  We see that if we define $\gamma_1^{I}-\gamma_1^{I+b}\equiv
  K_I$ then $K_I=K_{I-a-b}$, so we have $\gamma$ independent $K_I$s. And we see that
  $\sum_{{\cal S}_{b}^I}K_I=0$.

  We can look on $\sum_{{\cal S}_{b}^I,{\cal S}_{-a-b}^I}K_I={k\over\gamma}\sum_{{\cal
  S}_{b}^I}K_I$. The left hand side is obviously invariant
  under $I\to I+a,I+b,I-a-b$, thus we conclude that if $\sum_{{\cal
  S}_{b}^I}K_I=0$ for some {\bf I} then it is true for any {\bf
  I}. And also by the same arguments if $\sum_{{\cal
  S}_{b}^I}K_I=0$ then also $\sum_{{\cal
  S}_{a}^I}K_I=0$.
  Thus essentially we have one constraint on $\gamma$
  $K_I$s.

  Further we get that:

  \begin{eqnarray}\label{rels_gamma}
   \gamma_1^{I}+K_I&=&\gamma_1^{I+b}\nonumber\\
    \gamma_2^{I}+K_I&=&\gamma_2^{I+a}\nonumber\\
     -\gamma_3^{I}&=&\gamma_1^{I-a}+\gamma_2^{I-a-b}.
  \end{eqnarray}

 So finally we conclude that we have $\beta$ independent
$\gamma_1^I$s, $\alpha$ independent $\gamma_2^I$s and $\gamma-1$
independent $K_I$s $\to$   we can have $(\alpha+\beta+\gamma-1)$
  independent $\gamma$-functions as expected.

 Let's now write the one loop conditions for vanishing $\beta$-functions:

 \begin{eqnarray}
   \gamma_{Q^I_1}&=&{\cal A}(h^{I2}_{123}+h^{I2}_{132})-{\cal
   B}(g_{I}^2+g_{I+a}^2)=\gamma^I_1\nonumber\\
 \gamma_{Q^I_2}&=&{\cal A}(h^{I2}_{231}+h^{I2}_{213})-{\cal
   B}(g_{I}^2+g_{I+b}^2)=\gamma^I_2\nonumber\\
 \gamma_{Q^I_3}&=&{\cal A}(h^{I2}_{312}+h^{I2}_{321})-{\cal
   B}(g_{I}^2+g_{I-a-b}^2)=\gamma^I_3.
\end{eqnarray}

And so we get:

\begin{eqnarray}\label{1ll}
  \gamma_{Q^I_1}&=&{\cal A}(h^{2}_{I}+h^{'2}_{I})-{\cal
   B}(g_{I}^2+g_{I+a}^2)=\gamma^I_1\nonumber\\
 \gamma_{Q^I_2}&=&{\cal A}(h^{2}_{I-a}+h^{'2}_{I+b})-{\cal
   B}(g_{I}^2+g_{I+b}^2)=\gamma^I_2\nonumber\\
 \gamma_{Q^I_3}&=&{\cal A}(h^{2}_{I-a-b}+h^{'2}_{I-a})-{\cal
   B}(g_{I}^2+g_{I-a-b}^2)=\gamma^I_3.
\end{eqnarray}

From here we define:

 \begin{eqnarray}
  A_{I+a}&\equiv&{\cal A}h_I^2-{\cal B}g_{I+a}^2\quad\quad B_I\equiv
  {\cal A}h^{'2}_I-{\cal B}g_I^2\nonumber\\
  {1\over{\cal B}}C_I&\equiv&
  g_{I-a}^2+g_{I-b}^2-g_I^2-g_{I-a-b}^2.
 \end{eqnarray}

And further we obtain using (~\ref{rels_gamma}):

\begin{eqnarray}\label{one_l}
 \gamma_{Q_1^I}&=&A_{I+a}+B_I=\gamma_1^I\nonumber\\
 \gamma_{Q_2^I}&=&A_{I}+B_{I+b}=\gamma_2^I\nonumber\\
 \gamma_{Q_3^{I+a+b}}&=&A_{I+a}+B_{I+b}+C_{I+a+b}=-(\gamma_1^I+\gamma_2^I+K_I)
\end{eqnarray}

By subtracting the first equation from the third and summing over
${\cal S}_{b}^I$ we obtain that $\sum_{{\cal
S}_{b}^I}(2\gamma_1^I+\gamma_2^I+K_I)=\sum_{{\cal
S}_{b}^I}(2\gamma_1^I+\gamma_2^I)=0$, and further by using
(~\ref{rels_gamma}): $-{2k\over \beta}\gamma_1^I=\sum_{{\cal
S}_{b}^I}\gamma_2^I+2\sum_{j=0}^{{k\over\beta}-1}(({k\over\beta}-1-j)+1)K_{I+jb}$.

 By subtracting the second equation from the
third and summing over ${\cal S}_{a}^I$ we obtain that
$\sum_{{\cal S}_{a}^I}(\gamma_1^I+2\gamma_2^I+K_I)=\sum_{{\cal
S}_{a}^I}(\gamma_1^I+2\gamma_2^I)=0$, and further by using
(~\ref{rels_gamma}): $-{2k\over \alpha}\gamma_2^I=\sum_{{\cal
S}_{a}^I}\gamma_1^I+2\sum_{j=0}^{{k\over\alpha}-1}(({k\over\alpha}-1-j)+1)K_{I+ja}$.

Using the two relations above we get:

\begin{eqnarray}\label{condic}
{4k^2\over \alpha\beta}\gamma_1^I&=&\sum_{{\cal S}_{a}^I,{\cal
S}_{b}^I}\gamma_1^I+2\sum_{l,j}({k\over\alpha}-j)K_{I+ja+lb}
-{4k\over\alpha}\sum_{j=0}^{{k\over\beta}-1}({k\over\beta}-j)K_{I+jb}=\nonumber\\
&=&\sum_{{\cal S}_{a}^I,{\cal
S}_{b}^I}\gamma_1^I+{4k\over\alpha}\sum_{j=0}^{{k\over\beta}-1}jK_{I+jb}.
\end{eqnarray}

In last line we used $\sum_{{\cal S}^I_{a,b}}K^I=0$. Obviously
from here $\gamma_1^I+K_I=\gamma_1^{I+b}$ as we demand.

Now by subtracting the first two equations from one another in
(~\ref{one_l}) and summing over ${\cal S}_{a+b}^I$ we get
$\sum_{{\cal S}_{a+b}^{I+b}}\gamma_1^{I+b}=\sum_{{\cal
S}_{a+b}^I}\gamma_2^{I}$, and from here also
$\sum_I\gamma_1^I=\sum_I\gamma_2^I$ $\to$
$\sum_I\gamma_1^I=\sum_I\gamma_2^I=0$. From here and from
(~\ref{condic}) we see that
$\gamma_{1}^I=\beta\sum_{j=0}^{{k\over\beta}-1}jK_{I+jb}$ and
$\gamma_{2}^I=\alpha\sum_{j=0}^{{k\over\alpha}-1}jK_{I+ja}$. We
see that $\sum_{{\cal
S}_{a+b}^{I+b}}\gamma_1^{I+b}\propto\gamma_1^{I+b}$ and
$\sum_{{\cal S}_{a+b}^{I}}\gamma_2^{I}\propto\gamma_2^{I}$, thus
$\gamma_1^{I+b}=\gamma_2^I$.

We see also that $\gamma_1^I=\gamma_1^{I+a+b}$ and from
(~\ref{rels_gamma}) we get:

\begin{eqnarray}
\gamma_1^{I+a+b}&=&\gamma_2^{I+a}\nonumber\\
&\downarrow&\nonumber\\
\gamma_1^{I+a+b}&=&\gamma_2^I+K_I=\gamma_1^{I+b}+K_I\nonumber\\ &\downarrow&\nonumber\\
\gamma_1^{I+a+b}&=&\gamma_1^I+2K_I\nonumber\\ &\downarrow&\nonumber\\
K_I&=&0.
\end{eqnarray}

And so all $K^I$ have to vanish $\to$ all the $\gamma^I_{1,2,3}$
have to vanish. So at one loop order we can not turn on any non
vanishing anomalous dimensions. This is similar to what we found
in the ${\cal N}=2$ case.

 From the first and second equations in (~\ref{one_l}) we see that
 $A_I=A_{I+a+b}$, thus we will parameterize our solution by
 $\gamma$ $A_I$s. The $B_I$s are obtained from the $A_I$s. Now
 from the third and the first equations we get $
 A_{I+a}-A_I=-C_{I+a+b}$, from here we get the $C_I$s. From the
 definition of $C_I$s we get $\sum_{{\cal
S}_{b}^I}C_I,\sum_{{\cal S}_{a}^I}C_I=0$, thus there are only
$k+1-\alpha-\beta$ independent $C_I$s (the {\bf +1} is because
$\sum_IC_I$ follows from both constraints). From the $C_I$s we get
constraints on {\bf k} gauge couplings $\to$ we get
$\alpha+\beta-1$ independent gauge couplings. The $h_I$s and
$h'_I$s can be obtained from the $B_I$s, the $A_I$s and the gauge
couplings.

Thus to summarize, we have $\alpha+\beta+\gamma-1$ parameters for
our solution, as expected.

Let's now try to prove the all loop existence of the solutions
above. As in previous sections we divide our n'th order
contribution to the anomalous dimensions at one loop and the rest:

  \begin{eqnarray}
\gamma^{(n)}=\gamma^{(n)(1-loop)}+\gamma^{(n)(2..n-loops)}
  \end{eqnarray}

Again we say that the yet undetermined parameters appear only in
one loop.
 The one loop contribution will have the structure (~\ref{one_l}):

\begin{eqnarray}
 A_{I+a}^{(n)(1-loop)}+B_I^{(n)(1-loop)}&=&\gamma^{I(n)(1-loop)}_1\nonumber\\
 A_{I}^{(n)(1-loop)}+B_{I+b}^{(n)(1-loop)}&=&\gamma^{I(n)(1-loop)}_2\nonumber\\
 A_{I+a}^{(n)(1-loop)}+B_{I+b}^{(n)(1-loop)}+C_{I+a+b}^{(n)(1-loop)}&=&\gamma^{I(n)(1-loop)}_3.
\end{eqnarray}

We will try to parameterize the $\gamma^{(n)(2..n-loops)}$ in the
similar way:

\begin{eqnarray}\label{S}
  S_{I+a}+T_I-\tilde \gamma^I_1&=&\gamma^{I(n)(2..n-loop)}_1\nonumber\\
 S_{I}+T_{I+b}-\tilde \gamma^I_2&=&\gamma^{I(n)(2..n-loop)}_2\nonumber\\
 S_{I+a}+T_{I+b}+P_{I+a+b}+(\tilde \gamma^I_1+\tilde \gamma^I_2+\tilde
 K_{I})&=&\gamma^{I(n)(2..n-loop)}_3.
\end{eqnarray}

Our  aim is obviously to say that
$A_{I}=-S_{I},B_{I}=-T_{I},C_I=-P_I$ and the anomalous dimensions
at n'th order are $\tilde \gamma^I_j$. We just have to check the
consistency of this.

By subtracting the first equation from the third and summing over
${\cal S}_{b}^I$ we obtain that $\sum_{{\cal
S}_{b}^I}(2\tilde\gamma_1^I+\tilde\gamma_2^I+\tilde
K_I)=\sum_{{\cal
S}_{b}^I}(\gamma_3^{I(n)(2..n-loop)}-\gamma_1^{I(n)(2..n-loop)})$,
and further by using (~\ref{rels_gamma}):

\begin{eqnarray}
 {2k\over
\beta}\tilde\gamma_1^I=-\sum_{{\cal
S}_{b}^I}\gamma_2^I-2\sum_{j=0}^{{k\over\beta}-1}j\tilde
K_{I+jb}+\sum_{{\cal
S}_{b}^I}(\gamma_3^{I(n)(2..n-loop)}-\gamma_1^{I(n)(2..n-loop)}).
\end{eqnarray}

 By subtracting the second equation from the
third and summing over ${\cal S}_{a}^I$ we obtain that
$\sum_{{\cal S}_{a}^I}(\tilde\gamma_1^I+2\tilde\gamma_2^I+\tilde
K_I )=\sum_{{\cal
S}_{a}^I}(\gamma_3^{I(n)(2..n-loop)}-\gamma_2^{I(n)(2..n-loop)})$,
and further by using (~\ref{rels_gamma}):

\begin{eqnarray}
{2k\over \alpha}\tilde\gamma_2^I=-\sum_{{\cal
S}_{a}^I}\gamma_1^I-2\sum_{j=0}^{{k\over\alpha}-1}j\tilde
K_{I+ja}+\sum_{{\cal
S}_{a}^I}(\gamma_3^{I(n)(2..n-loop)}-\gamma_2^{I(n)(2..n-loop)}).
\end{eqnarray}

 From these two observations we see that:

\begin{eqnarray}
{4k^2\over \alpha\beta}\tilde\gamma_1^I&=&\sum_{{\cal
S}_{a}^I,{\cal S}_{b}^I}\tilde\gamma_1^I-\sum_{{\cal
S}_{a}^I,{\cal
S}_{b}^I}(\gamma_3^{I(n)(2..n-loop)}-\gamma_2^{I(n)(2..n-loop)})+{2k\over\alpha}\sum_{j=0}^{{k\over\beta}-1}j\tilde K_{I+jb}+\nonumber\\
&+&{4k\over \alpha}\sum_{{\cal
S}_{b}^I}(\gamma_3^{I(n)(2..n-loop)}-\gamma_1^{I(n)(2..n-loop)})\equiv\nonumber\\
&\equiv& \sum_{{\cal S}_{a}^I,{\cal
S}_{b}^I}\tilde\gamma_1^I+{4k\over\alpha}\sum_{j=0}^{{k\over\beta}-1}j\tilde
K_{I+jb}+F(\gamma_l^{I(n)(2..n-loop)}).
\end{eqnarray}

We see that again $\tilde\gamma_1^I+\tilde
K_I=\tilde\gamma_1^{I+b}$.

From the first and second equations in (~\ref{S}) we see that we
have to get:

\begin{eqnarray}
  \sum_{{\cal S}_{a+b}^I}(\tilde\gamma^I_2-\tilde\gamma_1^I)=
  \sum_{{\cal
  S}_{a+b}^I}(\gamma^{I(n)(2..n-loop)}_1-\gamma_2^{I(n)(2..n-loop)}).
\end{eqnarray}

From these equations we can determine $\tilde \gamma_{1,2}^I$ and
$\tilde K_I$.

 After we determined the anomalous dimensions we can proceed to
 determine other quantities as we did in one loop:

\begin{eqnarray}
  T_{I+b}-T_I&=&-P_{I+a+b}+\gamma^{I(n)(2..n-loop)}_3-\gamma^{I(n)(2..n-loop)}_1+(2\tilde\gamma_1^I+\tilde\gamma_2^I)\nonumber\\
  S_{I+a}-S_I&=&-P_{I+a+b}+\gamma^{I(n)(2..n-loop)}_3-\gamma^{I(n)(2..n-loop)}_2+(\tilde\gamma_1^I+2\tilde\gamma_2^I).
\end{eqnarray}

 It seems that we have here $\alpha+\beta+\gamma-1$ additional parameters
 like we did at one loop,
 but as a matter of fact these parameters are just redefinitions of
 our one loop parameters.

 Thus to conclude, we can extend our one loop solutions to higher
 loops. The price we will have to pay is turning on non zero
 anomalous dimensions.

  In our example of (1,2,4) $\mathbb{C}^3/\mathbb{Z}_7$ we have
  $\alpha=\beta=\gamma=1$. So we expect to have a {\bf 2} dimensional
  manifold of solutions. It is easy to see that in this case
  (~\ref{one_l}) implies that all the gauge couplings are equal. And from here all the $h'_I$s are equal (to h') and also the $h^I$s
are equal (to h).

The condition we get in this case at one loop is:

\begin{eqnarray}
   h^{'2}(g,h)=2{{\cal B}\over{\cal A}}g^2-h^2.
 \end{eqnarray}

 So here with general SU(N)
group we have only two marginal deformations, parameterized by
{\bf g} and {\bf h}.  For {\bf k} prime we always have only a {\bf
2} dimensional manifold of fixed points.

So we see that there are actually cases "saturating" our lower
bound for the number of marginal directions that we found in the
beginning of the chapter.

\subsection{$SU(N=3)$}

In the special case of $SU(N=3)^k$ gauge group we can add
additional interactions:

\begin{eqnarray}
{\rho^I_j\over
 3!}\epsilon_{lmn}\epsilon^{abc}(Q^I_j)^l_a(Q^I_j)^m_b(Q^I_j)^n_c.
\end{eqnarray}

Obviously these interactions don't contribute to any mixing. The
extra $\beta$-functions are:

\begin{eqnarray}
\beta_{\rho^I_j}\propto3\gamma_{Q^I_j}.
\end{eqnarray}

Thus, for vanishing $\beta$-functions we have to demand vanishing
anomalous dimensions. We have {\bf 3k} anomalous dimensions and
{\bf 6k} couplings $\to$ we expect a {\bf 3k} dimensional manifold
of fixed points.

The one loop expressions we found above (~\ref{1ll}) are modified:

\begin{eqnarray}
  \gamma_{Q^I_1}&=&{\cal A}(h^{2}_{I}+h^{'2}_{I})+{\cal C}\rho^{I2}_1-{\cal
   B}(g_{I}^2+g_{I+a}^2)=0\nonumber\\
 \gamma_{Q^I_2}&=&{\cal A}(h^{2}_{I-a}+h^{'2}_{I+b})+{\cal C}\rho^{I2}_2-{\cal
   B}(g_{I}^2+g_{I+b}^2)=0\nonumber\\
 \gamma_{Q^I_3}&=&{\cal A}(h^{2}_{I-a-b}+h^{'2}_{I-a})+{\cal C}\rho^{I2}_3-{\cal
   B}(g_{I}^2+g_{I-a-b}^2)=0
\end{eqnarray}

By the same procedure as above we obtain (${\cal
C}\rho^{I2}_j\equiv D^I_j$):

\begin{eqnarray}
 \gamma_{Q_1^I}&=&A_{I+a}+B_I+D^I_1=0\nonumber\\
 \gamma_{Q_2^I}&=&A_{I}+B_{I+b}+D_2^I=0\nonumber\\
 \gamma_{Q_3^{I+a+b}}&=&A_{I+a}+B_{I+b}+C_{I+a+b}+D_3^{I+a+b}=0.
\end{eqnarray}

By subtracting these equations one from another and summing over
{\bf i} we get that $\sum_{{\cal
S}^{I+a}_{a+b}}D^{I+a}_1=\sum_{{\cal
S}^{I}_{a+b}}D^{I}_2,\sum_{{\cal S}^{I}_{b}}D^{I}_1=\sum_{{\cal
S}^{I+a}_{b}}D^{I+a}_3,\sum_{{\cal S}^{I}_{a}}D^{I}_2=\sum_{{\cal
S}^{I+b}_{a}}D^{I+b}_3$. These relations put
$\gamma+\alpha+\beta-1$ constraints on the {\bf 3k} D's.

 We parameterize our solution by the D's. From the first two
 equations above we get that:

 \begin{eqnarray}
 A_{I+a+b}-A_I=D_2^I-D_1^{I+b}
 \end{eqnarray}

Thus we get $\gamma$ independent $A_I$s, and the $B_I$s are
determined from the D's and the $A_I$s. From the second and the
third equations we get:

\begin{eqnarray}
A_{I+a}-A_I=D_2^I-D_3^{I+a+b}-C_{I+a+b}.
\end{eqnarray}

From here we can determine the gauge couplings, as before we see
that there are $\alpha+\beta-1$ independent gauge couplings.

So totally we will have $\gamma$ $A_I$s, $\alpha+\beta-1$
independent gauge couplings and $3k-(\gamma+\alpha+\beta-1)$
independent $\gamma$'s $\to$ we have here a ${\bf 3k}$ dimensional
manifold of fixed points, exactly as in the ${\cal N}=2$ case. The
extension to all orders is done exactly as in the previous cases.

\newpage

\section{(a,a,-2a)}\label{aa}

Here we deal with the $\overrightarrow{a}\equiv(a,a,-2a)$ case.
As an example we will keep in mind the $\mathbb{C}^3/\mathbb{Z}_4$
(1,1,2) case which can be represented by the following quiver
diagram:

 $ $\\

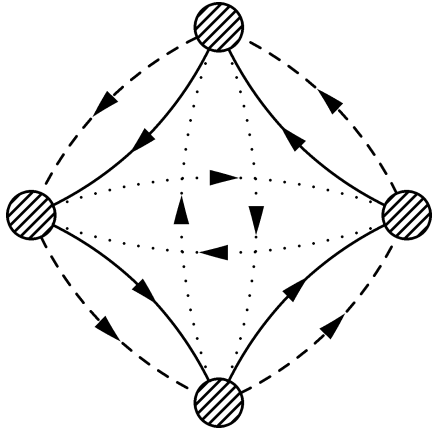
\begin{figure}[htbp]
\begin{fmffile}{quiqyIII}
\unitlength=1mm
\begin{center}
\begin{fmfchar*}(50,50)
 \fmfpen{thin}
 \fmfsurround{v1,v2,v3,v4}
 \fmf{dots_arrow,left=.2}{v1,v3,v1} \fmf{dots_arrow,left=.2}{v2,v4,v2}
 \fmf{plain_arrow,left=.2}{v1,v2,v3,v4,v1}
 \fmf{dashes_arrow,right=.2}{v1,v2,v3,v4,v1}
\fmfblob{150}{v1} \fmfblob{150}{v2} \fmfblob{150}{v3}
\fmfblob{150}{v4}
\end{fmfchar*}
\caption{$\mathbb{C}^3/\mathbb{Z}_4$ (1,1,2) quiver diagram}
\label{quiv5}
\end{center}
\end{fmffile}
\end{figure}

Here the three types of lines represent the three sectors of the
theory.

The possible interactions of the theory are the ones we saw for
general case, i.e $h^I_{123}$ etc, and also $h^I_{113},h^I_{223}$.
From the redundancy condition (~\ref{redu}) we see that:

\begin{eqnarray}
h^I_{123}&=&h^{I+a}_{231}=h^{I+2a}_{312}\nonumber\\
h^I_{132}&=&h^{I+a}_{321}=h^{I-a}_{213}\nonumber\\
h^I_{113}&=&h^{I+a}_{131}=h^{I+2a}_{311}\nonumber\\
h^I_{223}&=&h^{I+a}_{232}=h^{I+2a}_{322}.
\end{eqnarray}

Thus we see that all the couplings can be defined in terms of:
$h_I\equiv h^I_{123},h'_I\equiv h^I_{132},p_I\equiv h^I_{113}$ and
$s_I\equiv h^I_{223}$.

In this theory we have a global $SU(2)^k$ symmetry which rotates
$Q^I_1$ and $Q^I_2$. We had a very similar symmetry in the
 ${\cal N}=4$ case (there it was an SU(3) symmetry). In this case in
 general we expect the fields to mix:
 $\langle Q^{\dag I}_iQ^I_j\rangle\neq0$ for $i\neq j$. However, by using
 the $SU(2)^k$ symmetry we can make the anomalous dimensions
 diagonal
 $\langle Q^{\dag I}_iQ^I_j\rangle\to U^I_{li}U^{\dag I}_{mj}\langle Q^{\dag I}_iQ^I_j\rangle$.

 Let's now make the Leigh-Strassler analysis assuming that the
 anomalous dimensions are diagonal. The $\beta$-functions have to satisfy:

\begin{eqnarray}
   \beta_{h_I}&\propto& \gamma_1^I+\gamma_2^{I+a}+\gamma_3^{I+2a}\nonumber\\
   \beta_{h'_I}&\propto&
   \gamma_1^I+\gamma_3^{I+a}+\gamma_2^{I-a}\nonumber\\
   \beta_{g_I}&\propto&
   \gamma_1^I+\gamma_2^I+\gamma_3^I+\gamma_1^{I-a}+\gamma_2^{I-a}+\gamma_3^{I+2a}\nonumber\\
   \beta_{s_I}&\propto&\gamma_2^I+\gamma_2^{I+a}+\gamma_3^{I+2a}\nonumber\\
   \beta_{p_I}&\propto&\gamma_1^I+\gamma_1^{I+a}+\gamma_3^{I+2a}
 \end{eqnarray}

We see that from the first and the last equations, in order to
have vanishing $\beta$-functions we will have to have
$\gamma_2^{I+a}=\gamma_1^{I+a}\equiv\gamma^I$.\footnote{We see
that the anomalous dimensions are proportional to the identity in
the {\bf (1 2)} directions. No SU(2) rotation will change this
fact, so it will remain true on the fixed manifold for any
$SU(2)^k$ rotation, and we can use the $SU(2)^k$ symmetry again in
our discussion.} Assuming this the first two equations are
essentially the same.

 From the third equation we get that:

 \begin{eqnarray}
\gamma_1^{I}-\gamma_1^{I+a}=\gamma_1^{I-2a}-\gamma_1^{I-a}.
 \end{eqnarray}

Thus we see that: $\gamma_1^{I}-\gamma_1^{I+a}\equiv K_I$ where
$K_I=K_{I+2a}$. The largest common divisor of {\bf a} with {\bf k}
has to be {\bf 1} because otherwise also {\bf 2a} will have a non
trivial largest common divisor with {\bf k} $\to$ thus we see that
if {\bf k} is odd we will have only one $K_I$, and if {\bf k} is
even we will have two different $K_I$s (say A,B). Further, by
summing the definition of $K_I$ over {\bf I} we get that:

\begin{itemize}
 \item {\bf k} odd $\to$ $K_I=0$
 \item {\bf k} even $\to$ $A=-B$
\end{itemize}

From here in {\bf k} odd case we get that all the anomalous
dimensions $\gamma^I_1,\gamma^I_2$ are equal, and in {\bf k} even
case we get only two different anomalous dimensions (which we
denote by $\delta_1,\delta_2$):
$\gamma^{I+2a}_{1,2}=\gamma^{I}_{1,2}$. We also see that all the
$\gamma_3^I$s are equal and equal to $-(\delta_1+\delta_2)$.

So to summarize, we can parameterize our solutions here by one or
two parameters.

 Let's now write the one loop conditions for vanishing
 $\beta$-functions. First we write the diagonal contributions to the anomalous
dimensions:

 \begin{eqnarray}\label{conds_112}
   \gamma_{Q^I_1}&=&{\cal A}(h^{I2}_{123}+h^{I2}_{132})+{\cal C}(h^{I2}_{113}+h^{I2}_{131})-{\cal
   B}(g_{I}^2+g_{I+a}^2)=\gamma^I\nonumber\\
 \gamma_{Q^I_2}&=&{\cal A}(h^{I2}_{231}+h^{I2}_{213})+{\cal C}(h^{I2}_{223}+h^{I2}_{232})-{\cal
   B}(g_{I}^2+g_{I+a}^2)=\gamma^I\nonumber\\
 \gamma_{Q^I_3}&=&{\cal A}(h^{I2}_{312}+h^{I2}_{321})+{\cal C}(h^{I2}_{322}+h^{I2}_{311})-{\cal
   B}(g_{I}^2+g_{I-2a}^2)=-(\delta_1+\delta_2).
\end{eqnarray}

We can rewrite this as:

 \begin{eqnarray}\label{aa2}
   \gamma_{Q^I_1}&=&{\cal A}(h^{2}_{I}+h^{'2}_{I})+{\cal C}(p^{2}_{I}+p^{2}_{I-a})-{\cal
   B}(g_{I}^2+g_{I+a}^2)=\gamma^I\nonumber\\
 \gamma_{Q^I_2}&=&{\cal A}(h^{2}_{I-a}+h^{'2}_{I+a})+{\cal C}(s^{2}_{I}+s^{2}_{I-a})-{\cal
   B}(g_{I}^2+g_{I+a}^2)=\gamma^I\nonumber\\
 \gamma_{Q^I_3}&=&{\cal A}(h^{2}_{I-2a}+h^{'2}_{I-a})+{\cal C}(s^{2}_{I-2a}+p^{2}_{I-2a})-{\cal
   B}(g_{I}^2+g_{I-2a}^2)=-(\delta_1+\delta_2).\nonumber\\
\end{eqnarray}

From here:

\begin{eqnarray}
   h^{2}_I&=&-h^{'2}_I-{1\over{\cal A}}({\cal C}(p^{2}_I+p^{2}_{I-a})-{\cal
   B}(g_{I}^2+g_{I+a}^2))+\gamma^I\nonumber\\
  h^{2}_I&=&-h^{'2}_{I+2a}-{1\over{\cal A}}({\cal C}(s^{2}_{I+a}+s^{2}_{I})-{\cal
   B}(g_{I+2a}^2+g_{I+a}^2))+\gamma^{I+a}\nonumber\\
 h^{2}_I&=&-h^{'2}_{I+a}-{1\over{\cal A}}({\cal C}(s^{2}_{I}+p^{2}_{I})-{\cal
   B}(g_{I+2a}^2+g_{I}^2))+(\delta_1+\delta_2).
\end{eqnarray}

From here we can write:

\begin{eqnarray}\label{ee3}
-\gamma^I+h^{'2}_{I}+{1\over{\cal A}}({\cal
C}(p^{2}_{I}+p^{2}_{I-a})-{\cal
   B}g_{I}^2)&=&-\gamma^{I+a}+h^{'2}_{I+2a}+{1\over{\cal A}}({\cal C}(s^{2}_{I+a}+s^{2}_{I})-{\cal
   B}g_{I+2a}^2)\nonumber\\
   -\gamma^{I+a}+h^{'2}_{I+2a}+{1\over{\cal A}}({\cal C}s^{2}_{I+a}-{\cal
   B}g_{I+a}^2)&=&-(\delta_1+\delta_2)+h^{'2}_{I+a}+{1\over{\cal A}}({\cal C}p^{2}_{I}-{\cal
   B}g_{I}^2)\nonumber\\
   -(\delta_1+\delta_2)+h^{'2}_{I+a}+{1\over{\cal A}}({\cal C}s^{2}_{I}-{\cal
   B}g_{I+2a}^2)&=&-\gamma^{I}+h^{'2}_{I}+{1\over{\cal A}}({\cal C}p^{2}_{I-a}-{\cal
   B}g_{I+a}^2).\nonumber\\
\end{eqnarray}

Obviously the last two equations imply the first one. Now from the
last two equations we obtain:

\begin{eqnarray}
2(\delta_1+\delta_2-\gamma^I)+{\cal B}(g_{I-a}^2-g_{I}^2)= {\cal
B}(g_{I+a}^2-g_{I+2a}^2).
\end{eqnarray}

By summing this over {\bf I} we obtain $\delta_1=-\delta_2$, for
{\bf k} odd this immediately implies that all anomalous dimensions
vanish. From here:

\begin{eqnarray}
-2\gamma^I+{\cal B}(g_{I-a}^2-g_{I}^2)= {\cal
B}(g_{I+a}^2-g_{I+2a}^2).
\end{eqnarray}

Summing this over ${\cal S}^I_{2a}$ we obtain that also for even
{\bf k} the anomalous dimensions have to vanish. And in particular
we see that $(g_{I-a}^2-g_{I}^2)=(g_{I+a}^2-g_{I+2a}^2)$ and thus,
if {\bf k} is even we get two independent gauge couplings ({\bf
g}, ${\bf \tilde g}$), and if {\bf k} is odd we get only one
independent gauge coupling.

By adding the first and the second equations of (~\ref{ee3}) we
get:

\begin{eqnarray}\label{aa1}
   h^{'(I+a)2}-h^{'I2}={{\cal C}\over {\cal A}}(p_{I-a}^2-s_I^2)+{{\cal B}\over {\cal
   A}}(g_{I+2a}^2-g_{I+a}^2).
\end{eqnarray}

Thus we see that essentially we have here only $\alpha$(=1)
independent $h'_I$s, and all the others can be obtained from them.
We see that we also get a constraint on $s_I^2$ and $p_{I}^2$. By
summing over ${\cal S}^I_a$ the above equation we get that
$\sum_{{\cal S}^I_a}s_I^2=\sum_{{\cal S}^I_a}p_{I}^2$.

Now we write the non diagonal contributions:

\begin{eqnarray}\label{nondig}
  \gamma^I_{12}&\propto&h^I_{123}h^{*I}_{223}+h^I_{132}h^{*I}_{232}+h^I_{113}h^{*I}_{213}+h^I_{131}h^{*I}_{231}\nonumber\\
  &\downarrow &\nonumber\\
  \gamma^I_{12}&\propto&h^Is^{*I}+h^{'I}s^{*I-a}+p^Ih^{'*(I+a)}+p^{I-a}h^{*I-a},\nonumber\\
  \gamma^I_{21}&=&\gamma^{*I}_{12}.
\end{eqnarray}

Now let's assume that we are close to the orbifold theory, i.e.
$h_I=h+{\cal O}(\epsilon),h'_I=-h+{\cal O}(\epsilon),p_I={\cal
O}(\epsilon),s_I={\cal O}(\epsilon),g_I=g+{\cal O}(\epsilon)$
(where $\epsilon$ is some small parameter). In particular this
implies in the leading order for the non diagonal elements above:

\begin{eqnarray}
s^{*I}-s^{*I-a}-p^I+p^{I-a}=0.
\end{eqnarray}

 By using the
$SU(2)^k$ global symmetry (see Appendix ~\ref{apB}) we can set all
the $h'_I$'s to zero.

We will use the notations of the even {\bf k} case (the odd {\bf
k} case is
  obtained by simply putting $K_I=0$).
  We saw that in
 this case we get two gauge couplings: $g_I\equiv g$ for even {\bf I}
 and $g_I\equiv \tilde g$ for odd {\bf I}. Define $K_{I+a}\equiv {{\cal B}\over {\cal
 A}}(g_{I+2a}^2-g_{I+a}^2)$, obviously $K_I=-K_{I+a}$.
  From (~\ref{aa1}) we
 obtain:

\begin{eqnarray}
  {{\cal C}\over {\cal
   A}}(p_{I-a}^2-s_I^2)=-K_{I+a},
\end{eqnarray}

So we can write the $s_I$s in terms of the $p_I$s.

Now from (~\ref{nondig}) we see:

\begin{eqnarray}\label{aa10}
  \gamma^I_{12}&\propto& h^Is^{*I}+p^{I-a}h^{*I-a}=0
\end{eqnarray}

From here we see that:

\begin{eqnarray}\label{aa7}
  h^2_Is_{I}^2=p_{I-a}^2h_{I-a}^2
\end{eqnarray}

So now from (~\ref{aa2}) we get ($G\equiv {\cal B}(g_I^2+g_{I+a}^2)$):

\begin{eqnarray}\label{aa5}
{\cal A}h^{I2}+{\cal C}(p_{I}^{2}+p_{I-a}^{2})-G=0
\end{eqnarray}

From here we obtain: ${\cal A}h^{I2}=G-{\cal
  C}(p_{I}^{2}+p_{I-a}^{2})$. And by putting all our recent knowledge
into (~\ref{aa7}) we get:

\begin{eqnarray}
  p_{I-a}^2(p_I^2-p_{I-2a}^2)={{\cal A}\over {\cal C}}K_{I+a}({1\over {\cal
  C}}G-p_I^2-p_{I-a}^2).
  \end{eqnarray}

Obviously we have at least one solution because these equations are
linearly dependent by summing over {\bf I}. When we put the gauge
couplings to be equal ($K_{I}=0$), we get $p_I^2=p_{I+2a}^2$. and thus
in {\bf k} even case get {\bf 2} different $p_I$s and in {\bf k} odd
case get only one independent $p_I$.

 The extension to higher loops is as follows. The solution
 will be parameterized by {\bf $p_I$, $p_{I+a}$, g} as in the one-loop case. We notice that
 the solution we got necessarily has
 $\gamma_{ij}^I=\gamma_{ij}^{I+2a}$, thus we will now turn on
 interactions which will guarantee the existence of this property
 at any order of perturbation theory: we turn on $s_I=s_{I+2a}$,
 $h'_I=h'_{I+2a}$, $h_I=h_{I+2a}$. Note that at higher loops the
 anomalous dimensions don't necessarily vanish and so we can use
 the $SU(2)^k$ symmetry to make the anomalous dimensions diagonal
 as we did before, but now we can not take away $h'_I$s.

 So we see that we have here {\bf 6} couplings, {\bf 2} parameters
 from the anomalous dimensions ($\delta_1,\delta_2$) and we have {\bf
 8} different anomalous dimensions
 ($\gamma_{ii}^I,\gamma_{ii}^{I+a},\gamma_{12}^I,\gamma_{12}^{I+a}$)
 $\to$ we have {\bf 8} equations for {\bf 8} variables thus in
 general we
 get a solution. The global $SU(2)^k$ symmetry was already used to get rid of $h'_I$ in one loop
 and we can not use it again.

 So to summarize, we get here one additional marginal
deformation when all the gauge couplings are equal in the even
{\bf k} case. This extra operator cannot come from an operator in
the ${\cal N}=4$ theory because the marginal deformations which
come from there are obviously {\bf I} independent $\to$ this is a
new kind of deformation existing in this theory. The ${\cal
N}=2$ case with k=2 is a special case of the type discussed here (see
section (~\ref{sec_k_2})). The ${\cal N}=1$ theory can be seen as
an orbifold of the ${\cal N}=2$ theory, thus the new
\textit{exactly} marginal deformations we get here can be seen as
the descendants of the ${\cal N}=2$ deformations.

We can make an interesting observation here: if we put all the
$h_I$s and $h'_I$s to zero we are guaranteed to have diagonal
$\gamma$ matrices from the symmetry of the problem. The one loop
expressions become:

\begin{eqnarray}
   \gamma_{Q^I_1}&=&{\cal C}(p^{2}_I+p^{2}_{I-a})-{\cal
   B}(g_{I}^2+g_{I+a}^2)=\gamma^I\nonumber\\
 \gamma_{Q^I_2}&=&{\cal C}(s^{2}_I+s^{2}_{I-a})-{\cal
   B}(g_{I}^2+g_{I+a}^2)=\gamma^I\nonumber\\
 \gamma_{Q^I_3}&=&{\cal C}(s^{2}_{I-2a}+p^{2}_{I-2a})-{\cal
   B}(g_{I}^2+g_{I-2a}^2)=-(\delta_1+\delta_2).\nonumber\\
\end{eqnarray}

Further we get:

\begin{eqnarray}
{\cal C}(p^{2}_I+p^{2}_{I-a})&=&{\cal
   B}(g_{I}^2+g_{I+a}^2)+\gamma^I\nonumber\\
{\cal C}(p^{2}_I+p^{2}_{I-a})&=&{\cal
   B}(g_{I+2a}^2+g_{I-a}^2)-\gamma^I-2(\delta_1+\delta_2).
\end{eqnarray}

In the same way as before we obtain from here that the anomalous
dimensions should vanish. Further:

\begin{eqnarray}
g_{I+a}^2-g_{I-a}^2=g_{I+2a}^2-g_{I}^2,
\end{eqnarray}

and from here we see that all the gauge coupling should be equal
(say to {\bf g}).

From here we obtain that the solution is $p_I^2=s_I^2=g^2$. Thus
we see that we get a one dimensional family of conformal theories.
It can be easily proven that the solution extends to all orders of
the perturbation series. This solution is $SU(2)^k$ equivalent to
the solution without any $p_I$s or $s_I$s.

 \subsection{$SU(N=3)$}

 In addition to the (a,b,-a-b) case $SU(N=3)^k$ interactions we can have
 here also:

 \begin{eqnarray}
{\rho^I_{12}\over
 3!}\epsilon_{lmn}\epsilon^{abc}(Q^I_1)^l_a(Q^I_1)^m_b(Q^I_2)^n_c\nonumber\\
 {\rho^I_{21}\over
 3!}\epsilon_{lmn}\epsilon^{abc}(Q^I_2)^l_a(Q^I_2)^m_b(Q^I_1)^n_c.
\end{eqnarray}

These new interactions affect only $\gamma_{Q^I_1}$,
$\gamma_{Q^I_2}$ and the mixing terms.
 There are no additional constraints that we have to impose. We use the $SU(2)^k$ symmetry in order to cancel {\bf
k} couplings out of the $p_i,s_i$. Now we can take the {\bf k}
gauge couplings, {\bf 2k} $h_I$ and $h'_I$, the remaining {\bf k}
out of the $p_i,s_i$ and {\bf k} $\rho^I_{12}$ couplings as
parameters $\to$ $\rho^I_{21}$ will be set from the non diagonal
anomalous dimensions and the $\rho^I_i$ will be set from the
diagonal ones.
 So we get here a {\bf
5k} dimensional manifold of fixed points.

 From Leigh-Strassler analysis we get: here all anomalous dimensions have to vanish,
  we have here {\bf 3k+k} possibly non zero anomalous dimensions.
  We have {\bf 3k+2k+3k+2k=10k} couplings, {\bf k} of the couplings can be set to zero by $SU(2)^k$
  symmetry ( see Appendix ~\ref{apB}) $\to$ have a naive expectation
  for a
  {\bf 5k} dimensional  manifold of fixed points.

\newpage

\section{$(a,{k\over3}-a,-{k\over3})$, k=3k' }\label{k_3case}

In this case we can have an additional operator of the form
$h^I_{333}Q^I_3Q^{I+{2k\over3}}_3Q^{I+{4k\over3}}_3$.
 As we saw these theories are the same as $(a,{k\over3},-a-{k\over3})$ and we will give an example of
such a theory: the (1,2,3) $\mathbb{C}^3/\mathbb{Z}_6$ theory,

\begin{figure}[htbp]
\begin{fmffile}{quip}
\unitlength=1mm
\begin{center}
\begin{fmfchar}(50,50)
\fmfpen{thin}
 \fmfsurround{v1,v2,v3,v4,v5,v6}
 \fmf{dots_arrow}{v1,v2,v3,v4,v5,v6,v1}
 \fmf{dashes_arrow,left=.2}{v1,v4,v1} \fmf{dashes_arrow,left=.2}{v2,v5,v2}  \fmf{dashes_arrow,left=.2}{v3,v6,v3}
 \fmf{plain_arrow}{v1,v3,v5,v1} \fmf{plain_arrow}{v2,v4,v6,v2}
 \fmfblob{150}{v1} \fmfblob{150}{v2} \fmfblob{150}{v3}
 \fmfblob{150}{v4} \fmfblob{150}{v5} \fmfblob{150}{v6}
\end{fmfchar}
\caption{$\mathbb{C}^3/\mathbb{Z}_6$ (1,2,3) quiver diagram}
\end{center}
\end{fmffile}
\end{figure}
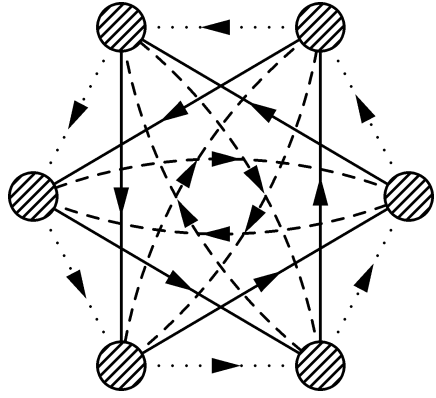

From the redundancy condition (~\ref{redu}) we see that:

\begin{eqnarray}
h^I_{123}&=&h^{I+a}_{231}=h^{I+k'}_{312}\nonumber\\
h^I_{132}&=&h^{I+a}_{321}=h^{I+a-k'}_{213}\nonumber\\
h^I_{333}&=&h^{I+2k'}_{333}=h^{I+k'}_{333}.
\end{eqnarray}

The general Leigh-Strassler analysis we did in the {\bf
(a,b,-a-b)} section is applicable here, only we'll have new
$\beta$-functions for the new interactions:

\begin{eqnarray}\label{spec1}
 \beta_{h^I_{333}}\propto
 \gamma^I_3+\gamma^{I+{k'}}_3+\gamma^{I+2k'}_3=0.
\end{eqnarray}

We see that this is essentially $\sum_{{\cal
S}^I_{2k\over3}}\gamma_3^I=0$. Thus we will use the {\bf
(a,b,-a-b)} notations with the new constraint. We add {\bf k'}
couplings to the theory but also have {\bf k'} new
$\beta$-functions $\to$ we do not expect any new marginal
directions.

 The perturbation theory calculations:

 \begin{eqnarray}\label{conds_123}
   \gamma_{Q^I_1}&=&{\cal A}(h^{I2}_{123}+h^{I2}_{132})-{\cal
   B}(g_{I}^2+g_{I+a}^2)=\gamma_1^I\nonumber\\
 \gamma_{Q^I_2}&=&{\cal A}(h^{I2}_{231}+h^{I2}_{213})-{\cal
   B}(g_{I}^2+g_{I+k'}^2)=\gamma_2^I\nonumber\\
 \gamma_{Q^I_3}&=&{\cal A}(h^{I2}_{312}+h^{I2}_{321})+{\cal C}h^{I2}_{333}-{\cal
   B}(g_{I}^2+g_{I-a-k'}^2)=\gamma_3^I.
\end{eqnarray}

And from here we get ($h^I_{333}\equiv s^I$,and $h_I,h'_I$ as in
previous sections):

 \begin{eqnarray}
   \gamma_{Q^I_1}&=&{\cal A}(h^{2}_I+h^{'2}_I)-{\cal
   B}(g_{I}^2+g_{I+a}^2)=\gamma_1^I\nonumber\\
 \gamma_{Q^I_2}&=&{\cal A}(h^{2}_{I-a}+h^{'2}_{I+k'-a})-{\cal
   B}(g_{I}^2+g_{I+k'-a}^2)=\gamma^I_2\nonumber\\
 \gamma_{Q^I_3}&=&{\cal A}(h^{2}_{I-k'}+h^{'2}_{I-a})+{\cal C}s^{I2}-{\cal
   B}(g_{I}^2+g_{I-k'}^2)=\gamma^I_3.
\end{eqnarray}

Defining $A_I,B_I,C_I,K_I$ as in the {\bf (a,b,-a-b)} section, and
$S_{I-a-b}\equiv {\cal C}s^{I2}$:

\begin{eqnarray}\label{as}
A_{I+a}+B_I&=&\gamma_1^I\nonumber\\
 A_I+B_{I+b}&=&\gamma_2^I\nonumber\\
 A_{I+a}+B_{I+b}+C_{I+a+b}+S_I&=&-(\gamma_1^I+\gamma^I_2+K_I).
\end{eqnarray}

By subtracting the first equation from the second and summing over
${\cal S}^I_{a+b}$ we get that:

\begin{eqnarray}\label{3k'}
 \sum_{{\cal S}^I_{a+b}}\gamma_1^I=\sum_{{\cal
 S}^I_{a+b}}\gamma_2^I.
\end{eqnarray}

But from (~\ref{spec1}) we see that( $a+b={k\over3}$):

\begin{eqnarray}
 \sum_{{\cal S}^I_{a+b}}\gamma_1^I+k'K_I=-\sum_{{\cal
 S}^I_{a+b}}\gamma_2^I.
\end{eqnarray}

And from here summing over ${\cal S}^I_{b}$ and using $\sum_{{\cal
S}^I_{b}}K_I=0$ we get:

\begin{eqnarray}
 \sum_{I}\gamma_1^I=-\sum_{I}\gamma_2^I.
\end{eqnarray}

From here and from (~\ref{3k'}) we get that
$\sum_{I}\gamma_1^I=\sum_{I}\gamma_2^I=0$.
 By subtracting the first equation from the
third in (~\ref{as}) and summing over {\bf I} we get that:

\begin{eqnarray}
\sum_IS_I=-\sum_I(\gamma_1^I+\gamma^I_2+K_I)=0
\end{eqnarray}

 Because $S_I$ is a positive definite
quantity this implies that $S_I=0$ and thus we can not turn on the
new interaction at one loop. Of course the question that arises is
about the possibility of turning it on at higher loops, like in
section (~\ref{CCC}).

\subsection{$SU(N=3)$}

This case is exactly like the {\bf (a,b,-a-b)} case, the new
interaction here does not add any new possibilities for the
$SU(N=3)^k$ interactions. But however lets look at the one loop
analysis ($\rho^{I2}_i$ is defined as in previous sections):

 \begin{eqnarray}
   \gamma_{Q^I_1}&=&{\cal A}(h^{2}_I+h^{'2}_I)+{\cal C}\rho^{I2}_1-{\cal
   B}(g_{I}^2+g_{I+a}^2)=0\nonumber\\
 \gamma_{Q^I_2}&=&{\cal A}(h^{2}_{I-a}+h^{'2}_{I+k'-a})+{\cal C}\rho^{I2}_2-{\cal
   B}(g_{I}^2+g_{I+k'-a}^2)=0\nonumber\\
 \gamma_{Q^I_3}&=&{\cal A}(h^{2}_{I-k'}+h^{'2}_{I-a})+{\cal C}s^{I2}+{\cal C}\rho^{I2}_3-{\cal
   B}(g_{I}^2+g_{I-k'}^2)=0.
\end{eqnarray}

We see that we can parameterize a solution by $h_I$, $h'_I$, $g_I$
and by $s_I$ $\to$ we get a {\bf 4k} dimensional manifold of fixed
points.

\newpage

\section{$(1,1,4)$}

 The (1,1,4) case can be represented by the
following diagram.

\begin{figure}[htbp]
\begin{fmffile}{qui}
\unitlength=1mm
\begin{center}
\begin{fmfchar}(50,50)
\fmfpen{thin}
 \fmfsurround{v1,v2,v3,v4,v5,v6}
 \fmf{dots_arrow,left=.1}{v1,v2,v3,v4,v5,v6,v1}
 \fmf{dashes_arrow,right=.1}{v1,v2,v3,v4,v5,v6,v1}
 \fmf{plain_arrow}{v1,v5,v3,v1} \fmf{plain_arrow}{v2,v6,v4,v2}
 \fmfblob{150}{v1} \fmfblob{150}{v2} \fmfblob{150}{v3}
 \fmfblob{150}{v4} \fmfblob{150}{v5} \fmfblob{150}{v6}
\end{fmfchar}
\caption{$\mathbb{C}^3/\mathbb{Z}_6$ (1,1,4) quiver diagram}
\label{quive}
\end{center}
\end{fmffile}
\end{figure}
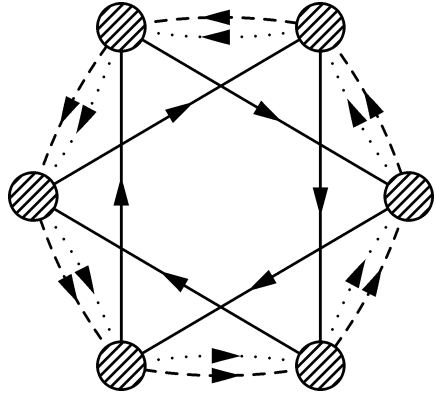

Essentially this case is in the intersection of
$(a,{k\over3}-a,-{k\over3})$ and (a,a,-2a). Thus we can treat this
case as (a,a,-2a) with the additional operator
$h^I_{333}Q^I_3Q^{I-2a}_3Q^{I-4a}_3$. As in the previous section
we have here $\sum_{2a}\gamma_3^I=0$, from the {\bf (a,a,-2a)}
case we know that $\gamma_3^I=-(\delta_1+\delta_2)$ and thus
$\delta_1=-\delta_2$ $\to$ $\gamma_3^I=0$,
$\gamma_I=-\gamma_{I+a}$.
 From section (~\ref{aa}) we
get($h^I_{333}\equiv t^I$):

 \begin{eqnarray}
   \gamma_{Q^I_1}&=&{\cal A}(h^{2}_{I}+h^{'2}_{I})+{\cal C}(p^{2}_{I}+p^{2}_{I-a})-{\cal
   B}(g_{I}^2+g_{I+a}^2)=\gamma^I\nonumber\\
 \gamma_{Q^I_2}&=&{\cal A}(h^{2}_{I-a}+h^{'2}_{I+a})+{\cal C}(s^{2}_{I}+s^{2}_{I-a})-{\cal
   B}(g_{I}^2+g_{I+a}^2)=\gamma^I\nonumber\\
 \gamma_{Q^I_3}&=&{\cal A}(h^{2}_{I-2a}+h^{'2}_{I-a})+{\cal C}(s^{2}_{I-2a}+p^{2}_{I-2a})+{\cal D}t_I^2-{\cal
   B}(g_{I}^2+g_{I-2a}^2)=0.\nonumber\\
\end{eqnarray}

From here ($\gamma_I\to{1\over {\cal A}}\gamma_I$):

\begin{eqnarray}
   h^{2}_{I}&=&-h^{'2}_{I}-{1\over{\cal A}}({\cal C}(p^{2}_{I}+p^{2}_{I-a})-{\cal
   B}(g_{I}^2+g_{I+a}^2))+\gamma^I\nonumber\\
  h^{2}_{I}&=&-h^{'2}_{I+2a}-{1\over{\cal A}}({\cal C}(s^{2}_{I+a}+s^{2}_{I})-{\cal
   B}(g_{I+2a}^2+g_{I+a}^2))+\gamma^{I+a}\nonumber\\
 h^{2}_{I}&=&-h^{'2}_{I+a}-{1\over{\cal A}}({\cal C}(s^{2}_{I}+p^{2}_{I})-{\cal D}t_{I+2a}^2-{\cal
   B}(g_{I+2a}^2+g_{I}^2)).
\end{eqnarray}

From here we can write:

\begin{eqnarray}
&h^{'2}_{I}&+{1\over{\cal A}}({\cal
C}(p^{2}_{I}+p^{2}_{I-a})-{\cal
   B}g_{I}^2)-\gamma_I=h^{'2}_{I+2a}+{1\over{\cal A}}({\cal C}(s^{2}_{I+a}+s^{2}_{I})-{\cal
   B}g_{I+2a}^2)-\gamma_{I+a}\nonumber\\
   &h^{'2}_{I+2a}&+{1\over{\cal A}}({\cal C}s^{2}_{I+a}-{\cal
   B}g_{I+a}^2)-\gamma_{I+a}=h^{'2}_{I+a}+{1\over{\cal A}}({\cal D}t_{I+2a}^2+{\cal C}p^{2}_{I}-{\cal
   B}g_{I}^2)\nonumber\\
   &h^{'2}_{I+a}&+{1\over{\cal A}}({\cal D}t_{I+2a}^2+{\cal C}s^{2}_{I}-{\cal
   B}g_{I+2a}^2)=h^{'2}_{I}+{1\over{\cal A}}({\cal C}p^{2}_{I-a}-{\cal
   B}g_{I+a}^2)-\gamma_I.\nonumber\\
\end{eqnarray}

Obviously the last two equations imply the first one. Now from the
last two equations we obtain:

\begin{eqnarray}
{\cal D}t_{I+a}^2-{\cal B}(g_{I-a}^2-g_{I}^2)= -{\cal
D}t_{I+2a}^2-{\cal B}(g_{I+a}^2-g_{I+2a}^2)-2\gamma_I.\nonumber\\
\end{eqnarray}

Again by summing both sides over {\bf I} we get
$\sum_It_I^2=-\sum_It_I^2$ $\to$ $t_I=0$. Thus we see that we are
essentially back to the {\bf (a,a,-2a)} case.

\subsection{$SU(N=3)$}

Here again exactly as in the {\bf (a,a,-2a)} case we can have a
solution parameterized by $h_I$, $h'_I$, {\bf k} of the $p_I$ and
$s_I$, the gauge couplings, $\rho^I_{12}$ and now also by the
$t_I$ $\to$ giving a total of a {\bf 6k=36} dimensional manifold
of
 fixed points.

\newpage

\section{(1,1,1)}

This case is the richest one - all the interactions we discussed
in previous sections can be turned on here. This theory can be
depicted by the following quiver diagram:

$ $\\

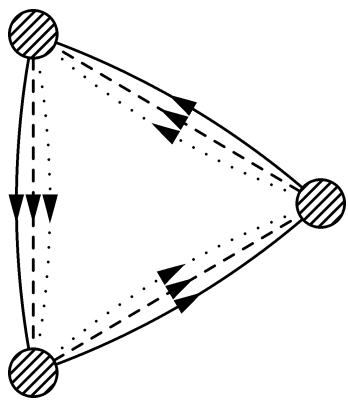
\begin{figure}[htbp]
\begin{fmffile}{quiqyUU}
\unitlength=1mm
\begin{center}
\begin{fmfchar*}(50,50)
 \fmfpen{thin}
 \fmfsurround{v1,v2,v3}
 \fmf{dots_arrow,left=.1}{v1,v2,v3,v1}
 \fmf{plain_arrow,right=.1}{v1,v2,v3,v1}
 \fmf{dashes_arrow}{v1,v2,v3,v1}
\fmfblob{150}{v1} \fmfblob{150}{v2} \fmfblob{150}{v3}
\end{fmfchar*}
\caption{$\mathbb{C}^3/\mathbb{Z}_3$ (1,1,1) quiver diagram}
\label{quiv2}
\end{center}
\end{fmffile}
\end{figure}

Here the different lines represent the three sectors we have in
our theory.

The (1,1,1) is the only $\mathbb{C}^3/\mathbb{Z}_3$ orbifold to
have ${\cal N}=1$ SUSY. We get a $U(N)^3$ gauge group with matter
content:

\begin{equation}
  3\times((N,\bar N,1)\oplus(1,N,\bar N)\oplus(\bar N,1,N))
\end{equation}

This theory was treated in~\cite{a:SilKach}. It was argued there
that the theory has a single marginal direction corresponding to
the gauge coupling. However in that analysis the cases were
classified by the global $S_3$ symmetry of the three complex
space-time coordinates. This overlooks the symmetries leading to
the additional operators that we saw in previous sections. So we
will now look for \textit{exactly} marginal deformations of this
 theory.

Obviously this case is a special case of the cases we considered
above, in particular of the {\bf (a,a,-2a)} case. There we found a
three dimensional manifold of fixed points, so here we expect  at
least this dimensionality.

 The most general marginal deformation here is\footnote{Here $a_l+a_m+a_n=0(mod{\bf k})$
 for any choice (l,m,n).}:

 \begin{eqnarray}
   h^I_{ijk}Q^I_iQ^{I+1}_jQ^{I+2}_k
 \end{eqnarray}

We have here a global $SU(3)^3$ symmetry rotating $Q^1_i$s,
$Q^2_i$s and $Q^3_i$s. By these rotations we can assume that the
anomalous dimensions are diagonal, and then:

\begin{eqnarray}
\beta_{h^I_{ijk}}\propto\gamma_{Q^I_i}+\gamma_{Q^{I+1}_j}+\gamma_{Q^{I+2}_k}
\end{eqnarray}

So we see that in general we have to put all the anomalous
dimensions to zero. In general we have ${\bf 3k\times 2}$
anomalous dimensions and ${\bf 3k\times 3+3}$ couplings, $\bf 3k$
of which can be set to zero by the global symmetry, and thus
expect for three dimensional manifold of \textit{exactly} marginal
deformations. However if we restrict ourselves to cases where we
will be guaranteed to have anomalous dimensions proportional to
the  identity matrix (by turning on only
$h^I_{123}=h^I_{231}=h^I_{312}$, $h^I_{213}=h^I_{132}=h^I_{321}$
and $h^I_{ijk}\propto\delta_{ij}\delta_{jk}$ interactions, and
keeping all the gauge couplings equal) we will get only {\bf 1}
independent anomalous dimension, and {\bf 4} couplings $\to$ we
expect a {\bf 3} dimensional manifold of fixed points and thus we
can deal only with this  restricted case\footnote{Obviously we
have other choices of the couplings satisfying this, but they all
are related by the global $SU(3)^3$ symmetry.} .

The anomalous dimensions still have to vanish, because
$\beta_{h^I_{lll}}\propto\gamma^I_l$.

Now the anomalous dimensions for the most general case are at
one-loop:

\begin{eqnarray}
\gamma^I_{ij}={\cal A} h^I_{ilm}h^{*I}_{jlm}-{\cal
B}(g_I^2+g^2_{I+1})\delta_{ij}
\end{eqnarray}

We see that we will have solutions to $\gamma=0$ only when
$h^I_{ilm}h^{*I}_{jlm}$ is proportional to the identity matrix, up
to the global $SU(3)^3$ rotations this implies the restrictions
above.

We now look at the one loop expressions in the restricted case
($h_I$s, $h'_I$s defined as in the previous sections and
$y_I\equiv h^I_{111}=h^I_{222}=h^I_{333}$, in our cases all the
{\bf h}s are equal and all the {\bf h'}s are equal):

\begin{eqnarray}
\gamma_I={\cal A}(h^2+h^{'2})+{\cal C}y_I^2-2{\cal B}g^2.
\end{eqnarray}

From here we easily see that all the $y_I$s have to be equal, and
we get one condition for {\bf 4} couplings $\to$ we get a {\bf 3}
dimensional manifold of fixed points.

As in previous sections, it is easy to construct the solutions in
all order of the perturbation series.

\subsection{$SU(N=3)$}

If we take the gauge group to be SU(3) there is a much larger
class of marginal operators one can add to the theory. All the
operators of the following form are possible:

\begin{eqnarray}
 {\rho^I_{ijk}\over
 3!}\epsilon_{lmn}\epsilon^{abc}(Q^I_i)^{l}_a(Q^I_j)^{m}_b(Q^I_k)^{n}_c.
\end{eqnarray}

Obviously $\rho^I_{ijk}$ has to be symmetric in the lower indices.
The Leigh-Strassler analysis teaches us in this case: we have
$6\times 3$ anomalous dimensions and $3\times3\times 3+10\times3+3$ couplings, have global $SU(3)^3$ symmetry
 with which we can fix $3\times3$ parameters
$\to$ naively we expect  a {\bf 11k} dimensional manifold of fixed
points.

Look now at the general one loop expression:

\begin{eqnarray}
\gamma^I_{ij}={\cal A} h^I_{ilm}h^{*I}_{jlm}+{\cal
C}\rho^I_{ilm}\rho^{*I}_{jlm}-{\cal
B}(g_I^2+g^2_{I+1})\delta_{ij}.
\end{eqnarray}

We will have a solution when ${\cal A} h^I_{ilm}h^{*I}_{jlm}+{\cal
C}\rho^I_{ilm}\rho^{*I}_{jlm}$ is proportional to the identity
matrix.

$\rho^I_{ijk}$ is symmetric we have only $10\times k$ $\rho$s. We
will take all the $h^I_{ijk}$s and gauge couplings to be
parameters. Also will take $\rho^I_{iii},\rho^I_{123}$ as
parameters $\to$ we have $3\times3\times3+3+3+3\times3=3\times 14$
parameters, and so $6\times 3$ of the couplings are still
undetermined. We have $6\times 3$ anomalous dimensions with which
we determine the yet undetermined couplings $\to$ we get $11\times
3$ dimensional manifold of fixed points (as expected) after
dividing by the $SU(3)^3$ global symmetry.

\newpage

\chapter{Summary and Discussion}

   First we summarize the results:
   \begin{itemize}
   \item ${\cal N}=4$\\
     We conclude that the only supersymmetric \textit{exactly} marginal deformations of ${\cal N}=4$ SYM are
   the superpotentials (other than changing the gauge coupling):

    \begin{eqnarray}
       \frac{i\delta\lambda\sqrt{2}}{3!}\epsilon_{ijk}Tr({\Phi}^i[{\Phi}^j,{\Phi}^k])\nonumber\\
       \sum_i\frac{h}{3!}Tr({\Phi}^i\left\{{\Phi}^i,{\Phi}^i\right\})\nonumber\\
    \frac{h_{123}}{3!}Tr({\Phi}^1\left\{{\Phi}^2,{\Phi}^3\right\}),
    \end{eqnarray}

  with one relation relating $\lambda$, $h_{123}$, h and the
  gauge coupling.
  These fixed points are IR stable. We saw that this theory
  is not asymptotically free for any choice of the coupling
  constants nor does it have (in perturbation theory) UV fixed points. These \textit{exactly} marginal
  deformations can be mapped to the strong coupling region by the
  S-duality transformation.

  In the strong coupling limit all our calculations are done with the assumption that we are close
  to the ${\cal N}=4$ line, because only there we can trust the S-duality
  transformation. An interesting question is whether there is a UV
  fixed point at strong coupling "far" from the ${\cal N}=4$ line.
  If there is such a UV fixed point then we can talk about the marginal deformations above not just
  at the conformal fixed point but at a larger set of points.

 \item ${\cal N}=2$

   We conclude that for the $\mathbb{C}^3/\mathbb{Z}_k$ orbifold theories we obtain:
   \begin{itemize}
   \item General k

      Here we are able to show the existence of one \textit{exactly} marginal direction
      (in addition to the gauge couplings) parameterized by a
      parameter {\bf X}. This direction can be seen at any order of
      perturbation theory. The X=0 case has ${\cal N}=2$ SUSY and all
      $\gamma$s vanish, however if $X\neq 0$ we can in principle have nonzero
      $\gamma$s. From Leigh and Strassler analysis we expect here
      to have
      another {\bf k-1} \textit{exactly} marginal directions which appear by turning on $Tr(\Phi_i\Phi_i\Phi_i)$ operators.
       We don't see these marginal directions up
      to three loops. This however does not necessarily prevent them from
      appearing at higher loops. The fate of these fixed points
      is in the hands of a linear combination of the $\gamma_{Q_i}$ and the $\gamma_{\Phi_i}$ from the case without
      the operators above: if it is positive or zero then these marginal
      directions are ruled out and if it is negative then we can have
      them. In any case the total number of \textit{exactly} marginal directions is at least
      {\bf k+1}.

    \item k=3

     In this case we have the {\bf X} direction like in general {\bf
     k}, and we can
     also have another {\bf 3} \textit{exactly} marginal directions. Here the result
     agrees with Leigh/Strassler analysis and we see all the
     marginal deformations already at one loop. So the total
     number of \textit{exactly} marginal directions here is {\bf 7}.

   \item SU(N=3)

     Here we have yet a larger space of deformations:
     we get {\bf 2k-1} additional deformations, which give for general {\bf
     k}
     a total of {\bf 3k} and for k=3 {\bf 3k+2=11} \textit{exactly} marginal directions. Again we see all the
     deformations already at one-loop and they agree with the Leigh/Strassler type analysis.
   \end{itemize}
$ $\\

 \item ${\cal N}=1$

    Here for general ${\bf (a_1,a_2,a_3)}$, $\mathbb{Z}_k$, we show that the number
    of \textit{exactly}
    marginal directions is (we denote the largest common divisor of
    $a_i$ with {\bf k} by $\alpha_i$)
    {\bf $\sum_i\alpha_i-1$.} In the case where two of the $a_i$s
    are equal and {\bf k} is even we get additional \textit{exactly} marginal directions.
    In the special case of SU(N=3) we get much larger manifolds of
    fixed points, ranging from dimension {\bf 3k} in the most general ${\bf (a_1,a_2,a_3)}$ $\mathbb{Z}_k$
    theory to {\bf 11k} in the k=3 case.

    There is also a special case here of $\mathbb{Z}_3$ in which
    we can turn on any interactions which we can turn on in
    principle in an ${\cal N}=1$ orbifold theory.

\end{itemize}

   There is a simple generalization of the results above. We are
   looking at a ${\bf (a_1,a_2,a_3)}$ $\mathbb{Z}_k$ orbifold theory. Thus we
   have in our theory, as we saw, three sectors: denote fields from
   each sector by $Q_i$. The number of marginal operators of the form
   $Q_1Q_2Q_3$ is $\sum_i\alpha_i-1$.

    First we can
   consider the ${\cal N}=4$ theory as a (0,0,0) orbifold theory. Then, it
   corresponds to the following quiver diagram:

\begin{figure}[htbp]
\begin{fmffile}{qq}
\unitlength=1mm
\begin{center}
\begin{fmfchar}(50,50)
\fmfpen{thin} \fmfsurround{a2,a3,a4}
 \fmf{phantom,tension=5}{a2,v2} \fmf{phantom,tension=5}{a3,v3} \fmf{phantom,tension=5}{a4,v4}
 \fmf{vanilla,left=.9}{v1,v2,v1} \fmf{dots,left=.9}{v1,v3,v1}
 \fmf{dashes,left=.9}{v1,v4,v1}
 \fmfblob{350}{v1}
\end{fmfchar}
\caption{${\cal N}=4$ (0,0,0) quiver diagram} \label{qq}
\end{center}
\end{fmffile}
\end{figure}
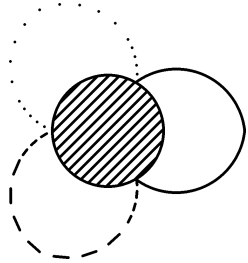

We found that there is always {\bf 1} additional marginal
   direction when we put all our gauge couplings equal (the {\bf X}
   solution in the ${\cal N}=2$ case and the {\bf h} solution in the
   ${\cal N}=1$ case). It is easy to
   understand its origin from the ``mother'' ${\cal N}=4$ theory: it
   is the $\mathbb{Z}_k$ projection of the $\Phi_1\Phi_2\Phi_3$ deformation
   appearing there. For general {\bf k} the other $\Phi^3$ deformation
   does not survive the orbifolding, however for the special {\bf k=3}
   case it does and we indeed see it in the reduced SUSY cases.
   From the analysis of~\cite{a:N_plan1,a:N_plan2} we know that the marginal
   deformations which survive the orbifold projection are known
   to be \textit{exactly} marginal at far as only the planar
   diagrams are concerned, however what we find is that even the
   non-planar diagrams don't prevent the surviving
   deformations from being \textit{exactly} marginal.

The more delicate point is with the marginal operators of the form
$\Phi_i^3$ in the general {\bf k} case, we have to deal with them
in each case when they appear independently. These deformations
are marginal and from Leigh-Strassler analysis some of them are expected to be
\textit{exactly} marginal. However we saw that they are prevented
from appearing at one loop (see sections
(~\ref{gen_kk},~\ref{k_3case}). Their fate has to be decided from
higher loop calculations.

   Another interesting observation is that in the ${\cal N}=2$ case the
   orbifold keeps a direction in $\mathbb{C}^3$ fixed, thus because the
   orbifold acts on the $S^5$ factor of the ${\bf AdS_5\times S^5}$ as it
   acts on the angular coordinates of the $\mathbb{R}^6\sim
   \mathbb{C}^3$, we will get fixed points on the sphere (actually a
   fixed circle). This enables the
    appearance of massless twisted sector states which can
   correspond to some \textit{exactly} marginal operators on the field theory side. And
   we see these twisted states $\to$ the {\bf (k-1)} blow up
   modes. Another case of \textit{exactly} marginal operators coming from the twisted sector are for example
   the {\bf 2} additional operators we get in the $\mathbb{Z}_3$ case.

   In the ${\cal N}=1$ case the only fixed point of the $\mathbb{Z}_k$ action is the
   origin of the $\mathbb{C}^3$. However we still can have massless
   twisted sector states. Remember that twisted sector strings are
   defined as strings which are closed up to the action of the twisting group
   $\Gamma$ ($\Gamma=\mathbb{Z}_k$ in our case). There are
   $\|\Gamma\|$ twisted sectors defined by every element of the
   twisting group. In our case the action of the orbifold is defined
   by vector $\overrightarrow{a}$:

   \begin{eqnarray}
  \wp \equiv
   \left(\begin{array}{ccc}
e^{{2\pi i\over k}a_1} &0  &0  \\
0 &e^{{2\pi i\over k}a_2}  &0  \\
0 &0  & e^{{2\pi i\over k}a_3}
\end{array}\right)
\end{eqnarray}

Now, assume $a_1$ has a largest common divisor larger than one
(say $\alpha$) with {\bf k}. Then, if we start with vector the
(1,0,0) we will get back to our starting point after ${k\over
\alpha}$ applications of ${\bf \wp}$. So, the
  $\wp^{k\over\alpha}$th twisted sector (which is not the identity
  $\equiv$ the untwisted sector)  has fixed points and could include massless
    states. There are $\alpha-1$ twisted sectors with massless
    states, and we find that each contributes one state which
    corresponds to an exactly
    marginal operator. Thus over all we have
    $\sum_i\alpha_i-3$ massless states from the twisted sector.

    Thus, we conclude that in the ${\cal N}=1$ case we can get massless
    states from the twisted sectors when the $a_i$ have non trivial
    largest common divisors with {\bf k} $\to$ and this is what we
    get. We got $\alpha+\beta+\gamma-3$
    marginal deformations that don't come from the ${\cal N}=4$
    theory $\to$ they have to come from the twisted sectors and we see
    that this is possible.

We see that the marginal deformations that we get are in agreement
with the string theory. The marginal deformations can be divided
to deformations that originate from the twisted sector ($\to$
there are always at least $\sum_i\alpha_i-3$ such deformations)
and to deformations which come from restrictions of the ${\cal
N}=4$ marginal deformations to orbifold group invariant parts. All
this is true for general SU({\bf N}) gauge group. However in
SU({\bf N}=3) we get a much larger space of deformations. These
deformations can not be related directly to the string theory,
because the field$\leftrightarrow$string theory correspondence is
well understood only in the large {\bf N} limit, if at all it is
true for finite {\bf N}.

The  $\sum_i\alpha_i-3$  \textit{exactly} marginal deformations
coming from the twisted sector are related to the ${\cal N}=2$
blow up modes. However we got several \textit{exactly} marginal
deformations which come from the twisted sector and are not
related to blow up modes: the {\bf 2} extra deformations of ${\cal
N}=2$ $\mathbb{Z}_3$ and the ${\bf (a,a,-2a)}$ extra deformations
for instance. These \textit{exactly} marginal deformations are
predictions of our analysis and their counterparts on the string
theory side have to be found. As was mentioned above the other
predictions are the SU({\bf N}=3) extra deformations, however they
are  hard to see in the string theory.

   We see that in each of the cases above we find a rich "zoo" of
   marginal deformations, giving a large set of conformal theories.
   From the {\bf AdS/CFT} correspondence, the \textit{exactly}
   marginal deformations should correspond to some fields on the
   string theory side, which are moduli of the theory with the SO(2,4) symmetry. The exact details
   of this correspondence have to be explored.

   The behavior of the marginal operators in the reduced SUSY cases
   under S-duality is another interesting question. We believe that
   there is an S-duality acting on these theories because we know that
   on the string theory side we have a type {\bf IIB} superstring on
   some background, which is believed to be self dual under S-duality.

\newpage

\appendix

\chapter{Three-loop calculation of $\gamma$}\label{Ap}

We calculate here the three-loop contribution to the $\gamma$
parameter as a function of $g^2$ and {\bf X}, which determine
$\alpha_i,\delta_i$ at leading order according to (~\ref{sol}). We
saw explicitly that at one loop order $\gamma$ is forced to be
zero. The one-loop solution assures us of having a two loop finite
theory, thus also at two loops $\gamma$ is zero. It is easy to
convince oneself of this. The only non finite two loop diagram,
not including one loop subdiagrams and  containing Yukawa type
interactions, is:

 
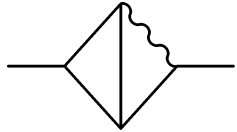
\begin{fmffile}{fifi}
 \unitlength=1mm
\begin{eqnarray}
\parbox{25mm}{
\begin{fmfchar}(30,30)
    \fmfpen{thin}
     \fmfleft{v1} \fmfright{v2} \fmftop{f1} \fmfbottom{f2}
     \fmf{phantom,tension=5}{f1,i1} \fmf{phantom,tension=5}{f2,i2}
     \fmf{vanilla,tension=2}{v1,o1}
     \fmf{vanilla,tension=2}{o2,v2}
    \fmf{vanilla}{o1,i1}\fmf{boson}{i1,o2}
    \fmf{vanilla}{o2,i2}\fmf{vanilla}{i2,o1}
    \fmf{vanilla}{i1,i2}
\end{fmfchar}
}
\end{eqnarray}
\end{fmffile}

 This diagram is proportional to $({N^2-1\over
N})^2(\alpha_i^2+\delta_i^2)(g_i^2+g_{i+1}^2)$ for the Q
propagator and to $N^2(N^2-1)(\alpha_i^2+\delta_{i-1}^2)g_i^2$ for
the $\Phi$ propagator. Both of these expressions don't contain the
{\bf X} parameter and thus they are the same as in the ${\cal
N}=2$ case, and there we know that the two-loop $\gamma$-function
vanishes.

 So the first non zero contribution to $\gamma$
is expected to appear at three loops. Three loops is also the
first order of appearance of non-planar diagrams. It was argued in
~\cite{a:N_plan1},~\cite{a:N_plan2} that the planar diagrams in
the orbifold theory are the same as in the ${\cal N}=4$ theory. So
we are tempted to assume that also when considering marginal
deformations the planar diagrams in both theories will be the same
(if the coupling of the different gauge groups are the same).
\footnote{The calculation of three loop $\beta$-function for
${\cal N}=4$ theory was done in ~\cite{a:Gris} and we will partly
use their results.}
 We  can assume
  that the $\gamma$ parameter is proportional to some power of the {\bf
  X}
  parameter, because when X=0 we are in the ${\cal N}=2$ case and we know that
  the anomalous dimensions vanish. Thus we are not interested in
  diagrams consisting of only gauge interactions. Also at three
  loops we can have diagrams with four gauge vertices and two
  matter vertices, these diagrams are proportional to
  $g_i^2g_{i+1}^2\alpha_i^2$ for instance.

  We argue that these diagrams don't contribute to the $\gamma$
  parameter. First we look at the Q propagator. At three loops the Q
  propagator superspace
diagrams are of a general structure:

$ $\\
$ $\\
$ $\\

\begin{figure}[htbp]
\begin{fmffile}{Q_lines}
\unitlength=1mm
\begin{center}
\begin{fmfchar*}(50,50)
    \fmfpen{thin}
    \fmfleft{v1} \fmfright{v2} \fmftop{o}
    \fmf{plain,tension=20,label=$Q$}{v1,a1} \fmf{plain,tension=20}{a1,a2} \fmf{plain,tension=20}{a2,a3}
    \fmf{plain,tension=20}{a3,v2}
    \fmfv{decor.shape=circle,decor.filled=empty,decor.size=400}{A}
    \fmf{phantom,tension=10}{o,A}
    \fmf{boson}{a1,A} \fmf{boson}{a2,A} \fmf{boson}{a3,A} \fmf{boson}{a4,A}
    \fmf{boson}{a5,A} \fmf{boson}{a6,A}
\end{fmfchar*}
\caption{A typical Q propagator}
\end{center}
\end{fmffile}
\end{figure}
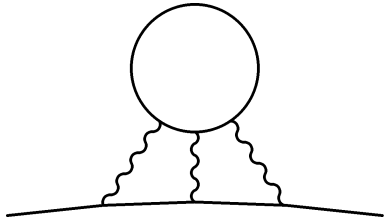

Here the horizontal line is a line of Q and $\tilde Q$ propagators
and the curled lines are the fields in adjoint ($\Phi$s and gauge
fields). By assumption we have only one $\Phi$ propagator and
assume it's $\Phi_i$. Then in case all six interactions happen on
the Q line we have a product of six matrices as the gauge factor.
Now, change $Q_i$ to $Q_{i-1}$ and $V_{i+1}$ to $V_{i-1}$. The
order of the product will be reversed because the fundamental
representation will be exchanged with the antifundamental, every
gauge vertex will receive a minus ( we have four such vertices, so
the overall sign won't change), and the diagram will have a factor
of $\delta_{i-1}^2g^2_{i,i-1}g^2_{i,i-1}$ if the original diagram
had $\alpha_{i}^2g^2_{i,i+1}g^2_{i,i+1}$. Now, if the diagram is
symmetric the reverse order of the product won't be important, and
so all the factors of these two diagrams are the same, and their
sum is proportional to
$\delta_{i-1}^2g^2_{i,i-1}g^2_{i,i-1}+\alpha_{i}^2g^2_{i,i+1}g^2_{i,i+1}$,
and thus after summation over {\bf i} the {\bf X} dependence
vanishes because $\alpha_i^2$ has {\bf +X} and $\delta_i^2$ has
{\bf -X}. If the diagram is not left right symmetric, there is
also its mirror image and we couple our new diagram with the
mirror image of the original one and obtain the same result.

If one of the gauge interactions happens not on the Q line we get
an extra factor of $f_{abc}$. Now reversing  the product changes
the sign of the expression, but we also now have an odd number of
gauge interactions on the Q line and so we have another sign flip
$\to$ again we find overall no X dependence. We can proceed in
this way by taking out the gauge vertices from the Q line and we
always get the same result.

 Now for the $\gamma_{\Phi_i}$ diagrams, here we have not a Q line but a Q loop and
 thus instead of a product of matrices we have a trace of such product,
 but these changes don't effect our considerations above because
 we always can take the trace in the end.

   Thus we
  conclude that the $\gamma$ parameter is proportional at least to
  the second power of {\bf X}, and comes from diagrams consisting of at
  most two gauge vertices.

  Now there are only four diagrams giving contributions
  that satisfy the conditions above (see ~\cite{a:Gris}):
$ $\\
$ $\\
$ $\\
$ $\\
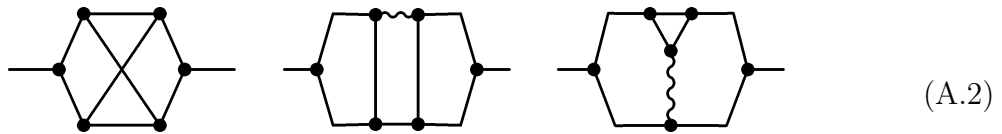
\begin{figure}[htbp]
\begin{fmffile}{rial_1}
\unitlength=1mm
\begin{eqnarray}\label{dags}
  \parbox{20mm}{
  \begin{fmfgraph}(30,30)
     \fmfpen{thin}
     \fmfleft{f1,v1,f2} \fmfright{f3,v2,f4}
     \fmf{phantom,tension=1}{f1,i1} \fmf{phantom,tension=1}{f3,i2}
     \fmf{phantom,tension=1}{f2,i3} \fmf{phantom,tension=1}{f4,i4}
     \fmf{vanilla,tension=1}{v1,o1} \fmf{vanilla}{o1,i1} \fmf{vanilla}{i1,i2}
     \fmf{vanilla}{i2,o2} \fmf{vanilla,tension=1}{o2,v2}
     \fmf{vanilla}{o1,i3} \fmf{vanilla}{i3,i4} \fmf{vanilla}{i4,o2}
     \fmf{vanilla,tension=0.01}{i1,i4} \fmf{vanilla,tension=0.01}{i2,i3}
     \fmfdot{i1} \fmfdot{i2} \fmfdot{i3} \fmfdot{i4}
     \fmfdot{o1} \fmfdot{o2}
  \end{fmfgraph}
  }\quad\quad\quad\quad
  \parbox{20mm}{
  \begin{fmfgraph}(30,30)
    \fmfpen{thin}
    \fmfleft{f1,v1,f2} \fmfright{f3,v2,f4}
    \fmf{phantom,tension=1}{f1,i1} \fmf{phantom,tension=1}{f3,i2}
    \fmf{phantom,tension=1}{f2,i3} \fmf{phantom,tension=1}{f4,i4}
    \fmf{vanilla,tension=1}{v1,o1} \fmf{vanilla}{o1,i1}
    \fmf{vanilla}{i1,s1} \fmf{vanilla}{s1,s2} \fmf{vanilla}{s2,i2}
     \fmf{vanilla}{i2,o2} \fmf{vanilla,tension=1}{o2,v2}
     \fmf{vanilla}{o1,i3} \fmf{vanilla}{i3,t1} \fmf{boson}{t1,t2}
     \fmf{vanilla}{t2,i4} \fmf{vanilla}{i4,o2}
     \fmf{vanilla,tension=0.01}{s1,t1} \fmf{vanilla,tension=0.01}{s2,t2}
     \fmfdot{t1} \fmfdot{t2} \fmfdot{s1} \fmfdot{s2} \fmfdot{o1} \fmfdot{o2}
  \end{fmfgraph}
  }\quad\quad\quad\quad
    \parbox{20mm}{
  \begin{fmfgraph}(30,30)
    \fmfpen{thin}
    \fmfleft{f1,v1,f2} \fmfright{f3,v2,f4}
    \fmf{phantom,tension=1}{f1,i1} \fmf{phantom,tension=1}{f3,i2}
    \fmf{phantom,tension=1}{f2,i3} \fmf{phantom,tension=1}{f4,i4}
    \fmf{vanilla,tension=1}{v1,o1} \fmf{vanilla}{o1,i1}
    \fmf{vanilla}{i1,s1} \fmf{vanilla}{s1,i2}
     \fmf{vanilla}{i2,o2} \fmf{vanilla,tension=1}{o2,v2}
     \fmf{vanilla}{o1,i3} \fmf{vanilla}{i3,t1} \fmf{vanilla}{t1,t2}
     \fmf{vanilla}{t2,i4} \fmf{vanilla}{i4,o2}
     \fmf{vanilla,tension=0.01}{t1,r} \fmf{vanilla,tension=0.01}{r,t2}
     \fmf{boson,tension=0.01}{r,s1}
     \fmfdot{t1} \fmfdot{t2} \fmfdot{s1} \fmfdot{r} \fmfdot{o1} \fmfdot{o2}
  \end{fmfgraph}
  }
\end{eqnarray}
\end{fmffile}
\end{figure}

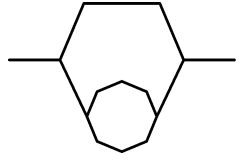
\begin{fmffile}{kuku}
\unitlength=1mm
\begin{eqnarray}\label{2_c}
\begin{fmfchar*}(30,30)
  \fmfpen{thin}
     \fmfleft{f1,v1,f2} \fmfright{f3,v2,f4}
     \fmf{phantom,tension=1}{f1,i1} \fmf{phantom,tension=1}{f3,i2}
     \fmf{phantom,tension=1}{f2,i3} \fmf{phantom,tension=1}{f4,i4}
     \fmf{vanilla,tension=1}{v1,o1} \fmf{vanilla}{o1,i1}
     \fmfpoly{circle,empty,label=$2$}{i1,h1,h2,h3,i2,q1,q2,q3}
     \fmf{vanilla}{i2,o2} \fmf{vanilla,tension=1}{o2,v2}
     \fmf{vanilla}{o1,i3} \fmf{vanilla}{i3,i4} \fmf{vanilla}{i4,o2}
\end{fmfchar*}
\end{eqnarray}
\end{fmffile}

Where:

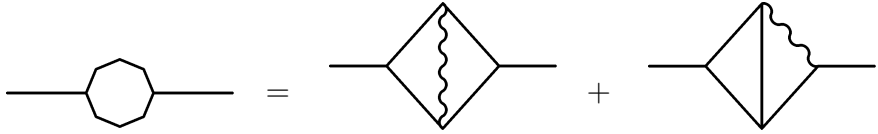
\begin{fmffile}{ro}
\unitlength=1mm
\begin{eqnarray}
\parbox{25mm}{
\begin{fmfchar*}(30,30)
  \fmfleft{v1} \fmfright{v2}
  \fmf{vanilla,tension=1}{v1,o1}\fmf{vanilla,tension=1}{o2,v2}
  \fmfpoly{circle,empty,label=$2$}{o1,h1,h2,h3,o2,q1,q2,q3}
\end{fmfchar*}
}\quad\quad =\quad
  \parbox{25mm}{
\begin{fmfchar}(30,30)
    \fmfpen{thin}
     \fmfleft{v1} \fmfright{v2} \fmftop{f1} \fmfbottom{f2}
     \fmf{phantom,tension=5}{f1,i1} \fmf{phantom,tension=5}{f2,i2}
     \fmf{vanilla,tension=2}{v1,o1}
     \fmf{vanilla,tension=2}{o2,v2}
    \fmf{vanilla}{o1,i1}\fmf{vanilla}{i1,o2}
    \fmf{vanilla}{o2,i2}\fmf{vanilla}{i2,o1}
    \fmf{boson}{i1,i2}
\end{fmfchar}
}\quad\quad +\quad
\parbox{25mm}{
\begin{fmfchar}(30,30)
    \fmfpen{thin}
     \fmfleft{v1} \fmfright{v2} \fmftop{f1} \fmfbottom{f2}
     \fmf{phantom,tension=5}{f1,i1} \fmf{phantom,tension=5}{f2,i2}
     \fmf{vanilla,tension=2}{v1,o1}
     \fmf{vanilla,tension=2}{o2,v2}
    \fmf{vanilla}{o1,i1}\fmf{boson}{i1,o2}
    \fmf{vanilla}{o2,i2}\fmf{vanilla}{i2,o1}
    \fmf{vanilla}{i1,i2}
\end{fmfchar}
}
\end{eqnarray}
\end{fmffile}

We begin with the first diagram.

 A calculation of a diagram consists of two main things: calculation
of the integrals and the symmetry factors and the calculation of
the algebra (gauge algebra, couplings etc.). The first part here
is the same for the Q propagator and for the $\Phi$ propagator and
we will neglect it. From the algebraic point of view: We calculate
$\sum_i\gamma_{Q_i}-{N^2-1\over N^2}\gamma_{\Phi_i}$, thus while
calculating $\sum_i\gamma_{Q_i}$ we can
  close the external legs of the diagram and multiply by $N^2$ (which is the gauge factor) and get the
desired quantity. When calculating $\sum_i\gamma_{\Phi_i}$  we can also close the external legs,
but now the gauge factor is $N^2-1$. Essentially we see that when we calculate $\sum_i\gamma_{\Phi_i}$ or
 $\sum_i\gamma_{Q_i}$ for the first diagram we calculate the same thing (when viewed as  closed diagrams in the sense above).
Thus, we get that
$N^2\sum_i\gamma_{Q_i}=(N^2-1)\sum_i\gamma_{\Phi_i}$  and  the
expression we want to calculate is zero.

  What remains to calculate now is the second and third  diagrams
  from (~\ref{dags}) and the diagram (~\ref{2_c}). The second and third
 diagrams from (~\ref{dags}) happen to be finite, thus we are left with the
 (~\ref{2_c})diagram. This
 diagram, although naively planar,
 is not necessarily such when considered in the double line
 notation. We will calculate it using the general three loop
 results given in ~\cite{a:beta}(see also ~\cite{a:parks}).

  In the notations of ~\cite{a:beta} we have a superpotential
  ${1\over6}Y_{ijk}\Theta_i\Theta_j\Theta_k$ and we define
  $C(R)^i_j\equiv(T^aT^a)^i_j$ where the $T^a$ are matrices of
  the representation {\bf R} of the gauge group. In our case we have k
  different gauge groups, but in the diagrams we are interested
  in,
  only one gauge group appears in each diagram, so we can use the
  results of ~\cite{a:beta} and simply sum over all gauge groups.
   In our case we have couplings of the form:

  \begin{eqnarray}\label{inter}
 Y\begin{pmatrix}a\\i\\\end{pmatrix}\begin{pmatrix}\alpha;l\\i\\\end{pmatrix}\begin{pmatrix}l;\gamma\\i\\\end{pmatrix}
 &\Phi_i^a&Q_{\alpha}^l\tilde Q_l^{\gamma}\nonumber\\
 &\downarrow&\\
  Y\begin{pmatrix}a\\i\\\end{pmatrix}\begin{pmatrix}\alpha;l\\i\\\end{pmatrix}\begin{pmatrix}l;\gamma\\i\\\end{pmatrix}
    &\equiv& \alpha_i(T^a)^{\alpha}_{\gamma}\nonumber\\
 Y\begin{pmatrix}a\\i+1\\\end{pmatrix}\begin{pmatrix}\alpha;l\\i\\\end{pmatrix}\begin{pmatrix}m;\alpha\\i\\\end{pmatrix}
    &\equiv& \delta_i(T^a)^m_l\nonumber
  \end{eqnarray}

 All other Y's vanish.

 \begin{eqnarray}
 (C(R)_Q)^i_j&=&{1\over2}{N^2-1\over N}\delta^i_j\nonumber\\
 (C(R)_{\Phi})^i_j&=&N\delta^i_j.
\end{eqnarray}

 In these notations in a one-loop and two-loop finite theory the
 three loop contribution to $\gamma$, proportional to $g^2$ is ${1\over(16\pi^2)^3}\Delta$, where:

 \begin{eqnarray}
 \Delta^i_j&\equiv&\kappa g^2 (YS_1Y)^i_j\nonumber\\
 S_{1j}^i&\equiv& Y^{imn}C(R)^p_mY_{jpn}\\
 (YS_1Y)^i_j&\equiv& Y^{imn}S^p_{1m}Y_{jpn}\nonumber\\
 \kappa&\equiv& 6\zeta(3)\nonumber.
 \end{eqnarray}

Here the indices represent each set of indices in (~\ref{inter}).
 Thus we should calculate the quantities above. We calculate for gauge group i and then sum over all gauge
 groups. First $S_1$ for a $\Phi$ index :

 \begin{eqnarray}
 (S_1)^{\Phi_i^a}_{\Phi_i^b}&=&
 2Y\begin{pmatrix}a\\i\\\end{pmatrix}\begin{pmatrix}\alpha;l\\i\\\end{pmatrix}\begin{pmatrix}l;\gamma\\i\\\end{pmatrix}C(R)_Q
 Y\begin{pmatrix}b\\i\\\end{pmatrix}\begin{pmatrix}\alpha;l\\i\\\end{pmatrix}\begin{pmatrix}l;\gamma\\i\\\end{pmatrix}+\nonumber\\
 &+&2Y\begin{pmatrix}a\\i\\\end{pmatrix}\begin{pmatrix}\alpha;l\\i-1\\\end{pmatrix}\begin{pmatrix}m;\alpha\\i-1\\\end{pmatrix}C(R)_Q
 Y\begin{pmatrix}b\\i\\\end{pmatrix}\begin{pmatrix}\alpha;l\\i-1\\\end{pmatrix}\begin{pmatrix}m;\alpha\\i-1\\\end{pmatrix}=\nonumber\\
 &=&2{1\over2}{N^2-1\over
 N}(\sum_l\alpha_i^2TrT^aT^b+\sum_{\alpha}\delta_{i-1}^2TrT^aT^b)=\nonumber\\
 &=&{1\over2}(N^2-1)(\alpha_i^2+\delta_{i-1}^2)\delta^a_b
\end{eqnarray}

 \begin{eqnarray}
 (S_1)^{\Phi_{i+1}^a}_{\Phi_{i+1}^b}&=&
2Y\begin{pmatrix}a\\i+1\\\end{pmatrix}\begin{pmatrix}\alpha;l\\i\\\end{pmatrix}\begin{pmatrix}m;\alpha\\i\\\end{pmatrix}C(R)_Q
 Y\begin{pmatrix}b\\i+1\\\end{pmatrix}\begin{pmatrix}\alpha;l\\i\\\end{pmatrix}\begin{pmatrix}m;\alpha\\i\\\end{pmatrix}=\nonumber\\
 &=&2{1\over2}{N^2-1\over N}\sum_\alpha\delta_i^2TrT^aT^b={1\over2}(N^2-1)\delta_i^2\delta^a_b
\end{eqnarray}

 \begin{eqnarray}
 (S_1)^{\Phi_{i-1}^a}_{\Phi_{i-1}^b}&=&
2Y\begin{pmatrix}a\\i-1\\\end{pmatrix}\begin{pmatrix}\alpha;l\\i-1\\\end{pmatrix}\begin{pmatrix}l;\gamma\\i-1\\\end{pmatrix}C(R)_Q
 Y\begin{pmatrix}b\\i-1\\\end{pmatrix}\begin{pmatrix}\alpha;l\\i-1\\\end{pmatrix}\begin{pmatrix}l;\gamma\\i-1\\\end{pmatrix}=\nonumber\\
 &=&2{1\over2}{N^2-1\over N}\sum_l\alpha_{i-1}^2TrT^aT^b={1\over2}(N^2-1)\alpha_{i-1}^2\delta^a_b
\end{eqnarray}

 There are two Qs coupled to gauge group i: $Q_i$ and $Q_{i-1}$, so we calculate $S_1$ for each one separately.
 First for $Q_i$:

 \begin{eqnarray}
 (S_1)^{(Q_i)^\alpha_l}_{(Q_i)^\beta_l}&=&
 Y\begin{pmatrix}\alpha;l\\i\\\end{pmatrix}\begin{pmatrix}a\\i\\\end{pmatrix}\begin{pmatrix}l;\gamma\\i\\\end{pmatrix}(C(R)_Q+C(R)_{\Phi})
 Y\begin{pmatrix}\beta;l\\i\\\end{pmatrix}\begin{pmatrix}a\\i\\\end{pmatrix}\begin{pmatrix}l;\gamma\\i\\\end{pmatrix}+\nonumber\\
 &+&Y\begin{pmatrix}\alpha;l\\i\\\end{pmatrix}\begin{pmatrix}a\\i+1\\\end{pmatrix}\begin{pmatrix}m;\alpha\\i\\\end{pmatrix}C(R)_Q
 Y\begin{pmatrix}\beta;l\\i\\\end{pmatrix}\begin{pmatrix}a\\i+1\\\end{pmatrix}\begin{pmatrix}m;\alpha\\i\\\end{pmatrix}=\nonumber\\
 &=&((N+{1\over2}{N^2-1\over N})\alpha_i^2T^aT^a+{1\over2}{N^2-1\over
 N}\delta_i^2T^aT^a)\delta^\alpha_\beta=\nonumber\\
 &=&(({1\over2}{N^2-1\over
 N})^2(\alpha_i^2+\delta_i^2)+{1\over2}(N^2-1)\alpha_i^2)\delta^\alpha_\beta
 \end{eqnarray}

 And the same for $Q_{i-1}$:

\begin{eqnarray}
 (S_1)^{(Q_{i-1})^\alpha_l}_{(Q_{i-1})^\beta_l}=
 (({1\over2}{N^2-1\over
 N})^2(\alpha_{i-1}^2+\delta_{i-1}^2)+{1\over2}(N^2-1)\delta_{i-1}^2)\delta^\alpha_\beta
 \end{eqnarray}

 Now we calculate $Y^*S_1Y$. First for $\Phi$ (We have contributions for the i'th
 gauge group for $\gamma_{\Phi_i},\gamma_{\Phi_{i+1}},\gamma_{\Phi_{i-1}}$):

 \begin{eqnarray}
(Y^*S_1Y)^{\Phi_{i}^a}_{\Phi_{i}^b}&=&
2Y\begin{pmatrix}a\\i\\\end{pmatrix}\begin{pmatrix}\alpha;l\\i\\\end{pmatrix}\begin{pmatrix}l;\gamma\\i\\\end{pmatrix}S_1^{Q_i}
 Y\begin{pmatrix}b\\i\\\end{pmatrix}\begin{pmatrix}\alpha;l\\i\\\end{pmatrix}\begin{pmatrix}l;\gamma\\i\\\end{pmatrix}+\nonumber\\
 &+&2Y\begin{pmatrix}a\\i\\\end{pmatrix}\begin{pmatrix}\alpha;l\\i-1\\\end{pmatrix}\begin{pmatrix}m;\alpha\\i-1\\\end{pmatrix}S_1^{Q_{i-1}}
 Y\begin{pmatrix}b\\i\\\end{pmatrix}\begin{pmatrix}\alpha;l\\i-1\\\end{pmatrix}\begin{pmatrix}m;\alpha\\i-1\\\end{pmatrix}=\nonumber\\
 &=&(({1\over2}{N^2-1\over
 N})^2(\alpha_i^2+\delta_i^2)+{1\over2}(N^2-1)\alpha_i^2)N\alpha_i^2\delta^a_b+\nonumber\\
 &+&(({1\over2}{N^2-1\over
 N})^2(\alpha_{i-1}^2+\delta_{i-1}^2)+{1\over2}(N^2-1)\delta_{i-1}^2)N\delta_{i-1}^2\delta^a_b
 \end{eqnarray}

 \begin{eqnarray}
   (Y^*S_1Y)^{\Phi_{i+1}^a}_{\Phi_{i+1}^b}&=&
2Y\begin{pmatrix}a\\i+1\\\end{pmatrix}\begin{pmatrix}\alpha;l\\i\\\end{pmatrix}\begin{pmatrix}m;\alpha\\i\\\end{pmatrix}S_1^{Q_i}
Y\begin{pmatrix}b\\i+1\\\end{pmatrix}\begin{pmatrix}\alpha;l\\i\\\end{pmatrix}\begin{pmatrix}m;\alpha\\i\\\end{pmatrix}=\nonumber\\
&=&(({1\over2}{N^2-1\over
 N})^2(\alpha_i^2+\delta_i^2)+{1\over2}(N^2-1)\alpha_i^2)N\delta_i^2\delta^a_b
\end{eqnarray}

 \begin{eqnarray}
   (Y^*S_1Y)^{\Phi_{i-1}^a}_{\Phi_{i-1}^b}&=&
2Y\begin{pmatrix}a\\i-1\\\end{pmatrix}\begin{pmatrix}\alpha;l\\i-1\\\end{pmatrix}\begin{pmatrix}l;\gamma\\i-1\\\end{pmatrix}S_1^{Q_{i-1}}
Y\begin{pmatrix}b\\i-1\\\end{pmatrix}\begin{pmatrix}\alpha;l\\i-1\\\end{pmatrix}\begin{pmatrix}l;\gamma\\i-1\\\end{pmatrix}=\nonumber\\
&=&(({1\over2}{N^2-1\over
 N})^2(\alpha_{i-1}^2+\delta_{i-1}^2)+{1\over2}(N^2-1)\delta_{i-1}^2)N\alpha_{i-1}^2\delta^a_b.
\end{eqnarray}

 From here we calculate the contribution of such diagrams to
$\sum_i\gamma_{\Phi_i}$, by multiplying the expression above by
$g_i^2$ and summing over {\bf i}, and because we are interested
only in the part proportional to $X^2$ and X we insert the
expressions (~\ref{sol}). We get:

 \begin{equation}
  \sum_i\gamma_{\Phi_i}|_{X^2}\propto
  N(N^2-1)(X^2-X^2)\sum_ig^2_i=0
 \end{equation}

 \begin{eqnarray}
\sum_i\gamma_{\Phi_i}|_{X}&\propto&\sum_i(({1\over2}{N^2-1\over
 N})^2((\alpha_i^2+\delta_i^2)(\alpha_i^2g_i^2+\delta_i^2g_{i+1}^2)+\nonumber\\
 &+&(\alpha_i^2+\delta_i^2)(\alpha_i^2g_{i+1}^2+\delta_i^2g_{i}^2))+{1\over2}N(N^2-1)\alpha_i^2\delta_i^2(g_{i+1}^2+g_{i}^2))=\nonumber\\
 &\propto&\sum_i(({1\over2}{N^2-1\over
 N})^2(g_{i+1}^2+g_{i}^2)(X(g_i^2-g_{i+1}^2)+g_i^2g_{i+1}^2(X-X))+\nonumber\\
 &+&{1\over2}N(N^2-1)(X(g_{i+1}^2+g_{i}^2)(g_{i+1}^2-g_{i}^2)))=0.
 \end{eqnarray}

Essentially the vanishing of $\sum_i\gamma_{\Phi_i}|_{X}$ can be
concluded by a similar argument to the one we presented in the
beginning of the section, so this calculation can be seen as an
explicit check of that argument.

 Now doing the same for the Q part we obtain:

\begin{eqnarray}
(Y^*S_1Y)^{(Q_{i})^\alpha_l}_{(Q_{i})^\beta_m}&=&
Y\begin{pmatrix}\alpha;l\\i\\\end{pmatrix}\begin{pmatrix}a\\i\\\end{pmatrix}\begin{pmatrix}l;\gamma\\i\\\end{pmatrix}(S_1^{Q_i}+S_1^{\Phi_i})
 Y\begin{pmatrix}\beta;l\\i\\\end{pmatrix}\begin{pmatrix}a\\i\\\end{pmatrix}\begin{pmatrix}l;\gamma\\i\\\end{pmatrix}+\nonumber\\
 &+&Y\begin{pmatrix}\alpha;l\\i\\\end{pmatrix}\begin{pmatrix}a\\i+1\\\end{pmatrix}\begin{pmatrix}m;\alpha\\i\\\end{pmatrix}(S_1^{Q_i}+S_1^{\Phi_{i+1}})
 Y\begin{pmatrix}\beta;l\\i\\\end{pmatrix}\begin{pmatrix}a\\i+1\\\end{pmatrix}\begin{pmatrix}m;\alpha\\i\\\end{pmatrix}=\nonumber\\
 &=&({1\over2}{N^2-1\over N})^3(\alpha_i^2+\delta_i^2)^2\delta^l_m\delta^\alpha_\beta+N(({1\over2}{N^2-1\over
 N})^2(\alpha_i^2(2\alpha_i^2+\delta_i^2+\delta_{i-1}^2)+\delta_i^4)\delta^l_m\delta^\alpha_\beta\nonumber\\
\end{eqnarray}

And similarly for $Q_{i-1},Q_{i-2},,Q_{i+1}$:

\begin{eqnarray}
(Y^*S_1Y)^{(Q_{i-1})^\alpha_l}_{(Q_{i-1})^\beta_m}&=&({1\over2}{N^2-1\over
N})^3(\alpha_{i-1}^2+\delta_{i-1}^2)^2\delta^l_m\delta^\alpha_\beta+\nonumber\\&+&N(({1\over2}{N^2-1\over
 N})^2(\delta_{i-1}^2(\alpha_i^2+2\delta_{i-1}^2+\alpha_{i-1}^2)+\alpha_{i-1}^4)\delta^l_m\delta^\alpha_\beta
 \end{eqnarray}

\begin{eqnarray}
(Y^*S_1Y)^{(Q_{i-2})^\alpha_l}_{(Q_{i-2})^\beta_m}&=&N({1\over2}{N^2-1\over
 N})^2\delta_{i-2}^2\alpha_{i-1}^2\delta^l_m\delta^\alpha_\beta
 \end{eqnarray}

 \begin{eqnarray}
(Y^*S_1Y)^{(Q_{i+1})^\alpha_l}_{(Q_{i+1})^\beta_m}&=&N({1\over2}{N^2-1\over
 N})^2\delta_{i}^2\alpha_{i+1}^2\delta^l_m\delta^\alpha_\beta
 \end{eqnarray}

Again we calculate $\sum_i\gamma_{Q_i}$ taking only the $X^2$
term:

\begin{eqnarray}
  \sum_i\gamma_{Q_i}|_{X^2}&\propto&
  N({N^2-1\over
  N})^2\sum_ig_i^2(\alpha_i^2(2\alpha_i^2+\delta_i^2+\delta_{i-1}^2)+\nonumber\\
   &+&\delta_{i-1}^2(\alpha_i^2+2\delta_{i-1}^2+\alpha_{i-1}^2)+\alpha_{i-1}^4+\delta_i^4+\delta_{i-2}^2\alpha_{i-1}^2
   +\delta_{i}^2\alpha_{i+1}^2)=\nonumber\\
&=& N({N^2-1\over
N})^2\sum_i(2(g_i^2\alpha_i^4+g_{i+1}^2\delta_i^4)+2g_i^2\alpha_i^2\delta_{i-1}^2+\nonumber\\
&+&(g_i^2+g_{i+1}^2)\alpha_i^2\delta_i^2+g_{i+1}^2\alpha_i^4+g_i^2\delta_i^4+g_i^2\delta_{i-2}^2\alpha_{i-1}^2+
g_i^2\delta_{i}^2\alpha_{i+1}^2)=\nonumber\\
&\to& N({N^2-1\over N})^2X^2\sum_ig_i^2(6-6)=0
\end{eqnarray}

The same calculation for the term proportional to {\bf X}  also
gives zero. Thus we conclude that for these diagrams and at three
loop order:

\begin{equation}
  \sum_i\gamma_{Q_i}-{N^2-1\over N^2}\gamma_{\Phi_i}=0.
\end{equation}

So the $\gamma$ parameter is zero up to three loops and we can not
turn on any non zero $h_i$s.

\newpage

\chapter{$SU(2)^k$ global symmetry}\label{apB}

 We will show here how one can use the $SU(2)^k$ global symmetry
 of the {\bf (a,a,-2a)} orbifold to get rid of the  $h'_I$s.

First we notice that the interactions we have can be written as:

\begin{eqnarray}
  \sum_IQ^{I+2a}_3(h^IQ_1^IQ^{I+a}_2+h^{'I+a}Q_2^IQ_1^{I+a}+p^IQ_1^IQ_1^{I+a}+s^IQ_2^IQ_2^{I+a}).
\end{eqnarray}

This can be rewritten as:

\begin{eqnarray}
\sum_IQ^{I+2a}_3
\begin{pmatrix}
  Q^I_1 &
  Q^I_2
\end{pmatrix}
\begin{pmatrix}
  p^I & h^I \\
  h^{'I+a} & s^I
\end{pmatrix}
\begin{pmatrix}
  Q^{I+a}_1 \\ Q^{I+a}_2
\end{pmatrix}.
\end{eqnarray}

By the global $SU(2)^k$ symmetry we can rotate $Q^I_1$ and $Q^I_2$
one into the other. Let's define:

\begin{eqnarray}
H^I\equiv \begin{pmatrix}
  p^I & h^I \\
  h^{'I+a} & s^I
\end{pmatrix}.
\end{eqnarray}

We will rotate the {\bf I}th set of Q's with a unitary matrix
$U^I$. Then the coupling matrix $H^I$  transforms as
$(U^I)^{-1}H^IU^{I+a}$.

Our goal is to get rid of the $h'_I$s, obviously we can get rid of
at least {\bf k} such couplings.

Now let's look at:

\begin{eqnarray}\label{expres}
(U^{I_0})^{-1}H^{I_0}U^{I_0+a}(U^{I_0+a})^{-1}H^{I_0+a}U^{I_0+2a}\cdots(U^{I_0+(k-1)a})^{-1}H^{I_0+(k-1)a}U^{I_0}
=(U^{I_0})^{-1}\prod_{l=0}^{k-1}H^{I_0+la}U^{I_0}\nonumber\\
\end{eqnarray}

Define: $U^{I_0+la}\equiv U_{l+1}$, $H^{I_0+la}\equiv H^{l}$,
$p_{I_0+la}\equiv p_l$, $s_{I_0+la}\equiv s_l$ .

Let's set $U_1$ so that:

\begin{eqnarray}\label{U11}
U_1^{-1}\prod_{l=0}^{k-1}H^{I+la}U_1=\begin{pmatrix}
  * & * \\
  0 & *
\end{pmatrix}
\end{eqnarray}

(It does not matter what the stars represent.) It is easy to
convince oneself that if we choose to be close to the orbifold
theory (see section (~\ref{aa})) we get that in the even {\bf k}
case:

\begin{eqnarray}
U_1^{-1}&\prod&_{l=0}^{k-1}H^{I+la}U_1=\nonumber\\&=&U_1^{-1}(-1)^{k\over2}\begin{pmatrix}
  h^k+{\cal O}(\epsilon) & h^{k-1}(\sum_{odd}p_l+\sum_{even}s_l)+{\cal O}(\epsilon^2)  \\
  -h^{k-1}(\sum_{odd}s_l+\sum_{even}p_l)+{\cal O}(\epsilon^2) & h^k+{\cal O}(\epsilon)\nonumber\\
\end{pmatrix}U_1
\end{eqnarray}

And in the odd {\bf k} case:

\begin{eqnarray}
U_1^{-1}&\prod&_{l=0}^{k-1}H^{I+la}U_1=\nonumber\\&=&U_1^{-1}(-1)^{k-1\over2}\begin{pmatrix}
   h^{k-1}(\sum_{odd}p_l+\sum_{even}s_l)+{\cal O}(\epsilon^2) & h^k+{\cal O}(\epsilon) \\
  -h^k+{\cal O}(\epsilon) & h^{k-1}(\sum_{odd}s_l+\sum_{even}p_l)+{\cal O}(\epsilon^2)\nonumber\\
\end{pmatrix}U_1
\end{eqnarray}

We see that the determinant of this expression is obviously non
zero ($h\gg s,p$), and thus on the right hand side of (~\ref{U11})
we don't get zeros on the diagonal.

 We will choose $U_{l+1}$ for $1\leq l< {k}$ so that:

 \begin{eqnarray}
U_{l}^{-1}H^{l}U_{l+1}=\begin{pmatrix}
  * & * \\
  0 & *
\end{pmatrix}.
 \end{eqnarray}

Here the point is that:

\begin{eqnarray}
\begin{pmatrix}
  * & * \\
  0 & *
\end{pmatrix}
\begin{pmatrix}
  * & * \\
  0 & *
\end{pmatrix}=
\begin{pmatrix}
  * & * \\
  0 & *
\end{pmatrix}
\end{eqnarray}

So that (~\ref{expres}) can be written as:

\begin{eqnarray}
\begin{pmatrix}
  * & * \\
  0 & *
\end{pmatrix}U_k^{-1}H_kU_1=\begin{pmatrix}
  * & * \\
  0 & *
\end{pmatrix}
\end{eqnarray}

And from here we easily see that:

\begin{eqnarray}
U_k^{-1}H_kU_1=\begin{pmatrix}
  * & * \\
  0 & *
\end{pmatrix}
\end{eqnarray}

And thus we succeeded in getting rid of all $h'_l$.

{}
\end{document}